\documentclass[aps,nofootinbib,12pt,tightenlines]{revtex4-1}

\usepackage[dvips]{graphicx}

\usepackage{amsmath,amssymb,mathrsfs}

\usepackage{bm}
\usepackage[hyperfootnotes=false]{hyperref}
\usepackage{times}
\usepackage{epsfig}
\usepackage{verbatim}
\usepackage{slashed}

\usepackage{graphics}
\usepackage{graphicx,epsfig,amssymb,amsmath,color,cancel,feyn,relsize}
\usepackage{placeins}
\usepackage{float}

\newcommand{\be}{\begin{equation}}
\newcommand{\ee}{\end{equation}}
\newcommand{\ba}{\begin{array}}
\newcommand{\ea}{\end{array}}
\newcommand{\bea}{\begin{eqnarray}}
\newcommand{\eea}{\end{eqnarray}}
\newcommand{\aem}{\alpha_{\rm em}}

\newcommand{\Pom}{\mathbb{P}}
\newcommand{\sigmap}{\sigma^{(+)}}
\newcommand{\sigmam}{\sigma^{(-)}}
\newcommand{\sigmapm}{\sigma^{(\pm)}}

\begin{document}

\title{Baryonic dark forces in electron-beam fixed-target experiments}

\author{Safa Ben Othman}

\author{Armita Jalooli}

\author{Sean Tulin}
\email{stulin@yorku.ca}

\affiliation{Department of Physics and Astronomy, York University, Toronto, ON, Canada, M3J 1P3}

\date{\today}

\begin{abstract}

New GeV-scale dark forces coupling predominantly to quarks offer novel signatures that can be produced directly and searched for at high-luminosity colliders.
We compute the photon-proton and electron-proton cross sections for producing a GeV-scale gauge boson arising from a $U(1)_B$ gauge symmetry.
Our calculation relies on vector meson dominance and a phenomenological model for diffractive scattering used for vector-meson photoproduction.
The parameters of our phenomenological model are fixed by performing a Markov Chain Monte Carlo fit to existing exclusive photoproduction data for $\omega$ and $\phi$ mesons.
Our approach can be generalized to other GeV-scale dark gauge forces.

\end{abstract}

\date{\today}

\maketitle

\section{Introduction}

Luminosity-frontier experiments have a unique niche for discovering new gauge forces that are light and weakly-coupled to the Standard Model (SM).
These searches have been motivated in part by the muon $(g\!-\!2)$ anomaly~\cite{Boehm:2003hm,Fayet:2007ua,Pospelov:2008zw}, the ATOMKI anomalies~\cite{Krasznahorkay:2015iga,Krasznahorkay:2019lyl}, and dark matter~\cite{Boehm:2003hm,Pospelov:2007mp,ArkaniHamed:2008qn,Pospelov:2008jd,Feng:2009mn,Hooper:2012cw}. 
Irrespective of particular anomalies, however, it is important to explore all possible ways new forces may couple to the SM. 
Most searches have focused on a dark photon arising via kinetic mixing~\cite{Holdom:1985ag}. 
Since the dark photon couples to the electromagnetic current, many experiments rely on its leptonic couplings to search for $e^+ e^-$ or $\mu^+ \mu^-$ resonances. 
Alternative approaches are also being pursued in case dark photons decay invisibly to light dark matter. 
(See~\cite{Jaeckel:2010ni,Essig:2013lka,Alexander:2016aln} for reviews.)

Different search strategies are needed to discover a leptophobic gauge force that couples predominantly to quarks. 
The simplest model is the $B$ boson that arises from a gauged $U(1)_B$ baryon number symmetry~\cite{Rajpoot:1989jb,Foot:1989ts,Nelson:1989fx,He:1989mi,Carone:1994aa,Bailey:1994qv,
Carone:1995pu,Aranda:1998fr,FileviezPerez:2010gw,Graesser:2011vj}. 
The interaction Lagrangian is
\be \label{eq:Lint}
\mathscr{L}_{\rm int} =  \left( \tfrac{1}{3} g_B + \varepsilon Q_q e \right) \bar q \gamma^\mu q B_\mu 
-\varepsilon e  \bar \ell \gamma^\mu \ell B_\mu  \, ,
\ee
where $g_B$ is the $U(1)_B$ gauge coupling. 
We also include dark-photon-like couplings where $\varepsilon$ is the kinetic-mixing parameter, $e$ is the usual proton electric charge, and $Q_q$ denotes quark electric charges in units of $e$.
Even if $\varepsilon=0$ at tree-level, a nonzero value arises via radiative corrections from heavy quarks~\cite{Carone:1995pu}.
Depending on $\varepsilon$, the $B$ boson is subject to many constraints from dark photon searches~\cite{Ilten:2018crw} and flavor physics~\cite{Dror:2017ehi,Dror:2017nsg}.
Since these effects are somewhat model-dependent~\cite{Gan:2020aco}, here we focus only on the leptophobic coupling $g_B$, i.e., taking the limit $g_B \gg \varepsilon e$.

In this case, the dominant $B$-boson decays are
\be
B \to \pi^0 \gamma \; , \quad B \to \pi^+ \pi^- \pi^0
\ee
for $m_B$ in the ranges 140$-$620 MeV and 620 MeV$-$1 GeV, respectively~\cite{Tulin:2014tya}. 
The subleading decay $B \to \pi^+ \pi^-$ is forbidden by $G$-parity but can proceed via isospin-breaking, e.g., due to $\rho$-$\omega$ mixing.
When $m_B < m_{\pi}$, we expect $B \to e^+ e^-$ via the small but nonzero kinetic mixing.

One discovery strategy is searching for leptophobic gauge bosons in light meson decays at meson factories.
For the $B$ boson, the two most promising channels are 
\be
\eta \to B \gamma \to \pi^0 \gamma \gamma \, , \quad \phi \to \eta B \to \eta \pi^0 \gamma \, .
\ee
These processes mimic rare decays in the SM~\cite{Nelson:1989fx,Tulin:2014tya} and are search targets for the Jefferson Eta Factory~\cite{Gan:2015nyc} and KLOE-2~\cite{AmelinoCamelia:2010me,delRio:2021xag}.
Belle has also searched for and found a null result for $\eta \to B \gamma \to \pi^+ \pi^- \gamma$~\cite{Won:2016pjz}.
However, the disadvantage of these searches is that the $B$-boson mass reach is limited by available phase space from the parent meson, $\sim 500 \; {\rm MeV}$.
Above this range, dijet decays in heavy-flavor quarkonia provide the strongest constraints~\cite{Aranda:1998fr}.

An alternative strategy is searching for new gauge bosons produced directly in collisions.
Naturally, the advantage is that more phase space is available (depending on beam energy) and more massive gauge bosons can be searched for.
In this work, we calculate the $B$-boson cross section via real and virtual photoproduction 
\be \label{eq:photoprod_process}
\gamma^{(*)} p \to B p \, ,
\ee
which can be searched for as a narrow $\pi^0 \gamma$ or $\pi^+ \pi^- \pi^0$ resonance in electron-proton collisions. (For simplicity, we consider a proton target only.)

The high-luminosity frontier for electron-proton collisions is of great interest for our understanding of strong dynamics and the structure of hadrons.
This includes fixed-target experiments, e.g., the current 12-GeV Continuous Electron Beam Accelerator Facility (CEBAF) at Jefferson Laboratory~\cite{Arrington:2021alx} and its proposed 22-GeV upgrade~\cite{Accardi:2023chb}, as well as various proposed electron-hadron colliders~\cite{Accardi:2012qut,LHeC:2020van,Bruning:2022hro}.
Here we focus on electron-beam fixed-target experiments and their potential for uncovering new leptophobic gauge bosons hidden in QCD.

Unfortunately, it is not possible to calculate photoproduction from first principles.
At large center-of-mass energies, $\sqrt{s} \gg {\rm GeV}$, photon-hadron interactions share the same soft diffractive behavior as purely hadronic interactions: namely, cross sections that grow weakly with $\sqrt{s}$ and are dominated by small momentum transfer, $\sqrt{|t|} \lesssim$ GeV~\cite{Bauer:1977iq}.\footnote{See also ``High Energy Soft QCD and Diffraction'' in the {\it Review of Particle Physics}~\cite{Workman:2022ynf} and references therein.}
Thus, we are led to a phenomenological model based on two assumptions.
First, the similarity between photon-hadron and hadron-hadron collisions points to the secretly hadronic nature of the photon. 
This is well-described by vector meson dominance (VMD), which assumes that external gauge fields couple to hadrons by mixing with light vector mesons (see e.g.~\cite{Bauer:1977iq,Schildknecht:2005xr}).
Second, the common scaling behavior observed in different hadronic processes points to a universal model to describe diffractive scattering.
This is well-described by the soft pomeron model~\cite{Donnachie:1984xq}.

We expect $B$-boson physics to follow similarly, provided $m_B \lesssim {\rm GeV}$.
Here we construct a phenomenological model for \eqref{eq:photoprod_process} based on the same physics used in the literature for describing photon-proton collisions in the Standard Model.
Our model has many phenomenological parameters, some of which we obtain from the literature.
The remainder we determine by calculating the photoproduction cross sections for $\omega,\phi$ mesons within our model and fitting them to experimental data.
Our fit spans center-of-mass energies $\sqrt{s} \sim 2-94 \; {\rm GeV}$, covering the full range of presently available data except for the threshold region ($\sqrt{s} \sim 1-2\; {\rm GeV})$.
Ultimately, we are able to predict the $B$-boson photoproduction cross section solely in terms of the new physics parameters $m_B$ and $\alpha_B = g_B^2/(4\pi)$.
The methods used can be adapted to other leptophobic gauge models as well, which we briefly discuss at the end.

Finally, we note related work by Fanelli and Williams~\cite{Fanelli:2016utb} that previously calculated the $B$-boson photoproduction cross section at fixed-target experiments.
Their calculation and ours both rely on VMD to model $B$-boson couplings to hadrons. 
However, they parametrize the remaining part of the amplitude simply in terms of SM cross sections for $\gamma p \to \omega p, \phi p$, whereas our work provides a direct calculation that is based on current theoretical models in the literature and calibrated to existing photoproduction data.
We also consider virtual photons in the initial state, which is needed to describe $B$-boson electroproduction.

The remainder of this work is organized as follows.
Sec.~\ref{sec:model} describes the phenomenological model used in our calculation.
Sec.~\ref{sec:fit} gives the results of our fit to $\omega,\phi$-meson photoproduction data, used to fit unknown parameters in our model.
Secs.~\ref{sec:Bboson} and \ref{sec:electro} present our results for $B$-bosons produced via real and virtual photoproduction, respectively. 
We also provide a comparison to the results of Ref.~\cite{Fanelli:2016utb}.
Our conclusions follow.
Additional material is provided in the appendices.

Analytic formulae in this work have been coded up in \texttt{Python} and are available at \url{https://github.com/dark-physics/baryonic-dark-forces}.


\section{Photoproduction model}
\label{sec:model}

To calculate the photoproduction cross section for the $B$ boson, we first use VMD to expand the matrix element as
\be \label{eq:cV}
\mathcal{M}(\gamma p \to B p) = \sum_{V=\rho^0,\omega,\phi} g_B\,  c_V(m_B^2) \, \mathcal{M}(\gamma p \to V p)
\ee
in terms of SM matrix elements for vector-meson photoproduction.
The coefficients $c_V$ are calculated below and neglecting isospin-violating effects (both in the SM and due to kinetic mixing) we need only consider $\omega,\phi$ mesons.

Next, we calculate the SM matrix elements for vector-meson photoproduction based on a phenomenological model, following previous literature~\cite{Berman:1963ef,Joos1964,Fraas:1972fj,Friman:1995qm,Zhao:1998fn,Oh:2000zi,Ewerz:2013kda}.
For $\omega$-photoproduction, it is long-known that the cross section is dominated by one-pion-exchange at low energies and diffractive scattering at high energies~\cite{Berman:1963ef,Fraas:1972fj}. 
For $\phi$-photoproduction, meson exchange is suppressed and diffractive scattering dominates~\cite{Barger:1970wk,Halpern:1972ab}.
In our setup, we take a $t$-channel model that includes exchange of both light pseudoscalars ($\pi^0$, $\eta$) and the pomeron.

It is customary to decompose the total matrix element as a sum of natural ($+$) and unnatural ($-$) parity amplitudes, i.e., 
\be
\mathcal{M}(\gamma p \to V p) = \mathcal{M}_+(\gamma p \to V p) + \mathcal{M}_-(\gamma p \to V p) \, ,
\ee
which in our setup arise, respectively, from pomeron ($+$) and pseudoscalar ($-$) exchange.
These contributions can be separately extracted from photoproduction with polarized photons~\cite{Schilling:1969um}.
Here we consider the unpolarized differential cross section
\be
\frac{d\sigma(\gamma p \to Vp)}{dt} = \frac{d\sigmap(\gamma p \to Vp)}{dt} + \frac{d\sigmam(\gamma p \to Vp)}{dt}
\ee
where 
\be
\frac{d\sigmapm(\gamma p \to Vp)}{dt} = \frac{1}{64\pi s^2} \sum_{\rm spins} \big|\mathcal{M}_\pm(\gamma p \to Vp)\big|^2 \, .
\ee
The two contributions do not interfere with one another in the limit $\sqrt{s} \gg m_p$.
Analogous formulae hold for $\gamma p \to B p$ as well.

\begin{figure}[t]
\begin{center}
\includegraphics[scale=0.75]{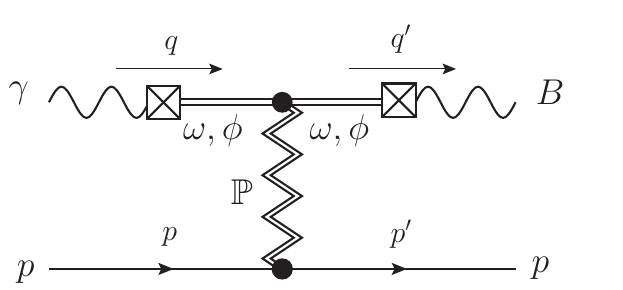}
\includegraphics[scale=0.75]{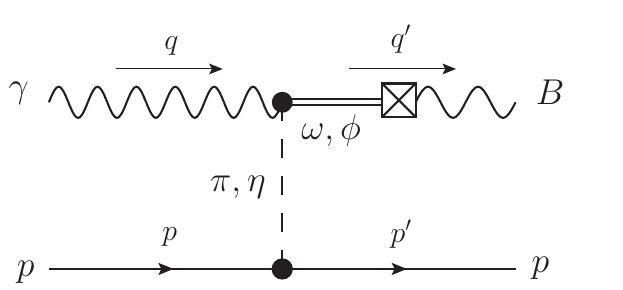}
\end{center}
\caption{\it Feynman diagrams are shown for $B$-boson photoproduction, with four-momenta as labeled. Left panel shows diffractive contribution to natural-parity amplitude from pomeron ($\Pom$) exchange. Right panel shows one-pseudoscalar-exchange contribution to unnatural-parity amplitude.
Crossed boxes ($\boxtimes$) denote mixing a la VMD. 
Dots ($\bullet$) denote vertices parametrized in our model.}
\label{fig:photo_feynman_diagram}
\end{figure}

Fig.~\ref{fig:photo_feynman_diagram} shows the Feynman diagrams for $B$-boson photoproduction in our $t$-channel model.
The left diagram is the natural-parity amplitude from pomeron exchange (jagged line), while the right diagram is the unnatural-parity amplitude from pseudoscalar-meson exchange (dashed line).
Black dots denote phenomenological vertices in our model, which we determine from experimental data.
Crossed boxes denote mixing between vector mesons and external gauge fields, a la VMD.
Without VMD mixing with the $B$ boson, the same diagrams and interactions correspond to $\omega,\phi$-meson photoproduction, which we also calculate.
In the remainder of this section, we go through the calculations in detail.

\subsection{Vector meson dominance}

Though VMD is well-known~\cite{Bauer:1977iq,Schildknecht:2005xr}, we summarize the basic idea here for completeness and to fix our notation.
For example, an amplitude with an external photon, with momentum $q$ and polarization $\lambda$, can be expressed as a sum over similar diagrams where the photon couples by mixing with a vector meson $V$:
\begin{center}
\includegraphics[scale=0.65]{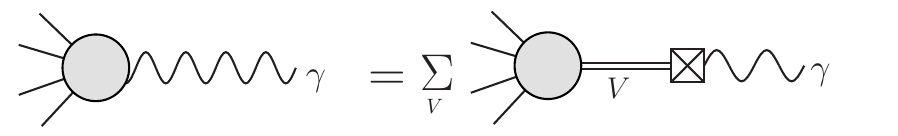} 
\end{center}
The $V$-$\gamma$ mixing part of the diagram introduces the factor
\be
V(q,\lambda) \; \feyn{m \mkern-5mu \boxtimes \mkern-5mu g} \; \gamma(q,\lambda) = - \frac{\sqrt{2} e f_V F_V(q^2) {\rm Tr}\left[\boldsymbol{T}_V \boldsymbol{Q} \right]}{m_V} \, ,
\ee
where $f_V$ is the meson decay constant and $F_V(q^2)$ is a form factor, normalized to $F_V(0) = 1$ for an on-shell photon.
For $\omega$ and $\phi$ mesons, it suffices to work in the narrow-width approximation, taking a simple Breit-Wigner form factor $F_{V}(q^2) = (1 - q^2/m_{V}^2 - i \Gamma_{V}/m_{V})^{-1}$. For the $\rho^0$ meson, its large width necessitates a more complicated form.

Additionally, $\boldsymbol{T}_V$ is the $U(3)$ flavor generator for a given meson and $\boldsymbol{Q} = {\rm diag}(\tfrac{2}{3}, - \tfrac{1}{3}, - \tfrac{1}{3})$ is the electric charge operator.
For ideal mixing, the lightest vector mesons that mix with the photon have generators
\be
\boldsymbol{T}_{\omega} = \tfrac{1}{2}\, {\rm diag}(1,1,0) \, , \quad 
\boldsymbol{T}_{\rho^0} = \tfrac{1}{2}\, {\rm diag}(1,-1,0) \, , \quad 
\boldsymbol{T}_{\phi} = \tfrac{1}{\sqrt{2}} \, {\rm diag}(0,0,1) \, \, .
\ee
The decay constant $f_V$ is defined by
\be
\langle 0 | \bar{q}^b \gamma^\mu q^a |V(q,\lambda) \rangle = \sqrt{2} f_V m_V \boldsymbol{T}_V^{ab} \varepsilon^\mu(q,\lambda) \, ,
\ee
where $\varepsilon^\mu$ is the polarization vector, following~\cite{Maris:1999nt}.
These are extracted from the measured partial widths $\Gamma(V \to e^+ e^-)$~\cite{Workman:2022ynf} according to
\be \label{eq:f}
f_{\rho} = \sqrt{\frac{3 m_{\rho} \Gamma(\rho^0 \to e^+e^-)}{2\pi \alpha_{\rm em}^2 } } \, , \quad
f_\omega = \sqrt{\frac{27 m_\omega \Gamma(\omega \to e^+e^-)}{2 \pi \alpha_{\rm em}^2 } } \, , \quad
f_\phi = \sqrt{\frac{ 27 m_\phi \Gamma(\phi \to e^+e^-)}{4 \pi \alpha_{\rm em}^2 } } \, .
\ee
The values are listed in Table~\ref{tab:f}.

\begin{center} 
\begin{table}[b] 
\setlength\tabcolsep{5pt}
\begin{tabular}{| c | c | c |} 
\hline
$f_\rho$ & $f_\omega$ & $f_\phi$ \\
\hline
$220 \pm 1 \; {\rm MeV}$ &
$201 \pm 3 \; {\rm MeV}$ &
$228 \pm 1 \; {\rm MeV}$ \\
\hline
\end{tabular}
\caption{Vector meson decay constants, from Eq.~\eqref{eq:f} using inputs from~\cite{Workman:2022ynf}.
\label{tab:f}}
\end{table}
\end{center}

For a $B$ boson (assumed to be on-shell), the VMD mixing factor is
\be \label{eq:VBmix}
V(q,\lambda) \; \feyn{m \mkern-5mu \boxtimes \mkern-5mu g} \; B(q,\lambda) = - \frac{\sqrt{2} f_V F_V(m_B^2) }{m_V} \,  {\rm Tr}\!\left[ \boldsymbol{T}_V \left( \tfrac{1}{3} g_B \boldsymbol{I} + \varepsilon e \boldsymbol{Q} \right)  \right]\, ,
\ee
where $\boldsymbol{I}$ is the identity matrix.
For $\varepsilon e \ll g_B$, $B$ bosons are produced predominantly via isoscalar $\omega,\phi$ mesons. 
Accordingly, the coefficients in Eq.~\eqref{eq:cV} are
\be
c_{\rho^0} = 0 \, , \quad 
c_\omega = - \frac{\sqrt{2} f_\omega F_\omega(m_B^2)}{3 m_\omega} \, , \quad 
c_\phi = - \frac{f_\phi F_\phi(m_B^2)}{3 m_\phi} \, .
\ee
The $B$-boson photoproduction matrix elements are related to those for $\omega,\phi$ mesons by the formula
\be \label{eq:Bboson_matrix_element}
\mathcal{M}_\pm(\gamma p \to B p) = - \frac{g_B}{3} \left( \frac{\sqrt{2} f_\omega F_\omega(m_B^2)}{m_\omega} \mathcal{M}_\pm(\gamma p \to \omega p)  + \frac{f_\phi F_\phi(m_B^2)}{m_\phi} \mathcal{M}_\pm(\gamma p \to \phi p)  \right)\, .
\ee

\subsection{Pseudoscalar-exchange}
\label{sec:pseudoscalar}

Pseudoscalar exchange is an important contribution to vector-meson photoproduction at low photon energies near threshold~\cite{Berman:1963ef,Joos1964,Friman:1995qm,Oh:2000zi}.  
Following Refs.~\cite{Friman:1995qm,Oh:2000zi}, we take a phenomenological Lagrangian 
\be \label{eq:Lpseudo}
\mathscr{L} \supset - g_{\pi NN} \bar{N} i \gamma_5 \tau^3 N \pi^0 - g_{\eta NN} \bar{N} i \gamma_5 N \eta - \sum_\varphi \sum_V \frac{e g_{\varphi \gamma V}}{m_V} \epsilon^{\mu\nu\alpha\beta} \partial_\mu A_\nu \partial_\alpha V_\beta \, \varphi
\, ,
\ee
to describe the interactions of pseudoscalar mesons $\varphi = \pi^0, \eta$, vector mesons $V=\rho^0,\omega,\phi$, nucleons $N = (p,n)$, and the photon field $A_\mu$. 

Each interaction in Eq.~\eqref{eq:Lpseudo} is parametrized by a coupling $g$.
The pion-nucleon coupling has been measured through $NN$ scattering to be $g_{\pi NN} \approx 13$~\cite{Machleidt:2001rw}, and $\tau^3 = {\rm diag}(1,-1)$ is the usual Pauli matrix.
In contrast, $g_{\eta NN}$ is not well-known by direct measurement~\cite{Pena:2000gb}. 
Indirect determinations (via the Goldberger-Treiman relation and $U(3)$ flavor symmetry) yield far more precise values: $g_{\pi NN} = 12.86 \pm 0.06$ and $g_{\eta NN} = 3.4 \pm 0.5$~\cite{Feldmann:1999uf}. 
Here we treat $g_{\pi NN}$ and $g_{\eta NN}$ as free parameters in our fit, assuming both have the same sign, following from $U(3)$ flavor symmetry.

The pseudoscalar-photon-vector-meson couplings $g_{\varphi \gamma V}$ are determined precisely from the measured partial widths $\Gamma(V \to \varphi \gamma)$~\cite{Workman:2022ynf}, according to the formula
\be \label{eq:g_P_gamma_V}
g_{\varphi \gamma V} = \sqrt{\frac{24 \Gamma(V \to \varphi \gamma)}{\alpha_{\rm em} m_V(1-m_\varphi^2/m_V^2)^3}} \, .
\ee
Using $U(3)$ flavor symmetry, we assume $g_{\rho \gamma \pi}$, $g_{\rho \gamma \eta}$, $g_{\omega \gamma \pi}$, $g_{\omega \gamma \eta}$, and $g_{\phi \gamma \eta}$ have the same relative sign.
The final coupling $g_{\phi \gamma \pi}$ arises through $\omega$-$\phi$ mixing away from the ideal limit.
Parametrized as $g_{\phi \gamma \pi} = - \tan \delta_V \, g_{\omega \gamma \pi}$, Ref.~\cite{Benayoun:1999fv} performed a global fit to meson decays and found an angle $\delta_V \approx -3^\circ$ relative to the ideal limit.
Hence, we take $g_{\phi \gamma \pi}$ to have the same relative sign as the others.
The numerical values of these couplings are in Table~\ref{tab:params}.

\begin{center} 
\begin{table}[b] 
\setlength\tabcolsep{5pt}
\begin{tabular}{| c | c | c | c | c|  c |} 
\hline
$g_{\pi \gamma \rho}$ & $g_{\eta \gamma \rho}$ & 
$g_{\pi \gamma \omega}$ & $g_{\eta \gamma \omega}$ &
$g_{\pi \gamma \phi}$ & $g_{\eta \gamma \phi}$ \\
\hline
$0.57 \pm 0.05$ &
$1.22 \pm 0.04$ &
$1.83 \pm 0.03$ &
$0.35 \pm 0.02$ &
$0.138 \pm 0.003$ &
$0.704 \pm 0.007$ \\
\hline
\end{tabular}
\caption{Pseudoscalar-photon-vector-meson couplings $g_{\varphi \gamma V}$ entering pseudoscalar-exchange amplitude, computed from Eq.~\eqref{eq:g_P_gamma_V} using inputs from~\cite{Patrignani:2016xqp}.
\label{tab:g_params}}
\end{table}
\end{center}
 
The matrix element for vector-meson photoproduction is
\be\label{eq:matrix_pseudo}
i \mathcal{M}_-(\gamma p \to V p) = i e \mathcal{A}_V(t)  \epsilon^{\mu\nu \alpha \beta} \epsilon_\mu(q^\prime) \epsilon_\nu(q)^* q_\alpha q^\prime_\beta \, \bar{u}(p^\prime) i \gamma_5 u(p) \, .
\ee
Here $u$ ($\bar u$) is the incoming (outgoing) proton spinor, and 
\be
\mathcal{A}_V(t) =  \frac{1}{m_V} \sum_{\varphi=\pi^0, \eta} \frac{ g_{\varphi NN} g_{\varphi \gamma V} \mathcal{F}(t,\Lambda_{\varphi NN},m_\varphi) \mathcal{F}(t,\Lambda_{\varphi \gamma V},m_\varphi) }{ t-m_\varphi^2 } \, .
\ee
Each vertex in Eq.~\eqref{eq:Lpseudo} is additionally dressed with a form factor
\be \label{eq:pseudo_form}
\mathcal{F}(t,\Lambda,m) = \frac{\Lambda^2 - m^2}{\Lambda^2 - t} \, ,
\ee
where $\Lambda$ represents a cutoff scale~\cite{Friman:1995qm}. 
From Eq.~\eqref{eq:matrix_pseudo}, it is straightforward to compute the differential cross section
\be \label{eq:sigVunnat}
\frac{d \sigmam(\gamma p \to V p)}{d t} = \frac{ \alpha_{\rm em} |t|  (t-m_V^2)^2 |\mathcal{A}_V(t)|^2}{16 s^2} \, ,
\ee
in the limit $\sqrt{s} \gg m_p$.

For $B$ boson photoproduction, we need consider processes involving only $\omega,\phi$-mesons.
Hence, we need six cutoffs in Eq.~\eqref{eq:pseudo_form}
\be \label{eq:cuts}
\Lambda_{\pi NN} \, , \; \Lambda_{\eta NN} \, , \; 
\Lambda_{\pi \gamma \omega} \, , \; \Lambda_{\eta \gamma \omega} \, , \;
\Lambda_{\pi \gamma \phi} \, , \; \Lambda_{\eta \gamma \phi} \; .
\ee
These are determined from photoproduction data. 
This has been done for $\omega$ photoproduction and the first four cutoffs in \eqref{eq:cuts} were determined to be in the range $0.5 - 1$ GeV~\cite{Oh:2000zi,Williams:2007zzg}.
Here we determine all six cutoffs through a joint fit to $\omega$ and $\phi$ photoproduction data, described below.

With the model parameters fixed, we calculate the unnatural parity contribution to the $B$-boson cross section following Eq.~\eqref{eq:Bboson_matrix_element}:
\be
\frac{d \sigmam(\gamma p \to B p)}{dt} = \frac{ \pi \alpha_{\rm em} \alpha_B |t| (t-m_B^2)^2}{36 s^2}  \left|\frac{\sqrt{2} \mathcal{A}_\omega(t) f_\omega F_\omega(m_B^2)}{m_\omega}
+ \frac{\mathcal{A}_\phi(t) f_\phi F_\phi(m_B^2) }{m_\phi} \right|^2\, . \label{eq:dsigdtBminus}
\ee

\subsection{Pomeron exchange}

\label{sec:pomeron}

At higher energy, the pomeron model gives a remarkably successful description of diffractive scattering between hadrons~\cite{Donnachie:1984xq,Donnachie:1992ny}.
Ewerz et al.~\cite{Ewerz:2013kda} have provided a field-theoretic formulation of the pomeron as an effective spin-2 interaction.\footnote{Diffractive models with a pomeron of spin-0 and spin-1 are disfavored from experimental data and on theoretical grounds, respectively~\cite{Ewerz:2016onn}.}
Here, we first compute the matrix element for $\mathcal{M}_+(V p \to V p)$ using their Feynman rules and then external gauge fields $\gamma,B$ are included through VMD.
One nice feature of~\cite{Ewerz:2013kda} is that there are no issues with VMD causing violation of Ward identities, as opposed to other treatments (e.g.,~\cite{Laget:1994ba,Pichowsky:1996tn,Oh:2000zi}).

Following Ref.~\cite{Ewerz:2013kda}, the matrix element is
\bea
i\mathcal{M}_+(V p \to V p) &=& -3i\beta_{\Pom NN} F_1(t) \bar{u}(p^\prime) \left\{ \frac{1}{2}\Big[ \gamma^\alpha (p + p^\prime)^\beta + \gamma^\beta (p + p^\prime)^\alpha \Big] - \frac{1}{4} g^{\alpha \beta} (\slashed{p} + \slashed{p}^\prime)\right\} u(p) \notag 
\\
&\;& \quad \times \; i F_M(t) \left[ 2 a_{\Pom VV} \Gamma^{(0)}_{\mu\nu\kappa\lambda}(q^\prime,-q) - b_{\Pom VV} \Gamma^{(2)}_{\mu\nu\kappa\lambda}(q^\prime,-q) \right] \epsilon^\nu(q) \epsilon^\mu(q^\prime)^* \notag \\
&\;& \quad \times \; \frac{1}{4s} \left(\frac{-is}{s_0}\right)^{\alpha_\Pom(t)-1} \left( g_{\alpha\kappa} g_{\beta \lambda} + g_{\alpha\lambda} g_{\beta \kappa} - \frac{1}{2} g_{\alpha\beta} g_{\kappa \lambda} \right) \, . \label{eq:VpVp}
\eea
On the right-hand side of Eq.~\eqref{eq:VpVp}, the first line represents the pomeron-nucleon vertex, which is parametrized by $\beta_{\Pom NN} F_1(t)$, where $\beta_{\Pom NN}$ is the coupling constant and $F_1(t)$ is the Dirac form factor of the proton~\cite{Donnachie:1984xq}.
The second line represents the pomeron-vector-meson vertex, parametrized by coupling constants $a_{\Pom VV}, b_{\Pom VV}$, as well as a (common) meson form factor $F_M(t)$.
We also have the tensors~\cite{Ewerz:2013kda}
\bea
\Gamma_{\mu\nu\kappa\lambda}^{(0)}(p,q) &=& \big[ (p \cdot q) g_{\mu\nu} - q_\mu p_\nu \big] \left[ p_\kappa q_\lambda + p_\lambda q_\kappa - \frac{1}{2} (p \cdot q) g_{\kappa\lambda} \right]\\
\Gamma_{\mu\nu\kappa\lambda}^{(2)}(p,q) &=& 
(p \cdot q) \left[ g_{\mu\kappa} g_{\nu \lambda} + g_{\mu \lambda} g_{\nu\kappa} -  g_{\mu\nu}g_{\kappa\lambda} \right] + g_{\mu\nu} \big[ p_\kappa q_\lambda + p_\lambda q_\kappa \big] 
\notag \\
&& \; - g_{\mu\kappa} p_\nu q_\lambda - g_{\mu\lambda} p_\nu q_\kappa - g_{\nu\kappa} p_\lambda q_\mu - g_{\nu\lambda} p_\kappa q_\mu + g_{\kappa\lambda} p_\nu q_\mu \; .
\eea
The third line of Eq.~\eqref{eq:VpVp} represents the pomeron propagator, where the pomeron trajectory is 
\be
\alpha_\Pom(t) = 1 + \epsilon_\Pom + t/s_0 \, .
\ee
Fits to $pp$ and $p \bar{p}$ scattering data have determined~\cite{Nachtmann:2003ik}
\be \label{eq:pom_params}
\beta_{\Pom NN} = 1.87 \; \textrm{GeV}^{-1} \, , \quad \alpha_\Pom(0) = 1 + \epsilon_\Pom = 1.0808 \, , \quad s_0 = 4 \; {\rm GeV}^2 \, .
\ee
In contrast, the pomeron-vector-meson couplings are less well-known. 
First, we expect $a_{\Pom \omega \omega} \approx a_{\Pom \rho \rho}$ and $b_{\Pom \omega \omega} \approx b_{\Pom \rho \rho}$~\cite{Bolz:2014mya}.
Second, it is argued that the total (inclusive) cross section for $\rho p$ scattering is related to that of $\pi^\pm p$ scattering, which holds to a good approximation experimentally~\cite{Ewerz:2013kda}. This yields a relation
\be \label{eq:ab_rho}
2 m_\rho^2 a_{\Pom \rho \rho} + b_{\Pom \rho\rho} = 7.04 \; {\rm GeV}^{-1} \; 
\ee
(and similarly for the $\omega$ couplings). 
A similar argument relating $\phi p$ and $K^\pm p$ scattering yields~\cite{Lebiedowicz:2018eui}
\be \label{eq:ab_phi}
2 m_\phi^2 a_{\Pom \phi \phi} + b_{\Pom \phi\phi} = 5.28 \; {\rm GeV}^{-1} \; .
\ee
Here we do not impose Eqs.~\eqref{eq:ab_rho} or \eqref{eq:ab_phi}, but rather we impose weaker priors which are discussed in Appendix~\ref{app:optical}.

Next, we use VMD to calculate the photoproduction matrix element from Eq.~\eqref{eq:VpVp}:
\be
\mathcal{M}_+(\gamma p \to V p) = - \frac{\sqrt{2} e f_V F_V(q^2) {\rm Tr}\left[\boldsymbol{T}_V \boldsymbol{Q} \right]}{m_V} \mathcal{M}_+(V p \to V p) \, ,
\ee
where $q^2=0$ for an on-shell photon.\footnote{We assume pomeron-vector-meson vertices are diagonal, neglecting non-diagonal (transition) vertices that are considered elsewhere~\cite{Lebiedowicz:2019boz}.}
Retaining only the leading term in powers of $s$, the vector-meson photoproduction cross section is
\bea
\frac{d \sigmap (\gamma p \to V p)}{dt} &=& 
\frac{\aem f_V^2}{8 m_V^2} \beta_{\Pom NN}^2 \left[a_{\Pom VV}^2 (m_V^2 - t)^2 + 2 a_{\Pom VV} b_{\Pom VV} m_V^2 + b_{\Pom VV}^2 \right] \notag \\
&& \quad \times \; \left|F^{(\Pom)}_{pV}(t)\right|^2  
\left( \frac{s}{s_0} \right)^{2 \alpha_\Pom(t) - 2} 
\times \left\{ \begin{array}{cc} 
9 & {\rm for} \; V = \rho^0 \\
1 & {\rm for} \; V = \omega \\
2 & {\rm for} \; V=\phi \end{array} \right. \, .
\eea
where $F^{(\Pom)}_{pV}(t)$ is a joint form factor for both pomeron-proton and pomeron-vector-meson interactions.
In Eq.~\eqref{eq:VpVp}, this is $F^{(\Pom)}_{pV}(t) = F_1(t) F_M(t)$.
However, here we adopt a different ansatz
\be
F^{(\Pom)}_{pV}(t) = e^{B_V t/2} \, ,
\ee
with a different slope $B_V$ for each vector meson~\cite{Laget:1994ba}, which provides a reasonably good fit to our dataset.

For $B$-boson production, only $\omega,\phi$-processes contribute.
The matrix element, following from Eq.~\eqref{eq:Bboson_matrix_element}, is
\be \label{eq:Bboson_matrix_element_2}
\mathcal{M}_+(\gamma p \to B p) 
= \frac{g_B e}{9} \left( \frac{f_\omega^2 F_\omega(m_B^2)}{m_\omega^2} \mathcal{M}_+(\omega p \to \omega p) - \frac{f_\phi^2 F_\phi(m_B^2)}{m_\phi^2} \mathcal{M}_+(\phi p \to \phi p) \right)
\ee
where the relative minus between the two terms is fixed by VMD since the $\omega$ and $\phi$ amplitudes are proportional to
\be
{\rm Tr}\left[\boldsymbol{T}_V \boldsymbol{Q} \right] {\rm Tr}\left[\boldsymbol{T}_V \right] 
= \left\{ \begin{array}{cl} 1/6 &  \; {\rm for} \; V=\omega \\
- 1/6 & \; {\rm for} \;  V=\phi \end{array} \right. \, .
\ee
This reflects the fact that the pomeron-exchange amplitude for $\gamma p \to B p$ must vanish in the $SU(3)$-flavor-symmetric limit, since the photon and $B$ boson couple to orthogonal generators.

Next, it is helpful to introduce the following general linear combinations of couplings
\bea
a_{\Pom \gamma B}(t,q^2,q^{\prime2}) &=&  \mathcal{C}_\omega(t,q^2,q^{\prime2}) \, a_{\Pom \omega \omega} - \mathcal{C}_\phi(t,q^2,q^{\prime2}) \, a_{\Pom \phi \phi} \label{eq:a_PgB}\\
b_{\Pom \gamma B}(t,q^2,q^{\prime2}) &=&  \mathcal{C}_\omega(t,q^2,q^{\prime2}) \,  b_{\Pom \omega \omega} - \mathcal{C}_\phi(t,q^2,q^{\prime2}) \, b_{\Pom \phi \phi} \label{eq:b_PgB}
\eea
with coefficients
\be
\mathcal{C}_V(t,q^2,q^{\prime2}) = \frac{ f_V^2 F_V(q^2) F_V(q^{\prime 2}) F_{pV}^{(\Pom)}(t)}{m_V^2} \, .
\ee
In terms of these quantities, the natural-parity contribution to the cross section is
\bea 
\frac{d \sigmap(\gamma p \to B p)}{dt} &=&
\frac{\alpha_B \alpha_{\rm em} \pi }{9} \beta_{\Pom NN}^2 \left(\frac{s}{s_0}\right)^{2 \alpha_\Pom(t) -2} \Big\{\big|a_{\Pom \gamma B}(t,0,m_B^2)\big|^2 (m_B^2 - t)^2  \notag \\
&& + \; 2 \, {\rm Re}\big[ a_{\Pom \gamma B}(t,0,m_B^2) b_{\Pom\gamma B}^*(t,0,m_B^2)\big] m_B^2 + \big|b_{\Pom \gamma B}(t,0,m_B^2)\big|^2 \Big\} \label{eq:dsigdtBplus} \, ,
\eea
which is evaluated as a function of $\alpha_B$ and $m_B$ once the parameters of the phenomenological model are fixed.

In the present work, we treat the parameters
\be
a_{\Pom \omega\omega}, \; b_{\Pom \omega\omega} , \; a_{\Pom \phi \phi}  , \; b_{\Pom \phi \phi}  , \;
B_\omega , \; B_\phi  , \; \alpha_\Pom(0) 
\ee
as freely varying in our fit.
However, we keep $\beta_{\Pom NN}$ and $s_0$ as fixed in Eq.~\eqref{eq:pom_params}.
While we do not expect uncertainties in these latter parameters to be small, variation can be (to some extent) absorbed into the other parameters.
We defer a joint fit to both nucleon scattering and photoproduction to future work.

\section{Numerical fit}
\label{sec:fit}

To determine the $B$-boson photoproduction cross section, we perform a fit to experimental data to determine the phenomenological parameters entering our model.
Our dataset consists of differential cross section measurements for exclusive $\omega,\phi$-photoproduction.
These include high precision measurements with the CEBAF Large Acceptance Spectrometer (CLAS)~\cite{Williams:2009ab,Dey:2014tfa}, which go from threshold up to $\sqrt{s} \approx 2.8$ GeV; older measurements~\cite{Ballam:1972eq,Barber:1985fr,Busenitz:1989gq}, which extend up to $\sqrt{s} \approx 20$ GeV; and from ZEUS at much larger energies, $\sqrt{s} \approx 70-94$ GeV~\cite{Derrick:1996yt,Derrick:1996af,ZEUS:1999ptu}.
This latter is particularly important for determining the pomeron contribution in $\gamma p \to \omega p$, which is not well-constrained from low-energy data alone.
We restrict our dataset to lie in range
\be \label{eq:limit_dataset}
\sqrt{s} > 2.3 \; {\rm GeV} \, , \quad |t| < 1 \; \textrm{GeV}^2
\ee
since our model aims to give the leading contribution in the diffractive limit.
Outside this range, diffractive scattering and/or meson-exchange are no longer dominant~\cite{Williams:2007zzg,Dey:2014tfa} and subdominant processes neglected in our model can become important, e.g., $\sigma$-meson exchange~\cite{Joos1964}, nucleon excitations~\cite{Zhao:1998fn,Oh:2000zi}, or exchanges of additional reggeons~\cite{Ewerz:2013kda}.
Since our aim is to describe the leading contributions to $B$-boson production, it suffices to neglect these.
We also include a 10\% systematic error on all data points, added in quadrature with statistical errors.

In addition, we include data for the total photon-proton cross section measured by H1~\cite{H1:1995hmw} and ZEUS~\cite{ZEUS:2001wan} at $\sqrt{s} \approx 200 \; {\rm GeV}$.
Discussed in Appendix~\ref{app:optical}, this data is also important for constraining the $\omega$-pomeron coupling.

Our fit has a similar but complementary spirit to other photoproduction fits from previous literature~\cite{Oh:2000zi,Williams:2007zzg,Lebiedowicz:2019boz}.
The models adopted therein include many additional contributions needed to describe $\omega$ and $\phi$ photoproduction data across the full kinematic ranges in $t$, however, these fits are each limited to smaller ranges of $\sqrt{s}$.
For new gauge forces, a complete phenomenological model including all known contributions would be ultimately desirable, but we defer this to future work.

Fixed parameters are given in Tables~\ref{tab:f} and \ref{tab:g_params}.
Here we take the central values as input and do not propagate uncertainties in our analysis.
Other parameters are given in Table~\ref{tab:params}.
Here $\beta_{\Pom NN}, \alpha_\Pom^\prime(0)$ are fixed from $pp, p\bar p$ scattering data~\cite{Nachtmann:2003ik}.
We fit the remaining fifteen parameters from experimental data using a Markov Chain Monte Carlo analysis.
For some parameters, we adopt Gaussian priors to exclude values far from expectations (discussed in Sec.~\ref{sec:model}).

\begin{center} 
\begin{table}[b] 
\setlength\tabcolsep{5pt}
\begin{tabular}{| c | c | c || c | c|  c |} 
\hline
parameter & prior & fit value &
parameter & prior & fit value\\
\hline
$\beta_{\Pom NN}$ & $1.87 \; \textrm{GeV}^{-1}$ & $-$  &
$g_{\pi NN}$ & $13.0 \pm 1.0$ & $14.0 \pm 0.3$ \\
$\alpha_\Pom(0) $ & none &$1.091\pm0.007$ & 
$g_{\eta NN}$ & $4.0 \pm 1.0$ & $4.1\pm 1.0$ \\
$\alpha_\Pom^\prime(0) = s_0^{-1}$ & 0.25 GeV$^{-2}$ & $-$ &
$\Lambda_{\pi NN}/{\rm GeV}$ & $0.8 \pm 0.2$ &  $0.74^{+0.09}_{-0.07}$ \\
$a_{\Pom \omega \omega}/{\rm GeV}^{-3}$ & \eqref{eq:pom_constraints} & $< 0.4$ & 
$\Lambda_{\eta NN}/{\rm GeV}$ & $0.8 \pm 0.2$ & $0.7 \pm 0.3$ \\
$b_{\Pom \omega \omega}/{\rm GeV}^{-1}$ & \eqref{eq:pom_constraints} & $7.4 \pm 0.5$ &
$\Lambda_{\pi \gamma\omega}/{\rm GeV}$ & $0.8 \pm 0.2$ & $0.8 \pm 0.1$ \\
$a_{\Pom \phi \phi}/{\rm GeV}^{-3}$ & \eqref{eq:pom_constraints}  & $0.9^{+0.3}_{-0.4}$&
$\Lambda_{\eta \gamma\omega}/{\rm GeV}$ & $0.8 \pm 0.2$ & $0.8 \pm 0.2$ \\
$b_{\Pom \phi \phi}/{\rm GeV}^{-3}$ & \eqref{eq:pom_constraints}  & $2.0^{+0.4}_{-0.3}$&
$\Lambda_{\pi \gamma\phi}/{\rm GeV}$ & $0.8 \pm 0.2$ & $0.5^{+0.2}_{-0.1}$ \\
$B_\omega/{\rm GeV}^2$ & none & $7.4\pm 0.3$ &
$\Lambda_{\eta \gamma\phi}/{\rm GeV}$ & $0.8 \pm 0.2$ & $0.6^{+0.3}_{-0.2}$ \\
$B_\phi/{\rm GeV}^2$ & none &$3.1\pm0.2$ &
 &  &  \\
\hline
\end{tabular}
\caption{Summary of fixed and fitted parameters in our model. Uncertainties for fixed inputs are quoted where known, but are not propagated in our analysis. All hadron masses and widths, and $\aem$, are also fixed and are taken from \cite{Patrignani:2016xqp}. Fitted parameters are determined in our fit to experimental data, subject to a Gaussian prior. The quoted best-fit values include statistical uncertainties only.\label{tab:params}}
\end{table}
\end{center}

The results from our fit are given in Table~\ref{tab:params}. 
The fitted parameters quoted are medians and one-sigma intervals, except for $a_{\Pom \omega\omega}$ which is consistent with zero and a one-sigma upper limit is provided.
For the most part, our results are consistent with values in the literature.
A similar phenomenological fit to $\omega$ data from CLAS~\cite{Williams:2007zzg} yielded the following values
\be
\Lambda_{\pi NN} = 0.6 \, , \quad 
\Lambda_{\pi\gamma\omega} = 0.7 \, , \quad
\Lambda_{\eta NN} = 1.0 \, , \quad
\Lambda_{\eta \gamma\omega} = 0.9 \, ,
\ee
albeit with different assumptions for the pseudoscalar-nucleon couplings and pomeron-exchange amplitude,
which are in agreement with our results.
The pomeron trajectory intercept found in our fit is in good agreement with the value $\alpha_\Pom(0) = 1 + \epsilon_\Pom = 1.0808$ quoted in the literature, as extracted from $pp$ and $p \bar{p}$ scattering~\cite{Nachtmann:2003ik}.
Our vector-meson-pomeron couplings satisfy the following relations
\be
2 m_\omega^2 a_{\Pom \omega \omega} + b_{\Pom \omega\omega} = 7.9 \pm 0.5 \; {\rm GeV}^{-1} \, , \quad 
2 m_\phi^2 a_{\Pom \phi \phi} + b_{\Pom \phi\phi} = 3.8 \pm 0.4 \, ,
\ee
to be compared with Eqs.~\eqref{eq:ab_rho} and \eqref{eq:ab_phi}.

\begin{figure}[t]
\begin{center}
\includegraphics[width=0.96\textwidth]{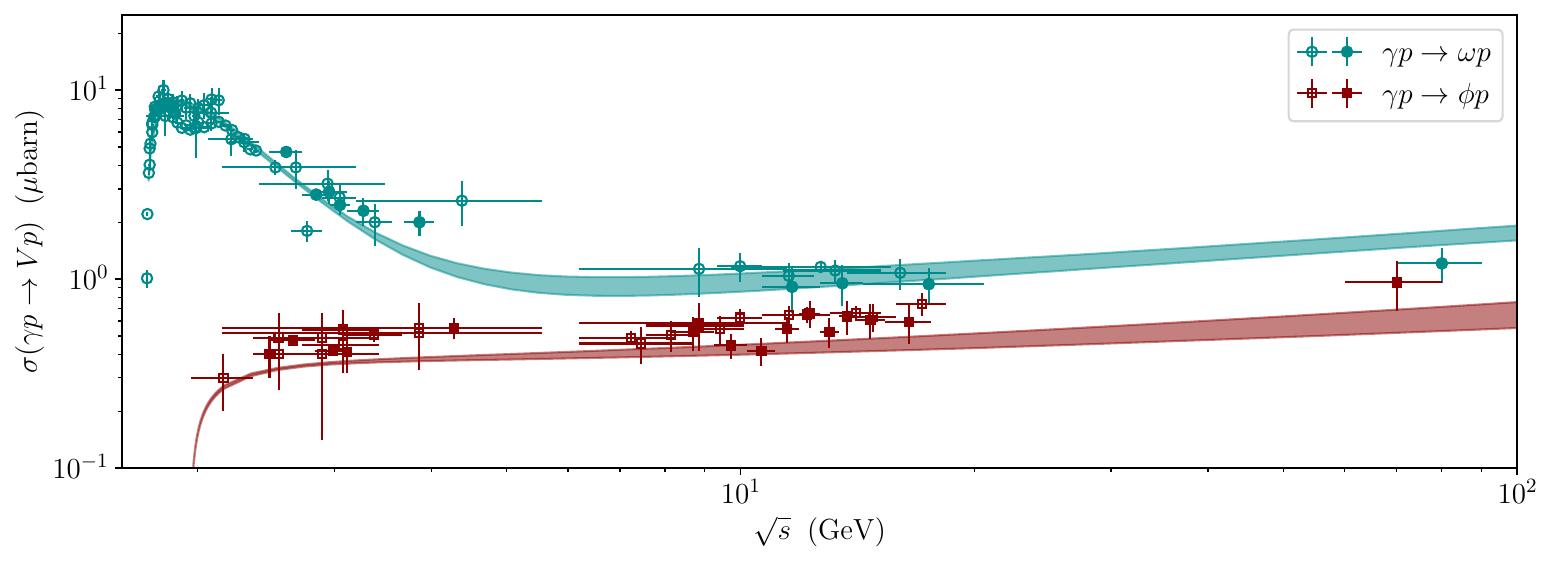}
\end{center}
\caption{\it Shaded bands show the exclusive $\omega,\phi$-photoproduction cross section from our model (90\% confidence intervals), as a function of center-of-mass energy $\sqrt{s}$.
Data points show total cross section values from experiments~\cite{OmegaPhoton:1983huz,Barber:1985fr,CBELSATAPS:2015wwn,Aachen-Hamburg-Heidelberg-Munich:1975jed,Busenitz:1989gq,Egloff:1979xg,Derrick:1996yt,Strakovsky:2014wja,BrownHarvardMITPadovaWeizmannInstituteBubbleChamberGroup:1967zz,Davier:1969nx,Ballam:1972eq,Breakstone:1981wk,Barth:2003kv,Barber:1981fj,OmegaPhoton:1984eqn,Egloff:1979mg,Bonn-CERN-EcolePoly-Glasgow-Lancaster-Manchester-Orsay-Paris-Rutherford-Sheffield:1980dwy,Derrick:1996af,HERAGroup:1987dng}, where the horizontal bar indicates the range of $\sqrt{s}$ covered.
}
\label{fig:vectormesoncrosssections}
\end{figure}

To illustrate the results of our fit, Fig.~\ref{fig:vectormesoncrosssections} shows the current world dataset for exclusive $\omega,\phi$-photoproduction cross sections as a function of $\sqrt{s}$ (data points).
Filled points represent datasets that were included in our fit, while open points were not, either because differential cross section data was not publicly available or because $\sqrt{s}$ was near the threshold region. 
The shaded bands are the results from our phenomenological model, where the band width represents a 90\% confidence interval in our fitted parameters.

Our model for $\omega$-photoproduction appears in good agreement with data.
However, our model for $\phi$-photoproduction appears to systematically under-predict experimentally-determined cross sections by $\mathcal{O}(30\%)$.
We emphasize that our model is fit to differential cross section data, whereas
determining the total cross section both on the theory and experimental sides requires extrapolating the differential cross section to the forward-angle limit, which may lead to additional systematic uncertainties.
In Appendix~\ref{sec:compare}, we provide a comparison between our model and experimental data for the differential photoproduction cross section included in our fit.

\section{$B$-boson photoproduction}
\label{sec:Bboson}

\begin{figure}[t]
\begin{center}
\includegraphics[width=0.99\textwidth]{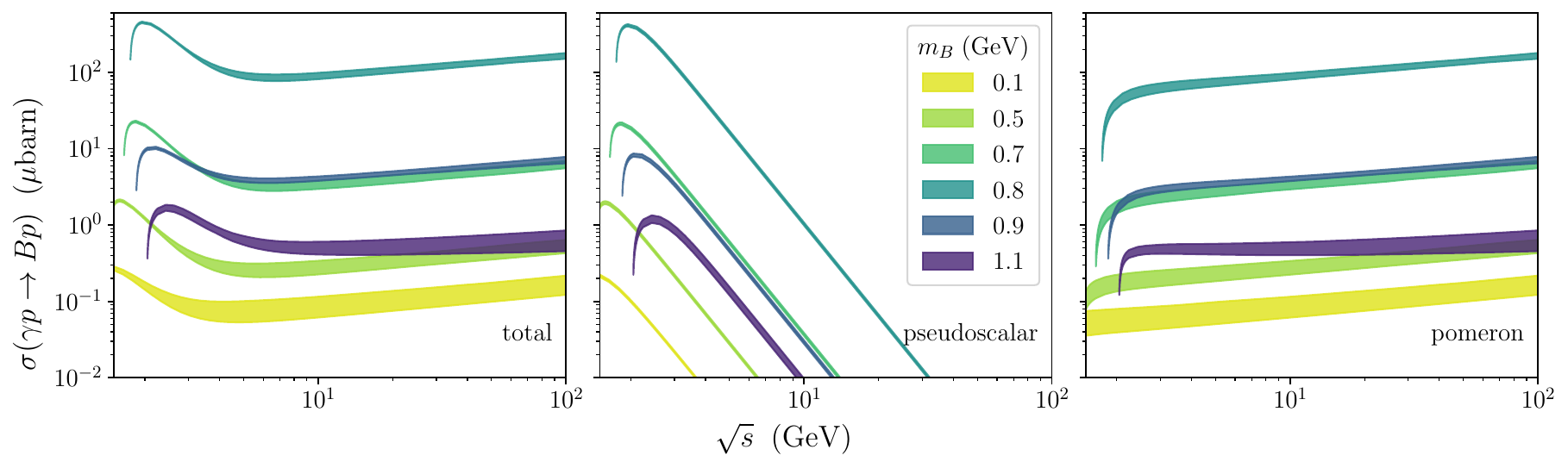}
\end{center}
\caption{\it Shaded bands show the $B$-boson photoproduction cross section from our model (90\% confidence intervals), as a function of center-of-mass energy $\sqrt{s}$ and for different masses $m_B$. Cross sections scale linearly with $\alpha_B$ and have been normalized to $\alpha_B = 1$.}
\label{fig:B-crosssection-vs-W}
\end{figure}

The differential cross section for $B$-boson photoproduction is
\be \label{eq:dsigdtB}
\frac{d \sigma(\gamma p \to B p)}{dt}
= \frac{d \sigmap(\gamma p \to B p)}{dt} + 
\frac{d \sigmam(\gamma p \to B p)}{dt} \, .
\ee
The unnatural ($-$) and natural ($+$) parity contributions are given by Eqs.~\eqref{eq:dsigdtBminus} and \eqref{eq:dsigdtBplus}, respectively.
In Fig.~\ref{fig:B-crosssection-vs-W}, we plot the $B$-boson photoproduction cross section as a function of $\sqrt{s}$, for various masses $m_B$.
The left panel shows the total cross section, while the center and right panels show the unnatural- and natural-parity cross sections separately from pseudoscalar and pomeron exchange, respectively.
For the parameter range shown, the cross section tends to be dominated the pomeron contribution which is only weakly-dependent on $\sqrt{s}$, characteristic of diffractive scattering, except for near threshold where pseudoscalar-exchange is important.\footnote{We  calculate the total cross section by integrating Eq.~\eqref{eq:dsigdtB} over the the full kinematic range for $t$, which includes large values of $|t|$ outside the diffractive regime. 
Limiting the integral to the diffractive range reduces the cross section, e.g., restricting $|t| < 1 \; {\rm GeV}$ leads to an $\mathcal{O}(10\%)$ reduction.}

\begin{figure}[t]
\begin{center}
\includegraphics[width=0.6\textwidth]{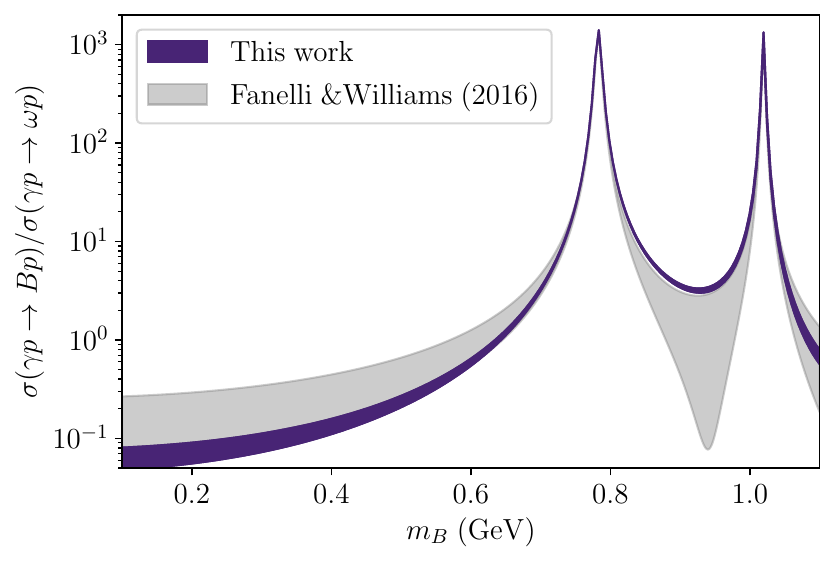}
\end{center}
\caption{\it Dark (purple) shaded band shows total cross section for $B$-boson photoproduction, relative to that for $\omega$ mesons, from our phenomenological model (90\% confidence intervals).
Light (gray) shaded band shows same quantity computed by Fanelli and Williams~\cite{Fanelli:2016utb}, where width comes from varying the unknown phase $\varphi_+$ in the range $[0,\pi]$.
Center-of-mass energy $\sqrt{s} = 4\; {\rm GeV}$ and coupling $\alpha_B = 1$ are fixed.
}
\label{fig:FWcomparison}
\end{figure}

The behavior of the cross section with $B$ boson mass is shown in Fig.~\ref{fig:FWcomparison}.
Here we show the photoproduction cross section relative to that for $\omega$-mesons, which in our model is $\sigma(\gamma p \to \omega p) \approx 1.2 \; {\rm \mu barn}$ for $\sqrt{s} = 4\; {\rm GeV}$
The darker (blue) band shows the prediction from our phenomenological model.
Due to vector-meson mixing, the cross section is strongly enhanced for $m_B$ near $m_\omega$ or $m_\phi$.

Next, we compare our results to those of Fanelli and Williams~\cite{Fanelli:2016utb}.
Their formula is
\begin{eqnarray}
\sigmapm(\gamma p \to B p) = \frac{4 \alpha_B \Phi(m_B)}{27}
\left( \frac{|F_\omega(m_B^2)|^2 \sigmapm(\gamma p \to \omega p)}{\Phi(m_\omega)} + 
\frac{|F_\phi(m_B^2)|^2 \sigmapm(\gamma p \to \phi p)}{2 \Phi(m_\phi)} \right. \notag \\
\left. + \; \frac{\cos\varphi_\pm |F_\omega(m_B^2)||F_\phi(m_B^2)|  \sqrt{ 2 \sigmapm(\gamma p \to \omega p) \sigmapm(\gamma p \to \phi p)}}{\sqrt{\Phi(m_\omega) \Phi(m_\phi)}}
\right) \, , \label{eq:Fanelli}
\end{eqnarray}
where the phase space factors are approximately $\Phi(m_B^2)/\Phi(m^2_{\omega,\phi}) \approx 1$ in the large-$s$ limit.
Similar to our work, Eq.~\eqref{eq:Fanelli} treats $B$ production using VMD via $\omega$ and $\phi$ mixing, with the added approximation $2m_\omega^2/f_\omega^2 \approx 2 m_\phi^2/f_\phi^2 \approx 12 \pi$ (which is true to better than $20\%$).
The matrix elements arising from $\omega$ or $\phi$ mixing are parametrized simply in terms of the respective SM cross sections, except for unknown relative phases $\varphi_{\pm}$.

Taking inputs from Ref.~\cite{Fanelli:2016utb}, the results of Eq.~\eqref{eq:Fanelli} are shown in Fig.~\ref{fig:FWcomparison} (light gray band).
The phase $\varphi_+$ is allowed to vary between $0$ and $\pi$, which sets the width of the band, while $\varphi_-$ does not enter Eq.~\eqref{eq:Fanelli} since $\sigma_-(\gamma p \to \phi p)$ is neglected.
The unknown phase $\varphi_+$ limits the precision of Eq.~\eqref{eq:Fanelli} in the absence of additional input (as the authors discuss).
On the other hand, our predictions are more precise and have no unknown phase, as it is fixed by the relative sign in Eq.~\eqref{eq:Bboson_matrix_element_2}. 
That is, we have $\cos\varphi_+ = -1$ for $m_B < m_\omega$ or $m_B > m_\phi$, while $\cos\varphi_+ = +1$ for $m_\omega < m_B < m_\phi$.
Fixing this sign in Eq.~\eqref{eq:Fanelli} shows good agreement with our work.

\section{$B$-boson electroproduction}
\label{sec:electro}

\subsection{Theoretical preliminaries}

\begin{figure}[t]
\begin{center}
\includegraphics[scale=0.8]{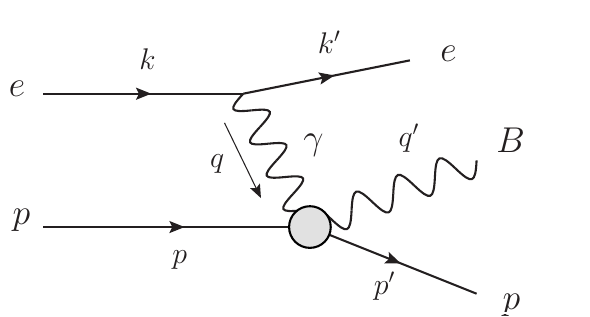}
\end{center}
\caption{Feynman diagram for $B$-boson electroproduction, with four-momenta as labeled. 
Gray blob denotes photoproduction vertex of our phenomenological model, shown in Fig.~\ref{fig:photo_feynman_diagram}.}
\label{fig:electro_feynman_diagram}
\end{figure}

$B$ bosons can be produced in electron-proton collisions, shown in Fig.~\ref{fig:electro_feynman_diagram}, analogous to virtual Compton scattering.
The particles involved have the following four-momenta: $k$ ($k^\prime$) and $p$ ($p^\prime$) for the incoming (outgoing) electron and proton, respectively, $q$ for the photon, and $q^\prime$ for the $B$ boson.
The usual kinematic variables for deep inelastic scattering are
\bea
Q^2 = - q^2  \, , \quad y = \frac{p\cdot q}{p \cdot k} \, , 
\quad \nu = \frac{p\cdot q}{m_p} \, , \quad W^2 = (p + q)^2 \; ,
\eea
where $Q^2$ is the momentum transfer, and $y$ is the fractional electron energy loss and $\nu$ is the photon energy in the lab frame (for a fixed proton target). 
The invariant mass-squared of the photon-proton system is $W^2$, which was defined as $s$ previously in Sec.~\ref{sec:model}. 
Here we denote $s_{\rm tot} = (p+k)^2$ as the total center-of-mass energy-squared for the entire electron-proton system, while $t = (q-q^\prime)^2$ is the same as previously defined for photoproduction.
Lastly, it is useful to note the following relations
\be
W^2 = y s_{\rm tot} + (1-y) m_p^2 - Q^2 
\ee
and
\be \label{eq:nu}
\nu = \frac{W^2 - m_p^2 + Q^2 }{2m_p}\; .
\ee

For unpolarized scattering and fixed beam energy, there are three independent kinematic variables, which we take to be $y$, $Q^2$, $t$. 
The triple-differential cross section for electroproduction can be expressed as
\be
\frac{d^3 \sigma(e p \to e p B) }{dy dQ^2 dt} = \Gamma_T(Q^2,y) \frac{d\sigma_T(\gamma^* p \to B p)}{dt} + \Gamma_L(Q^2,y) \frac{d\sigma_L(\gamma^* p \to B p)}{dt} \, .
\ee
It is customary to follow Hand's convention~\cite{Hand:1963bb} to express the electron-photon part of the cross section (following standard quantum electrodynamics) as an effective flux of transverse ($T$) and longitudinal ($L$) virtual photons
\bea
\Gamma_T(Q^2,y) &=& \frac{\aem W^2}{2\pi Q^2 y(W^2 + Q^2)} \left( 1 + (1-y)^2 - \frac{2 Q^2_{\rm min}(1-y)}{Q^2} \right) \\
\Gamma_L(Q^2,y) &=& \frac{\aem W^2 (1-y)}{\pi Q^2 y (W^2 + Q^2)}  
\eea
which multiply the corresponding photoproduction cross sections, in the limit $W \gg m_p$.\footnote{Following the standard formula~\cite{Peskin:1995ev}, the cross section for a virtual photon scattering on a proton at rest into a generic final state $X$, labeled by momenta $k_i$, is
\be
\sigma(\gamma^* p \to X) = \frac{1}{4 m_p |\mathbf q| } \prod_{i} \int \frac{d^3 k_i}{(2\pi)^3 2 E_{k_i}} (2\pi)^4 \delta^4(p+q - \textstyle \sum_i k_i) \big|\mathcal{M}(\gamma^* p \to X)\big|^2 \, .
\ee
Under Hand's convention, the photon momentum $|\mathbf{q}|$ in the flux prefactor is replaced by $\nu - Q^2/(2m_p)$, which is the three-momentum for an equivalent real photon that would give the same total center-of-mass energy $W$ as the virtual photon.
The electroproduction cross section does not depend on this choice provided different kinematic factors are absorbed into the definitions of the effective fluxes $\Gamma_{T,L}$~\cite{Bauer:1977iq}.
}
Also, we have $Q_{\rm min}^2 = m_e^2 y^2/(1-y)$.

Before we proceed further, let us provide the virtual-photoproduction cross sections for vector mesons in our model.
These formulae are not needed here, but are included for completeness and may be used in future work for constraining our model with vector-meson electroproduction data away from the $Q^2=0$ limit.
First, we write the differential cross sections as a sum of natural and unnatural parity contributions
\be 
\frac{d\sigma_{T,L}(\gamma^* p \to V p)}{dt} = \frac{d\sigmap_{T,L}(\gamma^* p \to V p)}{dt} + \frac{d\sigmam_{T,L}(\gamma^* p \to V p)}{dt} \, .
\ee
Under Hand's convention, the differential cross section is evaluated as  
\be
\frac{d\sigmapm(\gamma_\lambda^* p \to Vp)}{dt} = \frac{1}{32\pi W^2(W^2 + Q^2)} \sum_{\rm spins} \big|\mathcal{M}_\pm(\gamma_\lambda^* p \to Vp)\big|^2 \, ,
\ee
where the photon polarization $\lambda$ is excluded from the sum over spins. 
Next, following Sec.~\ref{sec:model}, we evaluate the squared matrix elements in Eqs.~\eqref{eq:matrix_pseudo} and \eqref{eq:VpVp}.
Here, however, the definite photon polarization vector $\epsilon_\lambda^\mu$ enters explicitly and we are left with various Lorentz scalar products involving $\epsilon_\lambda^\mu$.
To proceed, we work the frame where the initial proton is at rest and the virtual photon momentum is aligned along the $z$-axis, i.e., 
\be
p^\mu = (m_p, 0, 0,0) \, , \quad q^\mu = \left(\nu, 0, 0, \sqrt{Q^2 + \nu^2}\right) \, .
\ee
The transverse and longitudinal polarizations take the form, respectively,
\be
\epsilon_\pm^\mu = \frac{1}{\sqrt{2}} (0,1,\pm i, 0) \, , \quad \epsilon_0^\mu = \frac{1}{\sqrt{Q^2}} \left(\sqrt{\nu^2 + Q^2}, 0, 0, \nu\right) \, ,
\ee
satisfying (for space-like photon momentum $q^\mu$)
\be
\epsilon_\pm \cdot \epsilon_\pm^* = -1 \, , \quad
\epsilon_0 \cdot \epsilon_0^* = + 1 \, .
\ee
For the longitudinal cross section $d\sigmapm_L(\gamma p \to V p)/dt$, the Lorentz scalar products needed are
\be
\epsilon_0 \cdot p = \frac{1}{2\sqrt{Q^2}} \left( W^2 + Q^2 \right) \, , \quad 
\epsilon_0 \cdot q^\prime = \frac{1}{2\sqrt{Q^2}} \sqrt{(m_V^2 - t - Q^2)^2 + 4 Q^2(m_V^2 - t)}
\, , 
\ee
taking the limit $W \gg m_p$, as well as $\epsilon_0 \cdot q = 0$ and $\epsilon_0 \cdot p^\prime = \epsilon_0 \cdot (p - q^\prime)$.
For the transverse cross section, we average over transverse polarizations
\be \label{eq:sigma_trans}
\frac{d\sigmapm_T(\gamma^* p \to V p)}{dt} = \frac{1}{2} \left( \frac{d\sigmapm(\gamma^*_+ p \to V p)}{dt} + \frac{d\sigmapm(\gamma^*_- p \to V p)}{dt}\right) \, .
\ee
The identity 
\be
\epsilon_+^\mu \epsilon_+^{*\nu} + \epsilon_-^\mu \epsilon_-^{*\nu} = \epsilon_0^\mu \epsilon_0^{*\nu} - 
g^{\mu \nu} - \frac{q^\mu q^\nu}{Q^2}
\ee
is useful to express the sum over transverse polarizations in Eq.~\eqref{eq:sigma_trans} in terms of Lorentz scalar products given above.

With these manipulations, the positive-parity cross sections from pomeron exchange are
\bea 
\frac{d \sigmap_T(\gamma^* p \to V p)}{dt} &=& 
\frac{\aem f_V^2}{8 m_V^2} \beta_{\Pom NN}^2 \left( 1 + \frac{Q^2}{2W^2} \right)^2
\notag \\
&& \; \times \;  \Big\{a_{\Pom VV}^2 \big[(m_V^2 - t)^2 + Q^4 - 2 m_V^2 Q^2 \big]  + 2 a_{\Pom VV} b_{\Pom VV} (m_V^2 - Q^2) 
+ \; b_{\Pom VV}^2 \Big\}  
\notag \\
&& \; \times \; \left|F^{(\Pom)}_{pV}(t)\right|^2  
\left( \frac{W^2}{s_0} \right)^{2 \alpha_\Pom(t) - 2} 
\times \left\{ \begin{array}{cc} 
9 & {\rm for} \; V = \rho^0 \\
1 & {\rm for} \; V = \omega \\
2 & {\rm for} \; V=\phi \end{array} \right. \, .
\label{eq:sigma_trans_nat}\\
\frac{d \sigmap_L(\gamma^* p \to V p)}{dt} &=& 
\frac{\aem f_V^2}{2 m_V^2} \beta_{\Pom NN}^2 \left( 1 + \frac{Q^2}{2W^2} \right)^2 a_{\Pom VV}^2  Q^2 (m_V^2 - t) 
\notag \\
&& \; \times \; \left|F^{(\Pom)}_{pV}(t)\right|^2 \left( \frac{W^2}{s_0} \right)^{2 \alpha_\Pom(t) - 2} 
\times \left\{ \begin{array}{cc} 
9 & {\rm for} \; V = \rho^0 \\
1 & {\rm for} \; V = \omega \\
2 & {\rm for} \; V=\phi \end{array} \right. \, .
\label{eq:sigma_long_nat}
\eea
The negative-parity cross sections from pseudoscalar exchange are
\bea \label{eq:sigma_trans_unnat}
\frac{d \sigmam_T(\gamma^* p \to V p)}{dt} &=&  \frac{\alpha_{\rm em} |t|}{16 W^2 (W^2 + Q^2)} \left|\mathcal{A}_V(t)\right|^2 \left( (t-m_V^2)^2 + Q^4 + 2m_V^2 Q^2 \right)\, , \\
\label{eq:sigma_long_unnat}
\frac{d \sigmam_L(\gamma^* p \to V p)}{dt} &=&  \frac{\alpha_{\rm em} t^2 Q^2}{4 W^2 (W^2 + Q^2)} \left|\mathcal{A}_V(t)\right|^2 \, . 
\eea
We retain only the leading terms in powers of $W$, assuming $W \gg m_p, m_B, \sqrt{|t|}$. 
(We do not assume that $Q^2$ is small compared to $W^2$.)
In the $Q^2 = 0$ limit, the transverse cross sections reduce to those for photoproduction given above, while the longitudinal cross sections vanish as expected.

Next, we turn to $B$-boson virtual-photoproduction.
As above, the cross section is a sum of natural and unnatural parity contributions
\be 
\frac{d\sigma_{T,L}(\gamma^* p \to B p)}{dt} = \frac{d\sigmap_{T,L}(\gamma^* p \to B p)}{dt} + \frac{d\sigmam_{T,L}(\gamma^* p \to B p)}{dt} \, .
\ee
Using the Feynman rules of our model and the same manipulations given above, we have the following results.
The pseudoscalar-exchange contributions are
\bea
\frac{d \sigmam_T(\gamma^* p \to B p)}{dt} &=& \frac{\pi \alpha_{\rm em} \alpha_B}{36 W^2(W^2 + Q^2)} |t| \left( (t-m_B^2)^2 + Q^4 + 2Q^2 m_B^2 \right) \left| \mathcal{A}_B(t,Q^2))\right|^2  \\
\frac{d \sigmam_L(\gamma^* p \to B p)}{dt} &=& \frac{\pi \alpha_{\rm em} \alpha_B}{9 W^2(W^2 + Q^2)} t^2 Q^2 \left| \mathcal{A}_B(t,Q^2))\right|^2
\eea
where 
\be
\mathcal{A}_B(t,Q^2) = \frac{\sqrt{2} \mathcal{A}_\omega(t) f_\omega F_\omega(m_B^2) F_\omega(-Q^2)}{m_\omega}
+ \frac{\mathcal{A}_\phi(t) f_\phi F_\phi(m_B^2) F_\phi(-Q^2) }{m_\phi} \, .
\ee
The pomeron-exchange contributions are
\bea
\frac{d \sigmap_T(\gamma^* p \to B p)}{dt} &=&   \frac{\pi \alpha_B \alpha_{\rm em}}{9} \beta_{\Pom NN}^2 \left(\frac{W^2}{s_0}\right)^{2 \alpha_\Pom(t) -2} \left( 1 + \frac{Q^2}{2W^2} \right)^2 
\notag \\
&& \; \times \; \Big\{ \big|a_{\Pom \gamma B}(t,m_B^2,-Q^2)\big|^2 \big( (m_B^2 - t)^2 - 2 Q^2 m_B^2 +Q^4 \big)  \notag \\
&& \qquad + \; 2 (m_B^2-Q^2) \, {\rm Re}\big[ a_{\Pom \gamma B}(t,m_B^2,-Q^2) \, b_{\Pom\gamma B}^*(t,m_B^2,-Q^2) \big]
\notag \\
&& \qquad + \; \big|b_{\Pom \gamma B}(t,m_B^2,-Q^2)\big|^2 \Big\}  
\\
\frac{d \sigmap_L(\gamma^* p \to B p)}{dt} &=& \frac{4 \pi \alpha_B \alpha_{\rm em}}{9} \beta_{\Pom NN}^2 \left(\frac{W^2}{s_0}\right)^{2 \alpha_\Pom(t) -2} \left( 1 + \frac{Q^2}{2W^2} \right)^2 
\notag \\
&& \; \times \; \big|a_{\Pom \gamma B}(t,m_B^2,-Q^2)\big|^2 Q^2 (m_B^2 - t) 
\eea
where $a_{\Pom \gamma B}, b_{\Pom \gamma B}$ are defined in Eqs.~\eqref{eq:a_PgB} and \eqref{eq:b_PgB} and we take the limit $W \gg \sqrt{|t|}, m_p$.
Again, for $Q^2 = 0$, the transverse cross sections reduce to our previous results for real photoproduction given in Eqs.~\eqref{eq:dsigdtBminus} and \eqref{eq:dsigdtBplus}, while the longitudinal cross sections vanish as expected.

\begin{figure}[t]
\begin{center}
\includegraphics[width=\textwidth]{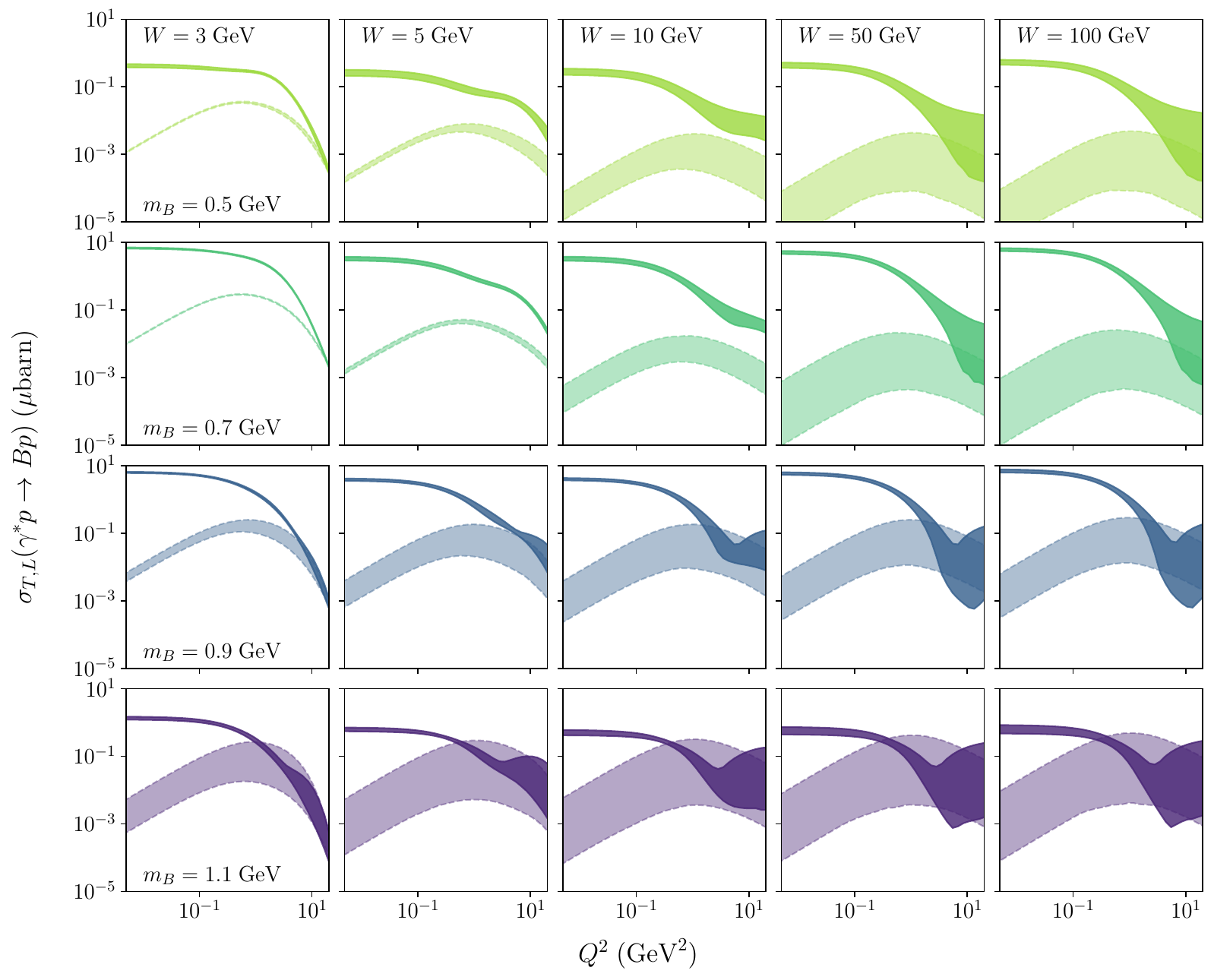}
\end{center}
\caption{\it $B$-boson photoproduction cross sections for transverse (darker band, solid lines) and longitudinal (lighter band, dashed lines) virtual photons are shown as a function of $Q^2$ for a grid of $m_B$ (rows) and $W$ (columns) values. 
Bands denote predictions from our phenomenogical model (90\% confidence intervals). Coupling $\alpha_B = 1$ is fixed.}
\label{fig:electro_B_diff}
\end{figure}

\subsection{Results}

Figure~\ref{fig:electro_B_diff} shows our predictions for the virtual-photoproduction cross sections for $B$ bosons, obtained by integrating the above formulas over $t$.
The dark band (solid lines) is the transverse cross section, while the lighter band (dashed lines) is the longitudinal cross section, as a function of $Q^2$.
Panels correspond to a grid of values for $m_B$ (rows) and $W$ (columns).
The width of each band represents the 90\% confidence interval from our parameter fit.

For $Q^2 < 0.1\; {\rm GeV}^2$, the effect of photon virtuality is neglible.
The transverse cross section is asymptotically equal to the real-photoproduction cross section, while the longitudinal cross section is comparatively suppressed.
In this regime, the electroproduction cross section is precisely determined by our phenomenological fit.
On the other hand, for larger $Q^2$, the longitudinal cross section may become comparable to the transverse one.
In this case, our fit does not well-constrain model predictions.
Presumably, this could be improved by including data from vector-meson electroproduction at larger $Q^2$ in our fits, but this remains for future work.

\begin{figure}[t]
\begin{center}
\includegraphics[width=0.6\textwidth]{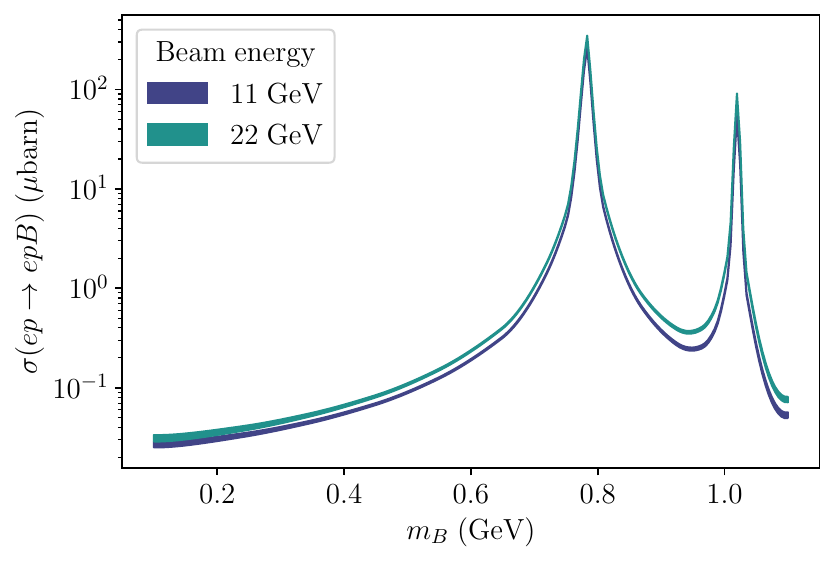}
\end{center}
\caption{\it Total $B$-boson electroproduction cross section is shown as a function of $m_B$ for electron beam energy $E_{\rm beam} = 11 \; {\rm GeV}$ and $22 \; {\rm GeV}$ on a fixed proton target, predicted from our phenomenogical model (90\% confidence intervals). Coupling $\alpha_B = 1$ is fixed.}
\label{fig:electro_B_tot}
\end{figure}

In Fig.~\ref{fig:electro_B_tot}, we show the total $B$-boson electroproduction cross section on a proton target
\be
\sigma(e p \to e p B) = \int dQ^2 dy \, \Big( \Gamma_T(Q^2,y) \sigma_T(\gamma^* p \to B p) + \Gamma_L(Q^2,y) \sigma_L(\gamma^* p \to B p) \Big) \, , 
\ee
integrated over the kinematically-allowed region, for electron beam energies $E_{\rm beam} = 11 \; {\rm GeV}$ and $22 \; {\rm GeV}$ on a fixed proton target.
This is representative of current and future beam energies at Jefferson Laboratory~\cite{Arrington:2021alx,Accardi:2023chb}.
Similar to $B$-boson photoproduction, our predictions vary most with $m_B$, but are less sensitive to the center-of-mass energy.
For fixed $m_B$, our predictions have relatively small uncertainties since the integral is dominated by the real-photoproduction region with small-$Q^2$, i.e., the diffractive region where our model is directly constrained by experiment.

For comparison, beam luminosities with the 22-GeV upgrade at Jefferson Laboratory are projected to be $\mathcal{L} \sim (10^{35} - 10^{38}) \; {\rm cm}^{-2} {\rm s}^{-1}$~\cite{Accardi:2023chb}.
According to Fig.~\ref{fig:electro_B_tot}, the total electroproduction cross section is in range $\sim (0.03 - 300) \times \alpha_B \; \mu{\rm barn}$.  
This translates into a total $B$-boson production rate of
\be
\mathcal{O}(30 - 300,000) \times \left(\frac{\alpha_B}{0.01}\right)
\left(\frac{\mathcal{L}}{10^{35} \; {\rm cm}^{-2} {\rm s}^{-1}}\right) \; {\rm s}^{-1} \, .
\ee
Here, we scale our predictions relative to $\alpha_B = 0.01$, which is the approximate upper limit applicable for most of $B$-boson parameter space in the $\sim 0.5 - 1\; {\rm GeV}$ mass range~\cite{Gan:2020aco}.

\begin{figure}[t]
\begin{center}
\includegraphics[width=\textwidth]{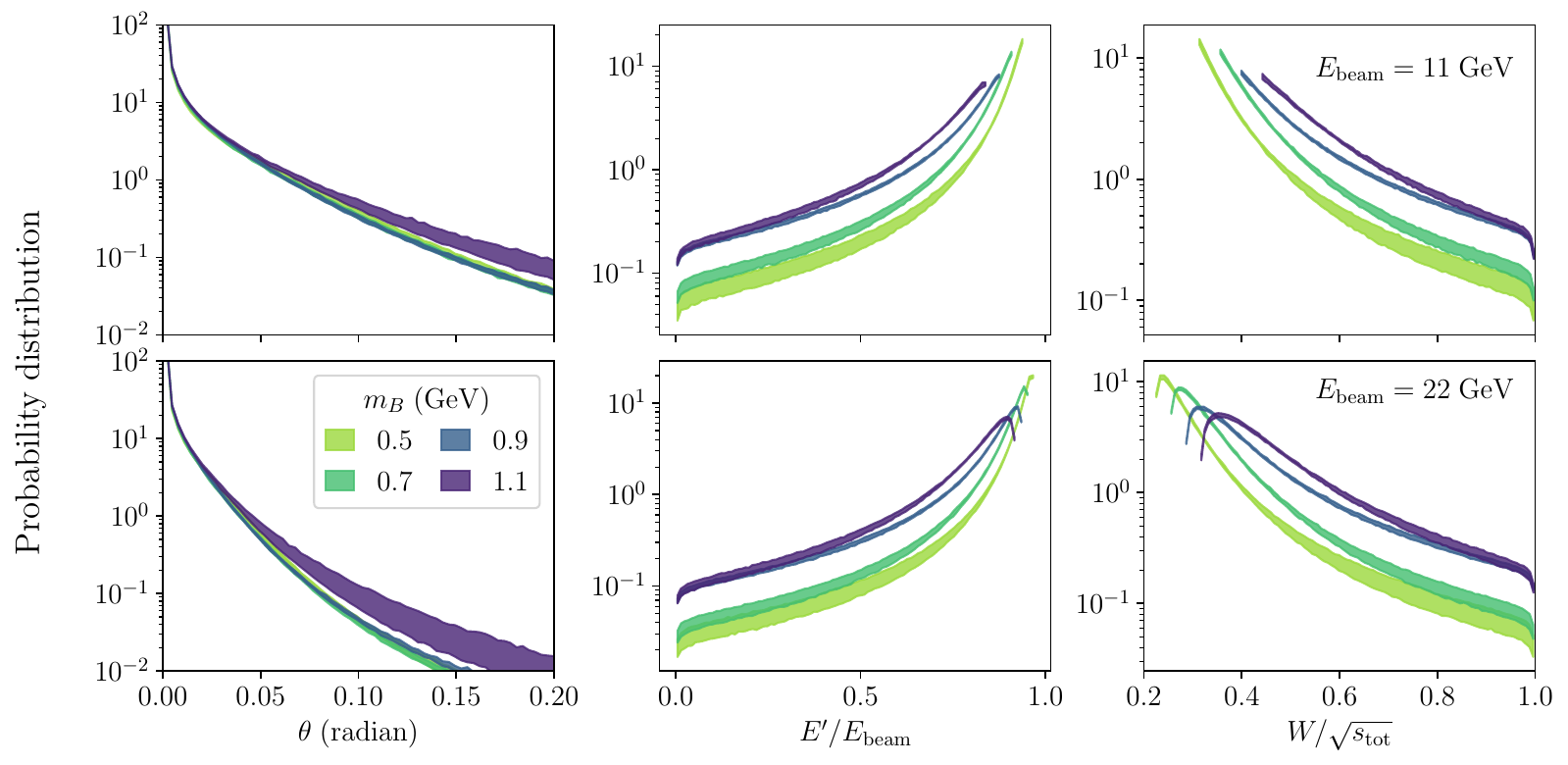}
\end{center}
\caption{\it Probability distributions of three kinematic quantities in $B$-boson electroproduction, electron scattering angle $\theta$ in the lab frame (left), outgoing electron energy $E^\prime$ relative to the beam energy $E_{\rm beam}$ (center), and photon-proton center-of-mass energy $W$ relative to the total electron-proton center-of-mass energy $\sqrt{s_{\rm tot}}$ (right), for $E_{\rm beam} = 11 \; {\rm GeV}$ (top) and $22 \; {\rm GeV}$ (bottom).
Different colored bands indicate predictions from our phenomenological model (90\% confidence intervals) for different $m_B$ values, from $m_B = 0.5\; {\rm GeV}$ (lightest) to $m_B = 1.1 \; {\rm GeV}$ (darkest).}
\label{fig:electro_B_pdfs}
\end{figure}

Of course, acceptance rates in real experiments are less than production rates due to incomplete coverage of kinematic phase space, i.e., from detector limitations or selection cuts.
Along these lines, Fig.~\ref{fig:electro_B_pdfs} shows the kinematic distributions of several quantities in $B$-boson electroproduction on a fixed proton target: scattering angle $\theta$ and relative energy of the scattered electron in the lab frame, and the center-of-mass energy of the photon-proton system relative to that of the total electron-proton system.
The vertical axis in these plots represents the marginalized probability distributions in these variables, with the total integral under the curves normalized to unity.
It is clear that the process is dominated by forward-scattering where the photon-proton system represented in Fig.~\ref{fig:photo_feynman_diagram} is near threshold.

\section{Conclusions}

Leptophobic gauge forces can provide different signatures compared to dark photons and other new states coupled to leptons.
As many experiments continue the search for new physics at the GeV scale, it is important to consider all possibilities.
The $B$ boson is the minimal model along these lines and is currently being searched for in rare meson decays.
However, direct production of GeV-scale $B$ bosons in colliders offers a complementary strategy that has not yet been searched for.

In this work, we calculated the real and virtual photoproduction cross sections for $B$ bosons on a proton target.
Combined with knowledge of $B$-boson decay channels~\cite{Tulin:2014tya}, these results can be used in experimental searches to make predictions and (in the absence of a discovery) set limits on $B$ boson parameter space.
Our calculation is based on phenomenological models for diffractive vector-meson photoproduction in the SM, as well as VMD.
We performed a comprehensive fit to the current world's dataset for $\omega$-meson and $\phi$-meson photoproduction to fix the phenomenological parameters of our model.
Our formulae contain complete kinematic information and can be used to determine acceptance efficiencies in experiments.

Our phenomenological approach can be generalized to other leptophobic $Z^\prime$ models as well.
For iso-singlet $Z^\prime$ bosons, this is straightforward modification of Eq.~\eqref{eq:Bboson_matrix_element}.
However, if the $Z^\prime$ boson is not an iso-singlet, one must consider $\rho^0$-meson mixing in addition to $\omega,\phi$-mixing considered here.
In this case, one must extend the $t$-channel model to include an additional state, e.g., $\sigma$-meson exchange, in order to reproduce experimental data for $\rho^0$-meson photoproduction~\cite{Joos1964}.
Extending our model along these lines and fitting to experimental data would be desirable, but we leave this to future work.

\begin{acknowledgments}

ST fondly remembers Martin Block for his guidance, encouragement, and kind hospitality in Aspen during the early stages of this work.
We are additionally indebted to Biplab Dey, Liping Gan, Ashot Gasparian, and Mike Williams for helpful discussions and correspondence.
We gratefully acknowledge use of \texttt{Jaxodraw} for preparing Feynman diagrams~\cite{Binosi:2003yf} and \texttt{FeynCalc}~\cite{Shtabovenko:2016sxi,Shtabovenko:2020gxv} for performing algebraic calculations in this work.

\end{acknowledgments}

\appendix

\section{Total cross sections and the optical theorem}
\label{app:optical}
The optical theorem allows us to relate the total (inclusive) vector-meson-proton cross section to the imaginary part of the elastic forward-scattering amplitude, namely, 
\be
V(q,\lambda) \; \; p(p,s) \longrightarrow V(q^\prime,\lambda^\prime) \; \; p(p^\prime,s^\prime)
\ee
where $s$ $(s^\prime)$ and $\lambda$ ($\lambda^\prime$) denote the helicities of the incoming (outgoing) proton and vector meson, respectively.
In the large-$s$ limit, the optical theorem takes the form~\cite{Peskin:1995ev} 
\be \label{eq:optical}
\sigma(V p \to {\rm all}) = \frac{1}{s} {\rm Im} \left[ \mathcal{M}(Vp \to Vp) \right]\Big|_{t=0} \, ,
\ee
where $s = (p + q)^2$ and the forward limit corresponds to $t=0$.

Here we calculate the total cross section for diffractive scattering.
Within our model, this is the positive-parity pomeron-exchange cross section, with amplitude given by Eq.~\eqref{eq:VpVp}.
In the forward limit, i.e., setting $p = p^\prime$, $q=q^\prime$, we have
\bea
i \mathcal{M}_+\big( V p \to V p\big) &=& \frac{3 \beta_{\Pom NN} F_{p V}^{(\Pom)}(0)}{s} \left( \frac{-i s}{s_0} \right)^{\alpha_\Pom(0) - 1} \notag \delta_{ss^\prime} \delta_{\lambda \lambda^\prime} \\
&& \qquad \times 
\Big[ b_{\Pom VV} \left( q^2 (\epsilon_\lambda \cdot p)^2 + \epsilon_\lambda^2 (q \cdot p)^2 \right) + 2 a_{\Pom VV} q^2 \epsilon_\lambda^2 (p \cdot q)^2 \Big] \, .
\eea
We have used the fact that $\bar u(p,s^\prime) \gamma^\mu u(p,s) = 2 p^\mu \delta_{ss^\prime}$ and neglected terms proportional to $m_p$.
We do not set $q^2 = m_V^2$, as we allow the vector meson to be off-shell.
Working in the centre-of-mass frame, the transverse polarization vectors $\epsilon_\pm$ satisfy
\be
\epsilon_{\pm} \cdot \epsilon_\pm^* = -1 \, , \quad \epsilon_\pm \cdot p = 0 \; ,
\ee
while the longitudinal polarization vector $\epsilon_0$ satisfies
\be
\epsilon_{0} \cdot \epsilon_0^* = -1 \, , \quad \epsilon_0 \cdot p = \frac{s}{2 \sqrt{q^2} } \; ,
\ee
provided $q$ is a time-like four-vector and $s \gg q^2$.

Using the optical theorem~\eqref{eq:optical}, we obtain the total cross sections
\bea
\sigma_+^{(T)}(V p \to {\rm all}) &=& 3 \beta_{\Pom NN} \left( b_{\Pom VV} + 2 a_{\Pom VV} q^2  \right) \cos\left(\frac{\pi \epsilon_\Pom }{2} \right) \left( \frac{s}{s_0} \right)^{\epsilon_\Pom} \label{eq:tot_T} \\
\sigma_+^{(L)}(V p \to {\rm all}) &=& 6 \beta_{\Pom NN}  a_{\Pom VV} q^2  \cos\left(\frac{\pi \epsilon_\Pom }{2} \right) \left( \frac{s}{s_0} \right)^{\epsilon_\Pom} \, \label{eq:tot_L}
\eea
for tranverse ($T$) and longitudinal ($L$) vector-meson polarizations.
Here we have used $F_{pV}^{(\Pom)}(0) = 1$ and $\epsilon_\Pom = \alpha_\Pom(0) - 1$.
Notably, different polarization states have different cross sections, as pointed out in Ref.~\cite{Ewerz:2013kda}.

For large $s$, it is expected that diffractive scattering dominates the total cross section.
Hence, Eqs.~\eqref{eq:tot_T} and \eqref{eq:tot_L} must be positive, which in turn imposes constraints on the couplings $a_{\Pom VV}$, $b_{\Pom VV}$.
For on-shell scattering ($q^2 = m_V^2$), we have
\be \label{eq:pom_constraints}
a_{\Pom VV} > 0 \; , \quad 
b_{\Pom VV} + 2 m_V^2 a_{\Pom VV} > 0 \; .
\ee

Next, we calculate the total cross section for photon-proton scattering.
Using the optical theorem and VMD, we have
\bea 
\sigma_+^{(T)}(\gamma p \to {\rm all}) &=& 2\pi \alpha_{\rm em} \left( \frac{f_\rho^2}{m_\rho^2} \sigma_+^{(T)}(\rho^0 p \to {\rm all}) + \frac{f_\omega^2}{9 m_\omega^2} \sigma_+^{(T)}(\omega p \to {\rm all}) + \frac{2f_\phi^2}{9m_\phi^2} \sigma_+^{(T)}(\phi p \to {\rm all}) \right) \notag 
\\
&=& 6\pi \alpha_{\rm em} \beta_{\Pom NN}  \cos\left(\frac{\pi \epsilon_\Pom }{2} \right) \left( \frac{s}{s_0} \right)^{\epsilon_\Pom}
\left( \frac{f_\rho^2}{m_\rho^2} b_{\Pom \rho \rho} + \frac{f_\omega^2}{9 m_\omega^2} b_{\Pom \omega \omega} + \frac{2f_\phi^2}{9m_\phi^2} b_{\Pom \phi \phi} \right) \label{eq:tot_gamma} \, ,
\eea
setting $q^2 = 0$ for a real photon.

The total photon-proton cross section has been measured by the H1~\cite{H1:1995hmw} and ZEUS~\cite{ZEUS:2001wan} collaborations to be
\be
\sigma(\gamma p \to {\rm all}) = \left\{ \begin{array}{ll} 
165 \pm 2 \pm 11 \; \mu{\rm barn} & {\rm at} \; \sqrt{s} = 200 \; {\rm GeV} \; {\rm [H1]} \\
174 \pm 1 \pm 13  \; \mu{\rm barn} & {\rm at} \; \sqrt{s} = 209 \; {\rm GeV} \; {\rm [ZEUS]} 
\end{array}\right.
\ee
where the uncertainties represent statistical and systematic errors, respectively.
Here we impose these measurements as an additional constraint on our model via Eq.~\eqref{eq:tot_gamma}, with the added assumption $b_{\Pom \rho \rho} = b_{\Pom \omega \omega}$.

\section{Comparison of model and data}

\label{sec:compare}

\begin{figure}[t]
\begin{centering}
\includegraphics[width=0.95\textwidth]{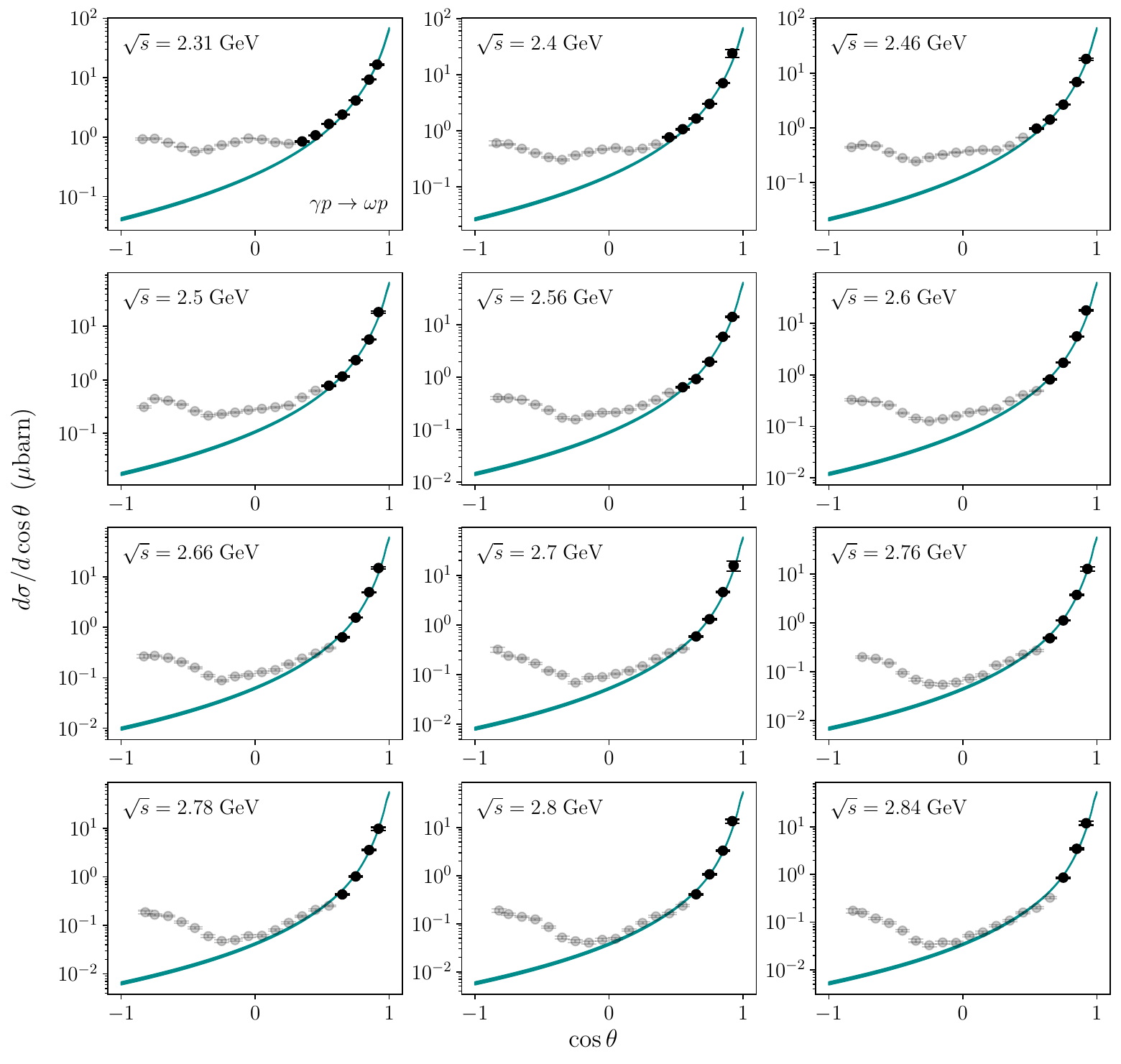}
\end{centering}
\caption{\it Shaded band shows our phenomenological model for the $\gamma p \to \omega p$ differential cross section (90\% confidence intervals), for different center-of-mass energies $\sqrt{s}$ and where $\theta$ is the scattering angle in the center-of-mass frame.
Data points are experimental results from CLAS~\cite{Williams:2009ab}.
Black points are included in our fit, gray points are not.}
\label{fig:Omegaplots}
\end{figure}

\begin{figure}[t]
\begin{centering}
\includegraphics[width=0.95\textwidth]{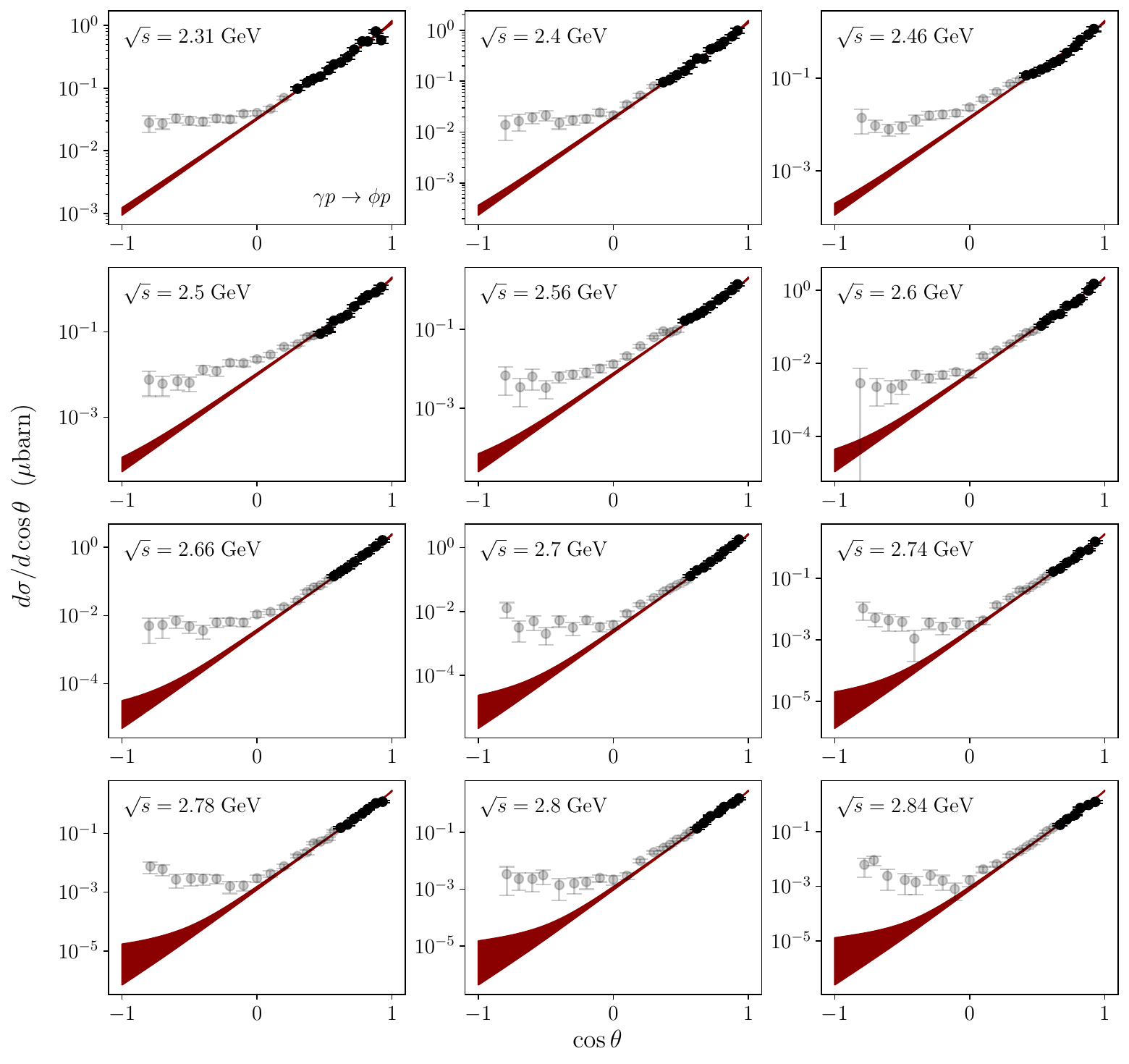}
\end{centering}
\caption{\it Shaded bands shows our phenomenological model for the $\gamma p \to \phi p$ differential cross section (90\% confidence intervals), for different center-of-mass energies $\sqrt{s}$ and where $\theta$ is the scattering angle in the center-of-mass frame..
Data points are experimental results from CLAS~\cite{Dey:2014tfa}.
Black points are included in our fit, gray points are not.}
\label{fig:Phiplots}
\end{figure}

\begin{figure}[t]
\begin{centering}
\includegraphics[width=0.96\textwidth]{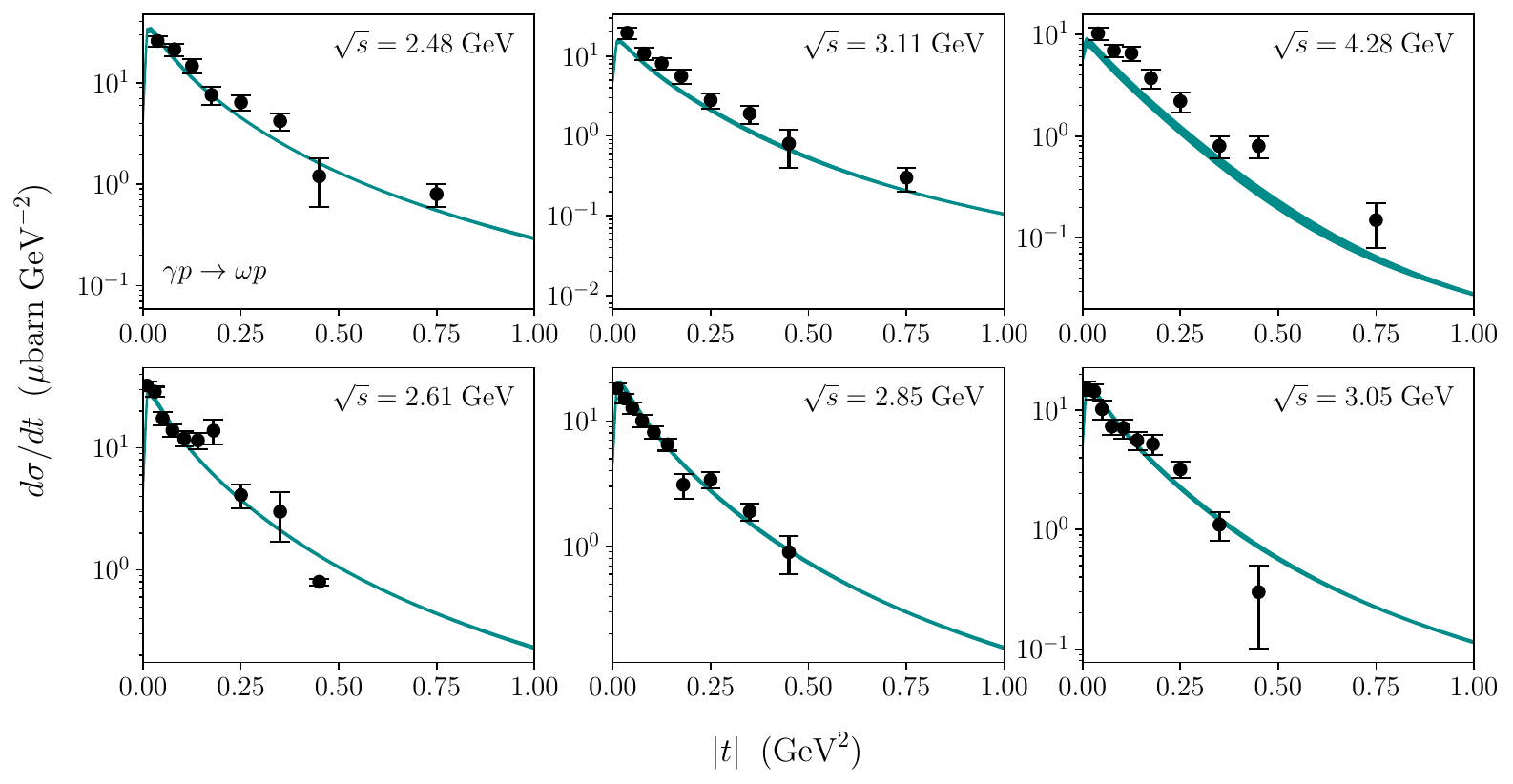}

\vspace{0.5cm}

\includegraphics[width=0.64\textwidth]{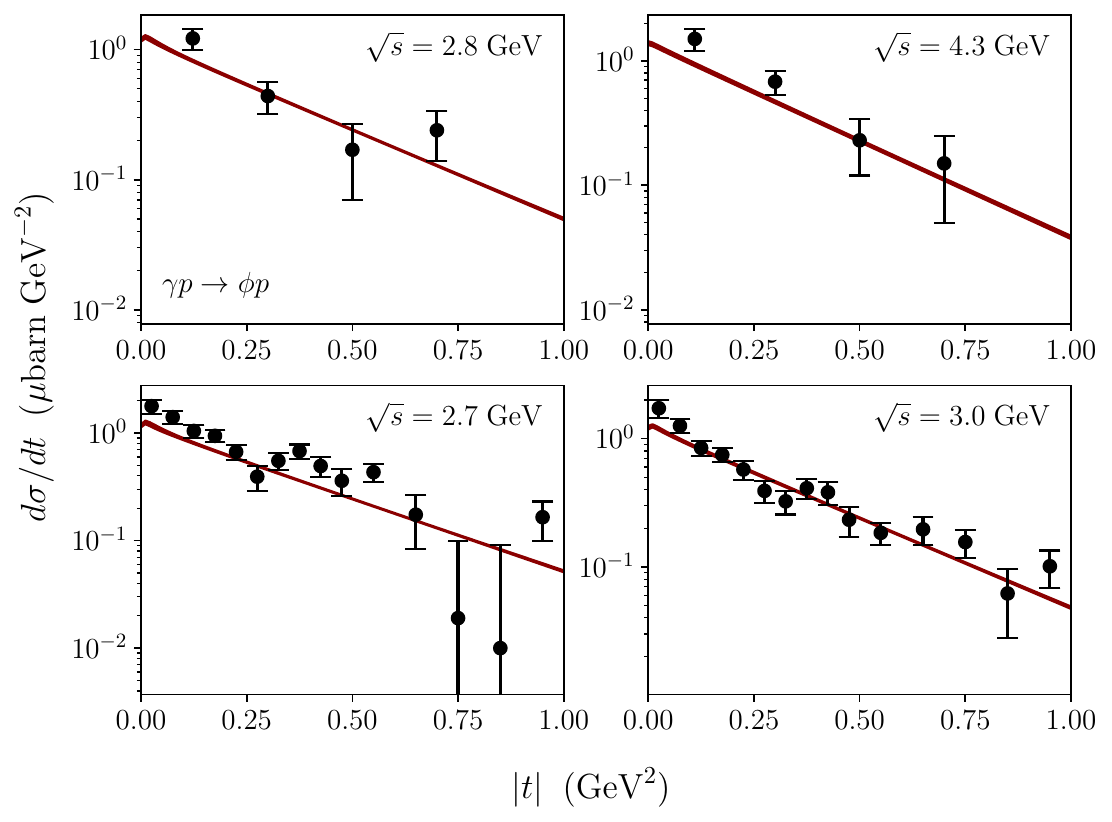}
\end{centering}
\caption{\it Shaded bands show our phenomenological model (90\% confidence intervals) for the differential cross sections for $\gamma p \to \omega p$ (top figure, blue bands) and $\gamma p \to \phi p$ (bottom figure, red bands) for different center-of-mass energies $\sqrt{s}$.
Data points are experimental results obtained at SLAC~\cite{Ballam:1972eq} (upper sets of panels) and Daresbury~\cite{Barber:1985fr,Barber:1981fj} (lower sets of panels).
All data points shown are included in our fit.}
\label{fig:oldplots}
\end{figure}

\begin{figure}[t]
\begin{centering}
\includegraphics[width=0.64\textwidth]{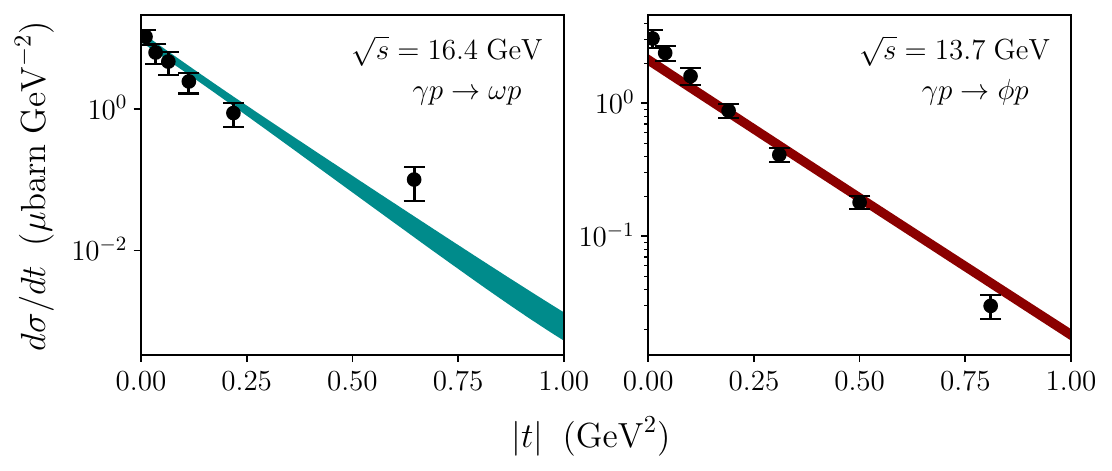}

\vspace{0.5cm}

\includegraphics[width=0.96\textwidth]{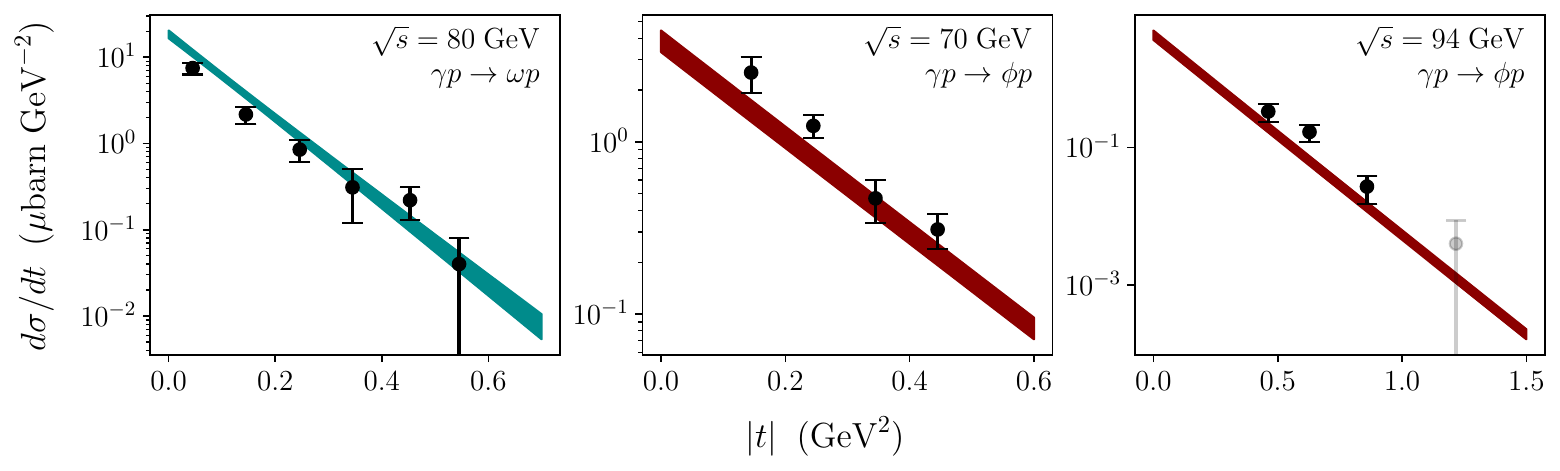}
\end{centering}
\caption{\it Shaded bands show our phenomenological model (90\% confidence intervals) for the $\gamma p \to \omega p, \phi p$ differential cross sections, for different center-of-mass energies $\sqrt{s}$.
Data points are experimental results obtained at Fermilab~\cite{Busenitz:1989gq} (upper panels) and from ZEUS (lower panels, left~\cite{Derrick:1996yt}, middle~\cite{Derrick:1996af}, and right~\cite{ZEUS:1999ptu}).
Black points are included in our fit, gray points are not.}
\label{fig:Busenitzplots}
\end{figure}

To illustrate the success of our fit, Figs.~\ref{fig:Omegaplots}, \ref{fig:Phiplots}, \ref{fig:oldplots}, \ref{fig:Busenitzplots} show the comparison between our phenomenological model and differential cross section measurements in our dataset. 
Shaded bands represent the 90\% confidence intervals from our fit for $\omega$-meson (blue) and $\phi$-meson (red) photoproduction.
Also, black data points are those included in our fit, while gray points are not, as per~\eqref{eq:limit_dataset}. 

Most of the data points in our fit come from CLAS~\cite{Williams:2009ab,Dey:2014tfa} with $\sqrt{s} \lesssim 2.8 \; {\rm GeV}$, shown in Figs.~\ref{fig:Omegaplots} and \ref{fig:Phiplots}.
However, it is reassuring that our model successfully describes other datasets as well.
These include data from SLAC~\cite{Ballam:1972eq} and Daresbury~\cite{Barber:1981fj,Barber:1985fr}, Fermilab~\cite{Busenitz:1989gq}, and even ZEUS~\cite{Derrick:1996af,Derrick:1996yt,ZEUS:1999ptu} at much larger energy (shown in Figs.~\ref{fig:oldplots} and \ref{fig:Busenitzplots}).
Thus, we have confidence in the accuracy of our model as a function of $\sqrt{s}$, except for near threshold.

On the other hand, our model does not account for the data outside the forward-scattering region.
While this could be improved by expanding the model to include contributions beyond $t$-channel exchange~\cite{Zhao:1998fn,Oh:2000zi}, we defer this to future study.
In any case, it is clear that photoproduction is in fact dominated by forward-scattering and our $t$-channel model provides a good description of the data.

\bibliography{Bphoto}

\begin{thebibliography}{97}%
\makeatletter
\providecommand \@ifxundefined [1]{%
 \@ifx{#1\undefined}
}%
\providecommand \@ifnum [1]{%
 \ifnum #1\expandafter \@firstoftwo
 \else \expandafter \@secondoftwo
 \fi
}%
\providecommand \@ifx [1]{%
 \ifx #1\expandafter \@firstoftwo
 \else \expandafter \@secondoftwo
 \fi
}%
\providecommand \natexlab [1]{#1}%
\providecommand \enquote  [1]{``#1''}%
\providecommand \bibnamefont  [1]{#1}%
\providecommand \bibfnamefont [1]{#1}%
\providecommand \citenamefont [1]{#1}%
\providecommand \href@noop [0]{\@secondoftwo}%
\providecommand \href [0]{\begingroup \@sanitize@url \@href}%
\providecommand \@href[1]{\@@startlink{#1}\@@href}%
\providecommand \@@href[1]{\endgroup#1\@@endlink}%
\providecommand \@sanitize@url [0]{\catcode `\\12\catcode `\$12\catcode
  `\&12\catcode `\#12\catcode `\^12\catcode `\_12\catcode `\%12\relax}%
\providecommand \@@startlink[1]{}%
\providecommand \@@endlink[0]{}%
\providecommand \url  [0]{\begingroup\@sanitize@url \@url }%
\providecommand \@url [1]{\endgroup\@href {#1}{\urlprefix }}%
\providecommand \urlprefix  [0]{URL }%
\providecommand \Eprint [0]{\href }%
\providecommand \doibase [0]{http://dx.doi.org/}%
\providecommand \selectlanguage [0]{\@gobble}%
\providecommand \bibinfo  [0]{\@secondoftwo}%
\providecommand \bibfield  [0]{\@secondoftwo}%
\providecommand \translation [1]{[#1]}%
\providecommand \BibitemOpen [0]{}%
\providecommand \bibitemStop [0]{}%
\providecommand \bibitemNoStop [0]{.\EOS\space}%
\providecommand \EOS [0]{\spacefactor3000\relax}%
\providecommand \BibitemShut  [1]{\csname bibitem#1\endcsname}%
\let\auto@bib@innerbib\@empty
\bibitem [{\citenamefont {Boehm}\ and\ \citenamefont
  {Fayet}(2004)}]{Boehm:2003hm}%
  \BibitemOpen
  \bibfield  {author} {\bibinfo {author} {\bibfnamefont {C.}~\bibnamefont
  {Boehm}}\ and\ \bibinfo {author} {\bibfnamefont {P.}~\bibnamefont {Fayet}},\
  }\href {\doibase 10.1016/j.nuclphysb.2004.01.015} {\bibfield  {journal}
  {\bibinfo  {journal} {Nucl. Phys.}\ }\textbf {\bibinfo {volume} {B683}},\
  \bibinfo {pages} {219} (\bibinfo {year} {2004})},\ \Eprint
  {http://arxiv.org/abs/hep-ph/0305261} {arXiv:hep-ph/0305261 [hep-ph]}
  \BibitemShut {NoStop}%
\bibitem [{\citenamefont {Fayet}(2007)}]{Fayet:2007ua}%
  \BibitemOpen
  \bibfield  {author} {\bibinfo {author} {\bibfnamefont {P.}~\bibnamefont
  {Fayet}},\ }\href {\doibase 10.1103/PhysRevD.75.115017} {\bibfield  {journal}
  {\bibinfo  {journal} {Phys.Rev.}\ }\textbf {\bibinfo {volume} {D75}},\
  \bibinfo {pages} {115017} (\bibinfo {year} {2007})},\ \Eprint
  {http://arxiv.org/abs/hep-ph/0702176} {arXiv:hep-ph/0702176 [HEP-PH]}
  \BibitemShut {NoStop}%
\bibitem [{\citenamefont {Pospelov}(2009)}]{Pospelov:2008zw}%
  \BibitemOpen
  \bibfield  {author} {\bibinfo {author} {\bibfnamefont {M.}~\bibnamefont
  {Pospelov}},\ }\href {\doibase 10.1103/PhysRevD.80.095002} {\bibfield
  {journal} {\bibinfo  {journal} {Phys. Rev.}\ }\textbf {\bibinfo {volume}
  {D80}},\ \bibinfo {pages} {095002} (\bibinfo {year} {2009})},\ \Eprint
  {http://arxiv.org/abs/0811.1030} {arXiv:0811.1030 [hep-ph]} \BibitemShut
  {NoStop}%
\bibitem [{\citenamefont {Krasznahorkay}\ \emph {et~al.}(2016)\citenamefont
  {Krasznahorkay} \emph {et~al.}}]{Krasznahorkay:2015iga}%
  \BibitemOpen
  \bibfield  {author} {\bibinfo {author} {\bibfnamefont {A.~J.}\ \bibnamefont
  {Krasznahorkay}} \emph {et~al.},\ }\href {\doibase
  10.1103/PhysRevLett.116.042501} {\bibfield  {journal} {\bibinfo  {journal}
  {Phys. Rev. Lett.}\ }\textbf {\bibinfo {volume} {116}},\ \bibinfo {pages}
  {042501} (\bibinfo {year} {2016})},\ \Eprint
  {http://arxiv.org/abs/1504.01527} {arXiv:1504.01527 [nucl-ex]} \BibitemShut
  {NoStop}%
\bibitem [{\citenamefont {Krasznahorkay}\ \emph {et~al.}(2019)\citenamefont
  {Krasznahorkay} \emph {et~al.}}]{Krasznahorkay:2019lyl}%
  \BibitemOpen
  \bibfield  {author} {\bibinfo {author} {\bibfnamefont {A.~J.}\ \bibnamefont
  {Krasznahorkay}} \emph {et~al.},\ }\href@noop {} {\  (\bibinfo {year}
  {2019})},\ \Eprint {http://arxiv.org/abs/1910.10459} {arXiv:1910.10459
  [nucl-ex]} \BibitemShut {NoStop}%
\bibitem [{\citenamefont {Pospelov}\ \emph {et~al.}(2008)\citenamefont
  {Pospelov}, \citenamefont {Ritz},\ and\ \citenamefont
  {Voloshin}}]{Pospelov:2007mp}%
  \BibitemOpen
  \bibfield  {author} {\bibinfo {author} {\bibfnamefont {M.}~\bibnamefont
  {Pospelov}}, \bibinfo {author} {\bibfnamefont {A.}~\bibnamefont {Ritz}}, \
  and\ \bibinfo {author} {\bibfnamefont {M.~B.}\ \bibnamefont {Voloshin}},\
  }\href {\doibase 10.1016/j.physletb.2008.02.052} {\bibfield  {journal}
  {\bibinfo  {journal} {Phys.Lett.}\ }\textbf {\bibinfo {volume} {B662}},\
  \bibinfo {pages} {53} (\bibinfo {year} {2008})},\ \Eprint
  {http://arxiv.org/abs/0711.4866} {arXiv:0711.4866 [hep-ph]} \BibitemShut
  {NoStop}%
\bibitem [{\citenamefont {Arkani-Hamed}\ \emph {et~al.}(2009)\citenamefont
  {Arkani-Hamed}, \citenamefont {Finkbeiner}, \citenamefont {Slatyer},\ and\
  \citenamefont {Weiner}}]{ArkaniHamed:2008qn}%
  \BibitemOpen
  \bibfield  {author} {\bibinfo {author} {\bibfnamefont {N.}~\bibnamefont
  {Arkani-Hamed}}, \bibinfo {author} {\bibfnamefont {D.~P.}\ \bibnamefont
  {Finkbeiner}}, \bibinfo {author} {\bibfnamefont {T.~R.}\ \bibnamefont
  {Slatyer}}, \ and\ \bibinfo {author} {\bibfnamefont {N.}~\bibnamefont
  {Weiner}},\ }\href {\doibase 10.1103/PhysRevD.79.015014} {\bibfield
  {journal} {\bibinfo  {journal} {Phys.Rev.}\ }\textbf {\bibinfo {volume}
  {D79}},\ \bibinfo {pages} {015014} (\bibinfo {year} {2009})},\ \Eprint
  {http://arxiv.org/abs/0810.0713} {arXiv:0810.0713 [hep-ph]} \BibitemShut
  {NoStop}%
\bibitem [{\citenamefont {Pospelov}\ and\ \citenamefont
  {Ritz}(2009)}]{Pospelov:2008jd}%
  \BibitemOpen
  \bibfield  {author} {\bibinfo {author} {\bibfnamefont {M.}~\bibnamefont
  {Pospelov}}\ and\ \bibinfo {author} {\bibfnamefont {A.}~\bibnamefont
  {Ritz}},\ }\href {\doibase 10.1016/j.physletb.2008.12.012} {\bibfield
  {journal} {\bibinfo  {journal} {Phys.Lett.}\ }\textbf {\bibinfo {volume}
  {B671}},\ \bibinfo {pages} {391} (\bibinfo {year} {2009})},\ \Eprint
  {http://arxiv.org/abs/0810.1502} {arXiv:0810.1502 [hep-ph]} \BibitemShut
  {NoStop}%
\bibitem [{\citenamefont {Feng}\ \emph {et~al.}(2009)\citenamefont {Feng},
  \citenamefont {Kaplinghat}, \citenamefont {Tu},\ and\ \citenamefont
  {Yu}}]{Feng:2009mn}%
  \BibitemOpen
  \bibfield  {author} {\bibinfo {author} {\bibfnamefont {J.~L.}\ \bibnamefont
  {Feng}}, \bibinfo {author} {\bibfnamefont {M.}~\bibnamefont {Kaplinghat}},
  \bibinfo {author} {\bibfnamefont {H.}~\bibnamefont {Tu}}, \ and\ \bibinfo
  {author} {\bibfnamefont {H.-B.}\ \bibnamefont {Yu}},\ }\href {\doibase
  10.1088/1475-7516/2009/07/004} {\bibfield  {journal} {\bibinfo  {journal}
  {JCAP}\ }\textbf {\bibinfo {volume} {07}},\ \bibinfo {pages} {004} (\bibinfo
  {year} {2009})},\ \Eprint {http://arxiv.org/abs/0905.3039} {arXiv:0905.3039
  [hep-ph]} \BibitemShut {NoStop}%
\bibitem [{\citenamefont {Hooper}\ \emph {et~al.}(2012)\citenamefont {Hooper},
  \citenamefont {Weiner},\ and\ \citenamefont {Xue}}]{Hooper:2012cw}%
  \BibitemOpen
  \bibfield  {author} {\bibinfo {author} {\bibfnamefont {D.}~\bibnamefont
  {Hooper}}, \bibinfo {author} {\bibfnamefont {N.}~\bibnamefont {Weiner}}, \
  and\ \bibinfo {author} {\bibfnamefont {W.}~\bibnamefont {Xue}},\ }\href
  {\doibase 10.1103/PhysRevD.86.056009} {\bibfield  {journal} {\bibinfo
  {journal} {Phys.Rev.}\ }\textbf {\bibinfo {volume} {D86}},\ \bibinfo {pages}
  {056009} (\bibinfo {year} {2012})},\ \Eprint {http://arxiv.org/abs/1206.2929}
  {arXiv:1206.2929 [hep-ph]} \BibitemShut {NoStop}%
\bibitem [{\citenamefont {Holdom}(1986)}]{Holdom:1985ag}%
  \BibitemOpen
  \bibfield  {author} {\bibinfo {author} {\bibfnamefont {B.}~\bibnamefont
  {Holdom}},\ }\href {\doibase 10.1016/0370-2693(86)91377-8} {\bibfield
  {journal} {\bibinfo  {journal} {Phys.Lett.}\ }\textbf {\bibinfo {volume}
  {B166}},\ \bibinfo {pages} {196} (\bibinfo {year} {1986})}\BibitemShut
  {NoStop}%
\bibitem [{\citenamefont {Jaeckel}\ and\ \citenamefont
  {Ringwald}(2010)}]{Jaeckel:2010ni}%
  \BibitemOpen
  \bibfield  {author} {\bibinfo {author} {\bibfnamefont {J.}~\bibnamefont
  {Jaeckel}}\ and\ \bibinfo {author} {\bibfnamefont {A.}~\bibnamefont
  {Ringwald}},\ }\href {\doibase 10.1146/annurev.nucl.012809.104433} {\bibfield
   {journal} {\bibinfo  {journal} {Ann. Rev. Nucl. Part. Sci.}\ }\textbf
  {\bibinfo {volume} {60}},\ \bibinfo {pages} {405} (\bibinfo {year} {2010})},\
  \Eprint {http://arxiv.org/abs/1002.0329} {arXiv:1002.0329 [hep-ph]}
  \BibitemShut {NoStop}%
\bibitem [{\citenamefont {Essig}\ \emph {et~al.}(2013)\citenamefont {Essig},
  \citenamefont {Jaros}, \citenamefont {Wester}, \citenamefont {Adrian},
  \citenamefont {Andreas} \emph {et~al.}}]{Essig:2013lka}%
  \BibitemOpen
  \bibfield  {author} {\bibinfo {author} {\bibfnamefont {R.}~\bibnamefont
  {Essig}}, \bibinfo {author} {\bibfnamefont {J.~A.}\ \bibnamefont {Jaros}},
  \bibinfo {author} {\bibfnamefont {W.}~\bibnamefont {Wester}}, \bibinfo
  {author} {\bibfnamefont {P.~H.}\ \bibnamefont {Adrian}}, \bibinfo {author}
  {\bibfnamefont {S.}~\bibnamefont {Andreas}},  \emph {et~al.},\ }\href@noop {}
  {\  (\bibinfo {year} {2013})},\ \Eprint {http://arxiv.org/abs/1311.0029}
  {arXiv:1311.0029 [hep-ph]} \BibitemShut {NoStop}%
\bibitem [{\citenamefont {Alexander}\ \emph {et~al.}(2016)\citenamefont
  {Alexander} \emph {et~al.}}]{Alexander:2016aln}%
  \BibitemOpen
  \bibfield  {author} {\bibinfo {author} {\bibfnamefont {J.}~\bibnamefont
  {Alexander}} \emph {et~al.}\ }(\bibinfo {year} {2016})\ \Eprint
  {http://arxiv.org/abs/1608.08632} {arXiv:1608.08632 [hep-ph]} \BibitemShut
  {NoStop}%
\bibitem [{\citenamefont {Rajpoot}(1989)}]{Rajpoot:1989jb}%
  \BibitemOpen
  \bibfield  {author} {\bibinfo {author} {\bibfnamefont {S.}~\bibnamefont
  {Rajpoot}},\ }\href {\doibase 10.1103/PhysRevD.40.2421} {\bibfield  {journal}
  {\bibinfo  {journal} {Phys.Rev.}\ }\textbf {\bibinfo {volume} {D40}},\
  \bibinfo {pages} {2421} (\bibinfo {year} {1989})}\BibitemShut {NoStop}%
\bibitem [{\citenamefont {Foot}\ \emph {et~al.}(1989)\citenamefont {Foot},
  \citenamefont {Joshi},\ and\ \citenamefont {Lew}}]{Foot:1989ts}%
  \BibitemOpen
  \bibfield  {author} {\bibinfo {author} {\bibfnamefont {R.}~\bibnamefont
  {Foot}}, \bibinfo {author} {\bibfnamefont {G.~C.}\ \bibnamefont {Joshi}}, \
  and\ \bibinfo {author} {\bibfnamefont {H.}~\bibnamefont {Lew}},\ }\href
  {\doibase 10.1103/PhysRevD.40.2487} {\bibfield  {journal} {\bibinfo
  {journal} {Phys.Rev.}\ }\textbf {\bibinfo {volume} {D40}},\ \bibinfo {pages}
  {2487} (\bibinfo {year} {1989})}\BibitemShut {NoStop}%
\bibitem [{\citenamefont {Nelson}\ and\ \citenamefont
  {Tetradis}(1989)}]{Nelson:1989fx}%
  \BibitemOpen
  \bibfield  {author} {\bibinfo {author} {\bibfnamefont {A.~E.}\ \bibnamefont
  {Nelson}}\ and\ \bibinfo {author} {\bibfnamefont {N.}~\bibnamefont
  {Tetradis}},\ }\href {\doibase 10.1016/0370-2693(89)90196-2} {\bibfield
  {journal} {\bibinfo  {journal} {Phys.Lett.}\ }\textbf {\bibinfo {volume}
  {B221}},\ \bibinfo {pages} {80} (\bibinfo {year} {1989})}\BibitemShut
  {NoStop}%
\bibitem [{\citenamefont {He}\ and\ \citenamefont {Rajpoot}(1990)}]{He:1989mi}%
  \BibitemOpen
  \bibfield  {author} {\bibinfo {author} {\bibfnamefont {X.-G.}\ \bibnamefont
  {He}}\ and\ \bibinfo {author} {\bibfnamefont {S.}~\bibnamefont {Rajpoot}},\
  }\href {\doibase 10.1103/PhysRevD.41.1636} {\bibfield  {journal} {\bibinfo
  {journal} {Phys.Rev.}\ }\textbf {\bibinfo {volume} {D41}},\ \bibinfo {pages}
  {1636} (\bibinfo {year} {1990})}\BibitemShut {NoStop}%
\bibitem [{\citenamefont {Carone}\ and\ \citenamefont
  {Murayama}(1995{\natexlab{a}})}]{Carone:1994aa}%
  \BibitemOpen
  \bibfield  {author} {\bibinfo {author} {\bibfnamefont {C.~D.}\ \bibnamefont
  {Carone}}\ and\ \bibinfo {author} {\bibfnamefont {H.}~\bibnamefont
  {Murayama}},\ }\href {\doibase 10.1103/PhysRevLett.74.3122} {\bibfield
  {journal} {\bibinfo  {journal} {Phys.Rev.Lett.}\ }\textbf {\bibinfo {volume}
  {74}},\ \bibinfo {pages} {3122} (\bibinfo {year} {1995}{\natexlab{a}})},\
  \Eprint {http://arxiv.org/abs/hep-ph/9411256} {arXiv:hep-ph/9411256 [hep-ph]}
  \BibitemShut {NoStop}%
\bibitem [{\citenamefont {Bailey}\ and\ \citenamefont
  {Davidson}(1995)}]{Bailey:1994qv}%
  \BibitemOpen
  \bibfield  {author} {\bibinfo {author} {\bibfnamefont {D.~C.}\ \bibnamefont
  {Bailey}}\ and\ \bibinfo {author} {\bibfnamefont {S.}~\bibnamefont
  {Davidson}},\ }\href {\doibase 10.1016/0370-2693(95)00052-M} {\bibfield
  {journal} {\bibinfo  {journal} {Phys.Lett.}\ }\textbf {\bibinfo {volume}
  {B348}},\ \bibinfo {pages} {185} (\bibinfo {year} {1995})},\ \Eprint
  {http://arxiv.org/abs/hep-ph/9411355} {arXiv:hep-ph/9411355 [hep-ph]}
  \BibitemShut {NoStop}%
\bibitem [{\citenamefont {Carone}\ and\ \citenamefont
  {Murayama}(1995{\natexlab{b}})}]{Carone:1995pu}%
  \BibitemOpen
  \bibfield  {author} {\bibinfo {author} {\bibfnamefont {C.~D.}\ \bibnamefont
  {Carone}}\ and\ \bibinfo {author} {\bibfnamefont {H.}~\bibnamefont
  {Murayama}},\ }\href {\doibase 10.1103/PhysRevD.52.484} {\bibfield  {journal}
  {\bibinfo  {journal} {Phys.Rev.}\ }\textbf {\bibinfo {volume} {D52}},\
  \bibinfo {pages} {484} (\bibinfo {year} {1995}{\natexlab{b}})},\ \Eprint
  {http://arxiv.org/abs/hep-ph/9501220} {arXiv:hep-ph/9501220 [hep-ph]}
  \BibitemShut {NoStop}%
\bibitem [{\citenamefont {Aranda}\ and\ \citenamefont
  {Carone}(1998)}]{Aranda:1998fr}%
  \BibitemOpen
  \bibfield  {author} {\bibinfo {author} {\bibfnamefont {A.}~\bibnamefont
  {Aranda}}\ and\ \bibinfo {author} {\bibfnamefont {C.~D.}\ \bibnamefont
  {Carone}},\ }\href {\doibase 10.1016/S0370-2693(98)01309-4} {\bibfield
  {journal} {\bibinfo  {journal} {Phys.Lett.}\ }\textbf {\bibinfo {volume}
  {B443}},\ \bibinfo {pages} {352} (\bibinfo {year} {1998})},\ \Eprint
  {http://arxiv.org/abs/hep-ph/9809522} {arXiv:hep-ph/9809522 [hep-ph]}
  \BibitemShut {NoStop}%
\bibitem [{\citenamefont {Fileviez~Perez}\ and\ \citenamefont
  {Wise}(2010)}]{FileviezPerez:2010gw}%
  \BibitemOpen
  \bibfield  {author} {\bibinfo {author} {\bibfnamefont {P.}~\bibnamefont
  {Fileviez~Perez}}\ and\ \bibinfo {author} {\bibfnamefont {M.~B.}\
  \bibnamefont {Wise}},\ }\href {\doibase 10.1103/PhysRevD.82.079901,
  10.1103/PhysRevD.82.011901} {\bibfield  {journal} {\bibinfo  {journal}
  {Phys.Rev.}\ }\textbf {\bibinfo {volume} {D82}},\ \bibinfo {pages} {011901}
  (\bibinfo {year} {2010})},\ \Eprint {http://arxiv.org/abs/1002.1754}
  {arXiv:1002.1754 [hep-ph]} \BibitemShut {NoStop}%
\bibitem [{\citenamefont {Graesser}\ \emph {et~al.}(2011)\citenamefont
  {Graesser}, \citenamefont {Shoemaker},\ and\ \citenamefont
  {Vecchi}}]{Graesser:2011vj}%
  \BibitemOpen
  \bibfield  {author} {\bibinfo {author} {\bibfnamefont {M.~L.}\ \bibnamefont
  {Graesser}}, \bibinfo {author} {\bibfnamefont {I.~M.}\ \bibnamefont
  {Shoemaker}}, \ and\ \bibinfo {author} {\bibfnamefont {L.}~\bibnamefont
  {Vecchi}},\ }\href@noop {} {\  (\bibinfo {year} {2011})},\ \Eprint
  {http://arxiv.org/abs/1107.2666} {arXiv:1107.2666 [hep-ph]} \BibitemShut
  {NoStop}%
\bibitem [{\citenamefont {Ilten}\ \emph {et~al.}(2018)\citenamefont {Ilten},
  \citenamefont {Soreq}, \citenamefont {Williams},\ and\ \citenamefont
  {Xue}}]{Ilten:2018crw}%
  \BibitemOpen
  \bibfield  {author} {\bibinfo {author} {\bibfnamefont {P.}~\bibnamefont
  {Ilten}}, \bibinfo {author} {\bibfnamefont {Y.}~\bibnamefont {Soreq}},
  \bibinfo {author} {\bibfnamefont {M.}~\bibnamefont {Williams}}, \ and\
  \bibinfo {author} {\bibfnamefont {W.}~\bibnamefont {Xue}},\ }\href {\doibase
  10.1007/JHEP06(2018)004} {\bibfield  {journal} {\bibinfo  {journal} {JHEP}\
  }\textbf {\bibinfo {volume} {06}},\ \bibinfo {pages} {004} (\bibinfo {year}
  {2018})},\ \Eprint {http://arxiv.org/abs/1801.04847} {arXiv:1801.04847
  [hep-ph]} \BibitemShut {NoStop}%
\bibitem [{\citenamefont {Dror}\ \emph
  {et~al.}(2017{\natexlab{a}})\citenamefont {Dror}, \citenamefont {Lasenby},\
  and\ \citenamefont {Pospelov}}]{Dror:2017ehi}%
  \BibitemOpen
  \bibfield  {author} {\bibinfo {author} {\bibfnamefont {J.~A.}\ \bibnamefont
  {Dror}}, \bibinfo {author} {\bibfnamefont {R.}~\bibnamefont {Lasenby}}, \
  and\ \bibinfo {author} {\bibfnamefont {M.}~\bibnamefont {Pospelov}},\ }\href
  {\doibase 10.1103/PhysRevLett.119.141803} {\bibfield  {journal} {\bibinfo
  {journal} {Phys. Rev. Lett.}\ }\textbf {\bibinfo {volume} {119}},\ \bibinfo
  {pages} {141803} (\bibinfo {year} {2017}{\natexlab{a}})},\ \Eprint
  {http://arxiv.org/abs/1705.06726} {arXiv:1705.06726 [hep-ph]} \BibitemShut
  {NoStop}%
\bibitem [{\citenamefont {Dror}\ \emph
  {et~al.}(2017{\natexlab{b}})\citenamefont {Dror}, \citenamefont {Lasenby},\
  and\ \citenamefont {Pospelov}}]{Dror:2017nsg}%
  \BibitemOpen
  \bibfield  {author} {\bibinfo {author} {\bibfnamefont {J.~A.}\ \bibnamefont
  {Dror}}, \bibinfo {author} {\bibfnamefont {R.}~\bibnamefont {Lasenby}}, \
  and\ \bibinfo {author} {\bibfnamefont {M.}~\bibnamefont {Pospelov}},\ }\href
  {\doibase 10.1103/PhysRevD.96.075036} {\bibfield  {journal} {\bibinfo
  {journal} {Phys. Rev. D}\ }\textbf {\bibinfo {volume} {96}},\ \bibinfo
  {pages} {075036} (\bibinfo {year} {2017}{\natexlab{b}})},\ \Eprint
  {http://arxiv.org/abs/1707.01503} {arXiv:1707.01503 [hep-ph]} \BibitemShut
  {NoStop}%
\bibitem [{\citenamefont {Gan}\ \emph {et~al.}(2022)\citenamefont {Gan},
  \citenamefont {Kubis}, \citenamefont {Passemar},\ and\ \citenamefont
  {Tulin}}]{Gan:2020aco}%
  \BibitemOpen
  \bibfield  {author} {\bibinfo {author} {\bibfnamefont {L.}~\bibnamefont
  {Gan}}, \bibinfo {author} {\bibfnamefont {B.}~\bibnamefont {Kubis}}, \bibinfo
  {author} {\bibfnamefont {E.}~\bibnamefont {Passemar}}, \ and\ \bibinfo
  {author} {\bibfnamefont {S.}~\bibnamefont {Tulin}},\ }\href {\doibase
  10.1016/j.physrep.2021.11.001} {\bibfield  {journal} {\bibinfo  {journal}
  {Phys. Rept.}\ }\textbf {\bibinfo {volume} {945}},\ \bibinfo {pages} {1}
  (\bibinfo {year} {2022})},\ \Eprint {http://arxiv.org/abs/2007.00664}
  {arXiv:2007.00664 [hep-ph]} \BibitemShut {NoStop}%
\bibitem [{\citenamefont {Tulin}(2014)}]{Tulin:2014tya}%
  \BibitemOpen
  \bibfield  {author} {\bibinfo {author} {\bibfnamefont {S.}~\bibnamefont
  {Tulin}},\ }\href {\doibase 10.1103/PhysRevD.89.114008} {\bibfield  {journal}
  {\bibinfo  {journal} {Phys. Rev.}\ }\textbf {\bibinfo {volume} {D89}},\
  \bibinfo {pages} {114008} (\bibinfo {year} {2014})},\ \Eprint
  {http://arxiv.org/abs/1404.4370} {arXiv:1404.4370 [hep-ph]} \BibitemShut
  {NoStop}%
\bibitem [{\citenamefont {Gan}(2015)}]{Gan:2015nyc}%
  \BibitemOpen
  \bibfield  {author} {\bibinfo {author} {\bibfnamefont {L.}~\bibnamefont
  {Gan}},\ }\bibfield  {booktitle} {\emph {\bibinfo {booktitle} {{Proceedings,
  8th International Workshop on Chiral Dynamics (CD15): Pisa, Italy, June
  29-July 3, 2015}}},\ }\href@noop {} {\bibfield  {journal} {\bibinfo
  {journal} {PoS}\ }\textbf {\bibinfo {volume} {CD15}},\ \bibinfo {pages} {017}
  (\bibinfo {year} {2015})}\BibitemShut {NoStop}%
\bibitem [{\citenamefont {Amelino-Camelia}\ \emph {et~al.}(2010)\citenamefont
  {Amelino-Camelia}, \citenamefont {Archilli}, \citenamefont {Babusci},
  \citenamefont {Badoni}, \citenamefont {Bencivenni} \emph
  {et~al.}}]{AmelinoCamelia:2010me}%
  \BibitemOpen
  \bibfield  {author} {\bibinfo {author} {\bibfnamefont {G.}~\bibnamefont
  {Amelino-Camelia}}, \bibinfo {author} {\bibfnamefont {F.}~\bibnamefont
  {Archilli}}, \bibinfo {author} {\bibfnamefont {D.}~\bibnamefont {Babusci}},
  \bibinfo {author} {\bibfnamefont {D.}~\bibnamefont {Badoni}}, \bibinfo
  {author} {\bibfnamefont {G.}~\bibnamefont {Bencivenni}},  \emph {et~al.},\
  }\href {\doibase 10.1140/epjc/s10052-010-1351-1} {\bibfield  {journal}
  {\bibinfo  {journal} {Eur.Phys.J.}\ }\textbf {\bibinfo {volume} {C68}},\
  \bibinfo {pages} {619} (\bibinfo {year} {2010})},\ \Eprint
  {http://arxiv.org/abs/1003.3868} {arXiv:1003.3868 [hep-ex]} \BibitemShut
  {NoStop}%
\bibitem [{\citenamefont {del Rio}(2021)}]{delRio:2021xag}%
  \BibitemOpen
  \bibfield  {author} {\bibinfo {author} {\bibfnamefont {E.~P.}\ \bibnamefont
  {del Rio}} (\bibinfo {collaboration} {KLOE-2}),\ }\href@noop {} {\  (\bibinfo
  {year} {2021})},\ \Eprint {http://arxiv.org/abs/2112.10110} {arXiv:2112.10110
  [hep-ex]} \BibitemShut {NoStop}%
\bibitem [{\citenamefont {Won}\ \emph {et~al.}(2016)\citenamefont {Won} \emph
  {et~al.}}]{Won:2016pjz}%
  \BibitemOpen
  \bibfield  {author} {\bibinfo {author} {\bibfnamefont {E.}~\bibnamefont
  {Won}} \emph {et~al.} (\bibinfo {collaboration} {Belle}),\ }\href {\doibase
  10.1103/PhysRevD.94.092006} {\bibfield  {journal} {\bibinfo  {journal} {Phys.
  Rev.}\ }\textbf {\bibinfo {volume} {D94}},\ \bibinfo {pages} {092006}
  (\bibinfo {year} {2016})},\ \Eprint {http://arxiv.org/abs/1609.05599}
  {arXiv:1609.05599 [hep-ex]} \BibitemShut {NoStop}%
\bibitem [{\citenamefont {Arrington}\ \emph {et~al.}(2022)\citenamefont
  {Arrington} \emph {et~al.}}]{Arrington:2021alx}%
  \BibitemOpen
  \bibfield  {author} {\bibinfo {author} {\bibfnamefont {J.}~\bibnamefont
  {Arrington}} \emph {et~al.},\ }\href {\doibase 10.1016/j.ppnp.2022.103985}
  {\bibfield  {journal} {\bibinfo  {journal} {Prog. Part. Nucl. Phys.}\
  }\textbf {\bibinfo {volume} {127}},\ \bibinfo {pages} {103985} (\bibinfo
  {year} {2022})},\ \Eprint {http://arxiv.org/abs/2112.00060} {arXiv:2112.00060
  [nucl-ex]} \BibitemShut {NoStop}%
\bibitem [{\citenamefont {Accardi}\ \emph {et~al.}(2023)\citenamefont {Accardi}
  \emph {et~al.}}]{Accardi:2023chb}%
  \BibitemOpen
  \bibfield  {author} {\bibinfo {author} {\bibfnamefont {A.}~\bibnamefont
  {Accardi}} \emph {et~al.},\ }\href@noop {} {\  (\bibinfo {year} {2023})},\
  \Eprint {http://arxiv.org/abs/2306.09360} {arXiv:2306.09360 [nucl-ex]}
  \BibitemShut {NoStop}%
\bibitem [{\citenamefont {Accardi}\ \emph {et~al.}(2016)\citenamefont {Accardi}
  \emph {et~al.}}]{Accardi:2012qut}%
  \BibitemOpen
  \bibfield  {author} {\bibinfo {author} {\bibfnamefont {A.}~\bibnamefont
  {Accardi}} \emph {et~al.},\ }\href {\doibase 10.1140/epja/i2016-16268-9}
  {\bibfield  {journal} {\bibinfo  {journal} {Eur. Phys. J. A}\ }\textbf
  {\bibinfo {volume} {52}},\ \bibinfo {pages} {268} (\bibinfo {year} {2016})},\
  \Eprint {http://arxiv.org/abs/1212.1701} {arXiv:1212.1701 [nucl-ex]}
  \BibitemShut {NoStop}%
\bibitem [{\citenamefont {Agostini}\ \emph {et~al.}(2021)\citenamefont
  {Agostini} \emph {et~al.}}]{LHeC:2020van}%
  \BibitemOpen
  \bibfield  {author} {\bibinfo {author} {\bibfnamefont {P.}~\bibnamefont
  {Agostini}} \emph {et~al.} (\bibinfo {collaboration} {LHeC, FCC-he Study
  Group}),\ }\href {\doibase 10.1088/1361-6471/abf3ba} {\bibfield  {journal}
  {\bibinfo  {journal} {J. Phys. G}\ }\textbf {\bibinfo {volume} {48}},\
  \bibinfo {pages} {110501} (\bibinfo {year} {2021})},\ \Eprint
  {http://arxiv.org/abs/2007.14491} {arXiv:2007.14491 [hep-ex]} \BibitemShut
  {NoStop}%
\bibitem [{\citenamefont {Br\"uning}\ \emph {et~al.}(2022)\citenamefont
  {Br\"uning}, \citenamefont {Seryi},\ and\ \citenamefont
  {Verd\'u-Andr\'es}}]{Bruning:2022hro}%
  \BibitemOpen
  \bibfield  {author} {\bibinfo {author} {\bibfnamefont {O.}~\bibnamefont
  {Br\"uning}}, \bibinfo {author} {\bibfnamefont {A.}~\bibnamefont {Seryi}}, \
  and\ \bibinfo {author} {\bibfnamefont {S.}~\bibnamefont {Verd\'u-Andr\'es}},\
  }\href {\doibase 10.3389/fphy.2022.886473} {\bibfield  {journal} {\bibinfo
  {journal} {Front. in Phys.}\ }\textbf {\bibinfo {volume} {10}},\ \bibinfo
  {pages} {886473} (\bibinfo {year} {2022})}\BibitemShut {NoStop}%
\bibitem [{\citenamefont {Bauer}\ \emph {et~al.}(1978)\citenamefont {Bauer},
  \citenamefont {Spital}, \citenamefont {Yennie},\ and\ \citenamefont
  {Pipkin}}]{Bauer:1977iq}%
  \BibitemOpen
  \bibfield  {author} {\bibinfo {author} {\bibfnamefont {T.~H.}\ \bibnamefont
  {Bauer}}, \bibinfo {author} {\bibfnamefont {R.~D.}\ \bibnamefont {Spital}},
  \bibinfo {author} {\bibfnamefont {D.~R.}\ \bibnamefont {Yennie}}, \ and\
  \bibinfo {author} {\bibfnamefont {F.~M.}\ \bibnamefont {Pipkin}},\ }\href
  {\doibase 10.1103/RevModPhys.50.261} {\bibfield  {journal} {\bibinfo
  {journal} {Rev. Mod. Phys.}\ }\textbf {\bibinfo {volume} {50}},\ \bibinfo
  {pages} {261} (\bibinfo {year} {1978})},\ \bibinfo {note} {[Erratum: Rev.
  Mod. Phys.51,407(1979)]}\BibitemShut {NoStop}%
\bibitem [{\citenamefont {Workman}\ and\ \citenamefont
  {Others}(2022)}]{Workman:2022ynf}%
  \BibitemOpen
  \bibfield  {author} {\bibinfo {author} {\bibfnamefont {R.~L.}\ \bibnamefont
  {Workman}}\ and\ \bibinfo {author} {\bibnamefont {Others}} (\bibinfo
  {collaboration} {Particle Data Group}),\ }\href {\doibase
  10.1093/ptep/ptac097} {\bibfield  {journal} {\bibinfo  {journal} {PTEP}\
  }\textbf {\bibinfo {volume} {2022}},\ \bibinfo {pages} {083C01} (\bibinfo
  {year} {2022})}\BibitemShut {NoStop}%
\bibitem [{\citenamefont {Schildknecht}(2006)}]{Schildknecht:2005xr}%
  \BibitemOpen
  \bibfield  {author} {\bibinfo {author} {\bibfnamefont {D.}~\bibnamefont
  {Schildknecht}},\ }\bibfield  {booktitle} {\emph {\bibinfo {booktitle}
  {{Lepton and photon interactions at high energies. Proceedings, 22nd
  International Symposium, LP 2005, Uppsala, Sweden, June 30-July 5, 2005}}},\
  }\href@noop {} {\bibfield  {journal} {\bibinfo  {journal} {Acta Phys.
  Polon.}\ }\textbf {\bibinfo {volume} {B37}},\ \bibinfo {pages} {595}
  (\bibinfo {year} {2006})},\ \Eprint {http://arxiv.org/abs/hep-ph/0511090}
  {arXiv:hep-ph/0511090 [hep-ph]} \BibitemShut {NoStop}%
\bibitem [{\citenamefont {Donnachie}\ and\ \citenamefont
  {Landshoff}(1984)}]{Donnachie:1984xq}%
  \BibitemOpen
  \bibfield  {author} {\bibinfo {author} {\bibfnamefont {A.}~\bibnamefont
  {Donnachie}}\ and\ \bibinfo {author} {\bibfnamefont {P.~V.}\ \bibnamefont
  {Landshoff}},\ }\bibfield  {booktitle} {\emph {\bibinfo {booktitle} {{NEW
  PARTICLE PRODUCTION. PROCEEDINGS, 19TH RENCONTRES DE MORIOND, HADRONIC
  SESSION, LA PLAGNE, FRANCE, MARCH 4-10, 1984}}},\ }\href {\doibase
  10.1016/0550-3213(84)90315-8} {\bibfield  {journal} {\bibinfo  {journal}
  {Nucl. Phys.}\ }\textbf {\bibinfo {volume} {B244}},\ \bibinfo {pages} {322}
  (\bibinfo {year} {1984})},\ \bibinfo {note} {[,813(1984)]}\BibitemShut
  {NoStop}%
\bibitem [{\citenamefont {Fanelli}\ and\ \citenamefont
  {Williams}(2017)}]{Fanelli:2016utb}%
  \BibitemOpen
  \bibfield  {author} {\bibinfo {author} {\bibfnamefont {C.}~\bibnamefont
  {Fanelli}}\ and\ \bibinfo {author} {\bibfnamefont {M.}~\bibnamefont
  {Williams}},\ }\href {\doibase 10.1088/0954-3899/44/1/014002} {\bibfield
  {journal} {\bibinfo  {journal} {J. Phys.}\ }\textbf {\bibinfo {volume}
  {G44}},\ \bibinfo {pages} {014002} (\bibinfo {year} {2017})},\ \Eprint
  {http://arxiv.org/abs/1605.07161} {arXiv:1605.07161 [hep-ph]} \BibitemShut
  {NoStop}%
\bibitem [{\citenamefont {Berman}\ and\ \citenamefont
  {Drell}(1964)}]{Berman:1963ef}%
  \BibitemOpen
  \bibfield  {author} {\bibinfo {author} {\bibfnamefont {S.~M.}\ \bibnamefont
  {Berman}}\ and\ \bibinfo {author} {\bibfnamefont {S.~D.}\ \bibnamefont
  {Drell}},\ }\href {\doibase 10.1103/PhysRev.133.B791} {\bibfield  {journal}
  {\bibinfo  {journal} {Phys. Rev.}\ }\textbf {\bibinfo {volume} {133}},\
  \bibinfo {pages} {B791} (\bibinfo {year} {1964})}\BibitemShut {NoStop}%
\bibitem [{\citenamefont {Joos}\ and\ \citenamefont {Kramer}(1964)}]{Joos1964}%
  \BibitemOpen
  \bibfield  {author} {\bibinfo {author} {\bibfnamefont {H.}~\bibnamefont
  {Joos}}\ and\ \bibinfo {author} {\bibfnamefont {G.}~\bibnamefont {Kramer}},\
  }\href {\doibase 10.1007/BF01379480} {\bibfield  {journal} {\bibinfo
  {journal} {Zeitschrift f{\"u}r Physik}\ }\textbf {\bibinfo {volume} {178}},\
  \bibinfo {pages} {542} (\bibinfo {year} {1964})}\BibitemShut {NoStop}%
\bibitem [{\citenamefont {Fraas}(1972)}]{Fraas:1972fj}%
  \BibitemOpen
  \bibfield  {author} {\bibinfo {author} {\bibfnamefont {H.}~\bibnamefont
  {Fraas}},\ }\href {\doibase 10.1016/0550-3213(72)90304-5} {\bibfield
  {journal} {\bibinfo  {journal} {Nucl. Phys. B}\ }\textbf {\bibinfo {volume}
  {36}},\ \bibinfo {pages} {191} (\bibinfo {year} {1972})}\BibitemShut
  {NoStop}%
\bibitem [{\citenamefont {Friman}\ and\ \citenamefont
  {Soyeur}(1996)}]{Friman:1995qm}%
  \BibitemOpen
  \bibfield  {author} {\bibinfo {author} {\bibfnamefont {B.}~\bibnamefont
  {Friman}}\ and\ \bibinfo {author} {\bibfnamefont {M.}~\bibnamefont
  {Soyeur}},\ }\href {\doibase 10.1016/0375-9474(96)00011-5} {\bibfield
  {journal} {\bibinfo  {journal} {Nucl. Phys.}\ }\textbf {\bibinfo {volume}
  {A600}},\ \bibinfo {pages} {477} (\bibinfo {year} {1996})},\ \Eprint
  {http://arxiv.org/abs/nucl-th/9601028} {arXiv:nucl-th/9601028 [nucl-th]}
  \BibitemShut {NoStop}%
\bibitem [{\citenamefont {Zhao}\ \emph {et~al.}(1998)\citenamefont {Zhao},
  \citenamefont {Li},\ and\ \citenamefont {Bennhold}}]{Zhao:1998fn}%
  \BibitemOpen
  \bibfield  {author} {\bibinfo {author} {\bibfnamefont {Q.}~\bibnamefont
  {Zhao}}, \bibinfo {author} {\bibfnamefont {Z.-p.}\ \bibnamefont {Li}}, \ and\
  \bibinfo {author} {\bibfnamefont {C.}~\bibnamefont {Bennhold}},\ }\href
  {\doibase 10.1103/PhysRevC.58.2393} {\bibfield  {journal} {\bibinfo
  {journal} {Phys. Rev.}\ }\textbf {\bibinfo {volume} {C58}},\ \bibinfo {pages}
  {2393} (\bibinfo {year} {1998})},\ \Eprint
  {http://arxiv.org/abs/nucl-th/9806100} {arXiv:nucl-th/9806100 [nucl-th]}
  \BibitemShut {NoStop}%
\bibitem [{\citenamefont {Oh}\ \emph {et~al.}(2001)\citenamefont {Oh},
  \citenamefont {Titov},\ and\ \citenamefont {Lee}}]{Oh:2000zi}%
  \BibitemOpen
  \bibfield  {author} {\bibinfo {author} {\bibfnamefont {Y.-s.}\ \bibnamefont
  {Oh}}, \bibinfo {author} {\bibfnamefont {A.~I.}\ \bibnamefont {Titov}}, \
  and\ \bibinfo {author} {\bibfnamefont {T.~S.~H.}\ \bibnamefont {Lee}},\
  }\href {\doibase 10.1103/PhysRevC.63.025201} {\bibfield  {journal} {\bibinfo
  {journal} {Phys. Rev.}\ }\textbf {\bibinfo {volume} {C63}},\ \bibinfo {pages}
  {025201} (\bibinfo {year} {2001})},\ \Eprint
  {http://arxiv.org/abs/nucl-th/0006057} {arXiv:nucl-th/0006057 [nucl-th]}
  \BibitemShut {NoStop}%
\bibitem [{\citenamefont {Ewerz}\ \emph {et~al.}(2014)\citenamefont {Ewerz},
  \citenamefont {Maniatis},\ and\ \citenamefont {Nachtmann}}]{Ewerz:2013kda}%
  \BibitemOpen
  \bibfield  {author} {\bibinfo {author} {\bibfnamefont {C.}~\bibnamefont
  {Ewerz}}, \bibinfo {author} {\bibfnamefont {M.}~\bibnamefont {Maniatis}}, \
  and\ \bibinfo {author} {\bibfnamefont {O.}~\bibnamefont {Nachtmann}},\ }\href
  {\doibase 10.1016/j.aop.2013.12.001} {\bibfield  {journal} {\bibinfo
  {journal} {Annals Phys.}\ }\textbf {\bibinfo {volume} {342}},\ \bibinfo
  {pages} {31} (\bibinfo {year} {2014})},\ \Eprint
  {http://arxiv.org/abs/1309.3478} {arXiv:1309.3478 [hep-ph]} \BibitemShut
  {NoStop}%
\bibitem [{\citenamefont {Barger}\ and\ \citenamefont
  {Cline}(1970)}]{Barger:1970wk}%
  \BibitemOpen
  \bibfield  {author} {\bibinfo {author} {\bibfnamefont {V.~D.}\ \bibnamefont
  {Barger}}\ and\ \bibinfo {author} {\bibfnamefont {D.}~\bibnamefont {Cline}},\
  }\href {\doibase 10.1103/PhysRevLett.24.1313} {\bibfield  {journal} {\bibinfo
   {journal} {Phys. Rev. Lett.}\ }\textbf {\bibinfo {volume} {24}},\ \bibinfo
  {pages} {1313} (\bibinfo {year} {1970})}\BibitemShut {NoStop}%
\bibitem [{\citenamefont {Halpern}\ \emph {et~al.}(1972)\citenamefont
  {Halpern}, \citenamefont {Prepost}, \citenamefont {Tompkins}, \citenamefont
  {Anderson}, \citenamefont {Gottschalk}, \citenamefont {Gustavson},
  \citenamefont {Ritson}, \citenamefont {Weitsch},\ and\ \citenamefont
  {Wiik}}]{Halpern:1972ab}%
  \BibitemOpen
  \bibfield  {author} {\bibinfo {author} {\bibfnamefont {H.~J.}\ \bibnamefont
  {Halpern}}, \bibinfo {author} {\bibfnamefont {R.}~\bibnamefont {Prepost}},
  \bibinfo {author} {\bibfnamefont {D.~H.}\ \bibnamefont {Tompkins}}, \bibinfo
  {author} {\bibfnamefont {R.~L.}\ \bibnamefont {Anderson}}, \bibinfo {author}
  {\bibfnamefont {B.}~\bibnamefont {Gottschalk}}, \bibinfo {author}
  {\bibfnamefont {D.}~\bibnamefont {Gustavson}}, \bibinfo {author}
  {\bibfnamefont {D.}~\bibnamefont {Ritson}}, \bibinfo {author} {\bibfnamefont
  {G.~A.}\ \bibnamefont {Weitsch}}, \ and\ \bibinfo {author} {\bibfnamefont
  {B.~H.}\ \bibnamefont {Wiik}},\ }\href {\doibase 10.1103/PhysRevLett.29.1425}
  {\bibfield  {journal} {\bibinfo  {journal} {Phys. Rev. Lett.}\ }\textbf
  {\bibinfo {volume} {29}},\ \bibinfo {pages} {1425} (\bibinfo {year}
  {1972})}\BibitemShut {NoStop}%
\bibitem [{\citenamefont {Schilling}\ \emph {et~al.}(1970)\citenamefont
  {Schilling}, \citenamefont {Seyboth},\ and\ \citenamefont
  {Wolf}}]{Schilling:1969um}%
  \BibitemOpen
  \bibfield  {author} {\bibinfo {author} {\bibfnamefont {K.}~\bibnamefont
  {Schilling}}, \bibinfo {author} {\bibfnamefont {P.}~\bibnamefont {Seyboth}},
  \ and\ \bibinfo {author} {\bibfnamefont {G.~E.}\ \bibnamefont {Wolf}},\
  }\href {\doibase 10.1016/0550-3213(70)90070-2} {\bibfield  {journal}
  {\bibinfo  {journal} {Nucl. Phys. B}\ }\textbf {\bibinfo {volume} {15}},\
  \bibinfo {pages} {397} (\bibinfo {year} {1970})},\ \bibinfo {note} {[Erratum:
  Nucl.Phys.B 18, 332 (1970)]}\BibitemShut {NoStop}%
\bibitem [{\citenamefont {Maris}\ and\ \citenamefont
  {Tandy}(1999)}]{Maris:1999nt}%
  \BibitemOpen
  \bibfield  {author} {\bibinfo {author} {\bibfnamefont {P.}~\bibnamefont
  {Maris}}\ and\ \bibinfo {author} {\bibfnamefont {P.~C.}\ \bibnamefont
  {Tandy}},\ }\href {\doibase 10.1103/PhysRevC.60.055214} {\bibfield  {journal}
  {\bibinfo  {journal} {Phys. Rev.}\ }\textbf {\bibinfo {volume} {C60}},\
  \bibinfo {pages} {055214} (\bibinfo {year} {1999})},\ \Eprint
  {http://arxiv.org/abs/nucl-th/9905056} {arXiv:nucl-th/9905056 [nucl-th]}
  \BibitemShut {NoStop}%
\bibitem [{\citenamefont {Machleidt}\ and\ \citenamefont
  {Slaus}(2001)}]{Machleidt:2001rw}%
  \BibitemOpen
  \bibfield  {author} {\bibinfo {author} {\bibfnamefont {R.}~\bibnamefont
  {Machleidt}}\ and\ \bibinfo {author} {\bibfnamefont {I.}~\bibnamefont
  {Slaus}},\ }\href {\doibase 10.1088/0954-3899/27/5/201} {\bibfield  {journal}
  {\bibinfo  {journal} {J. Phys.}\ }\textbf {\bibinfo {volume} {G27}},\
  \bibinfo {pages} {R69} (\bibinfo {year} {2001})},\ \Eprint
  {http://arxiv.org/abs/nucl-th/0101056} {arXiv:nucl-th/0101056 [nucl-th]}
  \BibitemShut {NoStop}%
\bibitem [{\citenamefont {Pena}\ \emph {et~al.}(2001)\citenamefont {Pena},
  \citenamefont {Garcilazo},\ and\ \citenamefont {Riska}}]{Pena:2000gb}%
  \BibitemOpen
  \bibfield  {author} {\bibinfo {author} {\bibfnamefont {M.~T.}\ \bibnamefont
  {Pena}}, \bibinfo {author} {\bibfnamefont {H.}~\bibnamefont {Garcilazo}}, \
  and\ \bibinfo {author} {\bibfnamefont {D.~O.}\ \bibnamefont {Riska}},\ }\href
  {\doibase 10.1016/S0375-9474(00)00449-8} {\bibfield  {journal} {\bibinfo
  {journal} {Nucl. Phys.}\ }\textbf {\bibinfo {volume} {A683}},\ \bibinfo
  {pages} {322} (\bibinfo {year} {2001})},\ \Eprint
  {http://arxiv.org/abs/nucl-th/0006011} {arXiv:nucl-th/0006011 [nucl-th]}
  \BibitemShut {NoStop}%
\bibitem [{\citenamefont {Feldmann}(2000)}]{Feldmann:1999uf}%
  \BibitemOpen
  \bibfield  {author} {\bibinfo {author} {\bibfnamefont {T.}~\bibnamefont
  {Feldmann}},\ }\href {\doibase 10.1142/S0217751X00000082} {\bibfield
  {journal} {\bibinfo  {journal} {Int. J. Mod. Phys.}\ }\textbf {\bibinfo
  {volume} {A15}},\ \bibinfo {pages} {159} (\bibinfo {year} {2000})},\ \Eprint
  {http://arxiv.org/abs/hep-ph/9907491} {arXiv:hep-ph/9907491 [hep-ph]}
  \BibitemShut {NoStop}%
\bibitem [{\citenamefont {Benayoun}\ \emph {et~al.}(1999)\citenamefont
  {Benayoun}, \citenamefont {DelBuono}, \citenamefont {Eidelman}, \citenamefont
  {Ivanchenko},\ and\ \citenamefont {O'Connell}}]{Benayoun:1999fv}%
  \BibitemOpen
  \bibfield  {author} {\bibinfo {author} {\bibfnamefont {M.}~\bibnamefont
  {Benayoun}}, \bibinfo {author} {\bibfnamefont {L.}~\bibnamefont {DelBuono}},
  \bibinfo {author} {\bibfnamefont {S.}~\bibnamefont {Eidelman}}, \bibinfo
  {author} {\bibfnamefont {V.~N.}\ \bibnamefont {Ivanchenko}}, \ and\ \bibinfo
  {author} {\bibfnamefont {H.~B.}\ \bibnamefont {O'Connell}},\ }\href {\doibase
  10.1103/PhysRevD.59.114027} {\bibfield  {journal} {\bibinfo  {journal} {Phys.
  Rev.}\ }\textbf {\bibinfo {volume} {D59}},\ \bibinfo {pages} {114027}
  (\bibinfo {year} {1999})},\ \Eprint {http://arxiv.org/abs/hep-ph/9902326}
  {arXiv:hep-ph/9902326 [hep-ph]} \BibitemShut {NoStop}%
\bibitem [{\citenamefont {Patrignani}\ \emph {et~al.}(2016)\citenamefont
  {Patrignani} \emph {et~al.}}]{Patrignani:2016xqp}%
  \BibitemOpen
  \bibfield  {author} {\bibinfo {author} {\bibfnamefont {C.}~\bibnamefont
  {Patrignani}} \emph {et~al.} (\bibinfo {collaboration} {Particle Data
  Group}),\ }\href {\doibase 10.1088/1674-1137/40/10/100001} {\bibfield
  {journal} {\bibinfo  {journal} {Chin. Phys.}\ }\textbf {\bibinfo {volume}
  {C40}},\ \bibinfo {pages} {100001} (\bibinfo {year} {2016})}\BibitemShut
  {NoStop}%
\bibitem [{\citenamefont {Williams}(2007)}]{Williams:2007zzg}%
  \BibitemOpen
  \bibfield  {author} {\bibinfo {author} {\bibfnamefont {M.}~\bibnamefont
  {Williams}},\ }\emph {\bibinfo {title} {{Measurement of differential cross
  sections and spin density matrix elements along with a partial wave analysis
  for $\gamma p \to p \omega$ using CLAS at Jefferson Lab}}},\ \href@noop {}
  {Ph.D. thesis},\ \bibinfo  {school} {Carnegie Mellon U.} (\bibinfo {year}
  {2007})\BibitemShut {NoStop}%
\bibitem [{\citenamefont {Donnachie}\ and\ \citenamefont
  {Landshoff}(1992)}]{Donnachie:1992ny}%
  \BibitemOpen
  \bibfield  {author} {\bibinfo {author} {\bibfnamefont {A.}~\bibnamefont
  {Donnachie}}\ and\ \bibinfo {author} {\bibfnamefont {P.~V.}\ \bibnamefont
  {Landshoff}},\ }\href {\doibase 10.1016/0370-2693(92)90832-O} {\bibfield
  {journal} {\bibinfo  {journal} {Phys. Lett.}\ }\textbf {\bibinfo {volume}
  {B296}},\ \bibinfo {pages} {227} (\bibinfo {year} {1992})},\ \Eprint
  {http://arxiv.org/abs/hep-ph/9209205} {arXiv:hep-ph/9209205 [hep-ph]}
  \BibitemShut {NoStop}%
\bibitem [{\citenamefont {Ewerz}\ \emph {et~al.}(2016)\citenamefont {Ewerz},
  \citenamefont {Lebiedowicz}, \citenamefont {Nachtmann},\ and\ \citenamefont
  {Szczurek}}]{Ewerz:2016onn}%
  \BibitemOpen
  \bibfield  {author} {\bibinfo {author} {\bibfnamefont {C.}~\bibnamefont
  {Ewerz}}, \bibinfo {author} {\bibfnamefont {P.}~\bibnamefont {Lebiedowicz}},
  \bibinfo {author} {\bibfnamefont {O.}~\bibnamefont {Nachtmann}}, \ and\
  \bibinfo {author} {\bibfnamefont {A.}~\bibnamefont {Szczurek}},\ }\href
  {\doibase 10.1016/j.physletb.2016.10.064} {\bibfield  {journal} {\bibinfo
  {journal} {Phys. Lett. B}\ }\textbf {\bibinfo {volume} {763}},\ \bibinfo
  {pages} {382} (\bibinfo {year} {2016})},\ \Eprint
  {http://arxiv.org/abs/1606.08067} {arXiv:1606.08067 [hep-ph]} \BibitemShut
  {NoStop}%
\bibitem [{\citenamefont {Laget}\ and\ \citenamefont
  {Mendez-Galain}(1995)}]{Laget:1994ba}%
  \BibitemOpen
  \bibfield  {author} {\bibinfo {author} {\bibfnamefont {J.~M.}\ \bibnamefont
  {Laget}}\ and\ \bibinfo {author} {\bibfnamefont {R.}~\bibnamefont
  {Mendez-Galain}},\ }\href {\doibase 10.1016/0375-9474(94)00428-P} {\bibfield
  {journal} {\bibinfo  {journal} {Nucl. Phys.}\ }\textbf {\bibinfo {volume}
  {A581}},\ \bibinfo {pages} {397} (\bibinfo {year} {1995})}\BibitemShut
  {NoStop}%
\bibitem [{\citenamefont {Pichowsky}\ and\ \citenamefont
  {Lee}(1997)}]{Pichowsky:1996tn}%
  \BibitemOpen
  \bibfield  {author} {\bibinfo {author} {\bibfnamefont {M.~A.}\ \bibnamefont
  {Pichowsky}}\ and\ \bibinfo {author} {\bibfnamefont {T.~S.~H.}\ \bibnamefont
  {Lee}},\ }\href {\doibase 10.1103/PhysRevD.56.1644} {\bibfield  {journal}
  {\bibinfo  {journal} {Phys. Rev.}\ }\textbf {\bibinfo {volume} {D56}},\
  \bibinfo {pages} {1644} (\bibinfo {year} {1997})},\ \Eprint
  {http://arxiv.org/abs/nucl-th/9612049} {arXiv:nucl-th/9612049 [nucl-th]}
  \BibitemShut {NoStop}%
\bibitem [{\citenamefont {Nachtmann}(2004)}]{Nachtmann:2003ik}%
  \BibitemOpen
  \bibfield  {author} {\bibinfo {author} {\bibfnamefont {O.}~\bibnamefont
  {Nachtmann}},\ }in\ \href {\doibase 10.1142/9789812702722_0023} {\emph
  {\bibinfo {booktitle} {{Proceedings, Ringberg Workshop on New Trends in HERA
  Physics 2003: Ringberg Castle, Tegernsee, Germany, September 28-October 3,
  2003}}}}\ (\bibinfo {year} {2004})\ pp.\ \bibinfo {pages} {253--267},\
  \Eprint {http://arxiv.org/abs/hep-ph/0312279} {arXiv:hep-ph/0312279 [hep-ph]}
  \BibitemShut {NoStop}%
\bibitem [{\citenamefont {Bolz}\ \emph {et~al.}(2015)\citenamefont {Bolz},
  \citenamefont {Ewerz}, \citenamefont {Maniatis}, \citenamefont {Nachtmann},
  \citenamefont {Sauter},\ and\ \citenamefont {Sch\"oning}}]{Bolz:2014mya}%
  \BibitemOpen
  \bibfield  {author} {\bibinfo {author} {\bibfnamefont {A.}~\bibnamefont
  {Bolz}}, \bibinfo {author} {\bibfnamefont {C.}~\bibnamefont {Ewerz}},
  \bibinfo {author} {\bibfnamefont {M.}~\bibnamefont {Maniatis}}, \bibinfo
  {author} {\bibfnamefont {O.}~\bibnamefont {Nachtmann}}, \bibinfo {author}
  {\bibfnamefont {M.}~\bibnamefont {Sauter}}, \ and\ \bibinfo {author}
  {\bibfnamefont {A.}~\bibnamefont {Sch\"oning}},\ }\href {\doibase
  10.1007/JHEP01(2015)151} {\bibfield  {journal} {\bibinfo  {journal} {JHEP}\
  }\textbf {\bibinfo {volume} {01}},\ \bibinfo {pages} {151} (\bibinfo {year}
  {2015})},\ \Eprint {http://arxiv.org/abs/1409.8483} {arXiv:1409.8483
  [hep-ph]} \BibitemShut {NoStop}%
\bibitem [{\citenamefont {Lebiedowicz}\ \emph {et~al.}(2018)\citenamefont
  {Lebiedowicz}, \citenamefont {Nachtmann},\ and\ \citenamefont
  {Szczurek}}]{Lebiedowicz:2018eui}%
  \BibitemOpen
  \bibfield  {author} {\bibinfo {author} {\bibfnamefont {P.}~\bibnamefont
  {Lebiedowicz}}, \bibinfo {author} {\bibfnamefont {O.}~\bibnamefont
  {Nachtmann}}, \ and\ \bibinfo {author} {\bibfnamefont {A.}~\bibnamefont
  {Szczurek}},\ }\href {\doibase 10.1103/PhysRevD.98.014001} {\bibfield
  {journal} {\bibinfo  {journal} {Phys. Rev. D}\ }\textbf {\bibinfo {volume}
  {98}},\ \bibinfo {pages} {014001} (\bibinfo {year} {2018})},\ \Eprint
  {http://arxiv.org/abs/1804.04706} {arXiv:1804.04706 [hep-ph]} \BibitemShut
  {NoStop}%
\bibitem [{\citenamefont {Lebiedowicz}\ \emph {et~al.}(2020)\citenamefont
  {Lebiedowicz}, \citenamefont {Nachtmann},\ and\ \citenamefont
  {Szczurek}}]{Lebiedowicz:2019boz}%
  \BibitemOpen
  \bibfield  {author} {\bibinfo {author} {\bibfnamefont {P.}~\bibnamefont
  {Lebiedowicz}}, \bibinfo {author} {\bibfnamefont {O.}~\bibnamefont
  {Nachtmann}}, \ and\ \bibinfo {author} {\bibfnamefont {A.}~\bibnamefont
  {Szczurek}},\ }\href {\doibase 10.1103/PhysRevD.101.094012} {\bibfield
  {journal} {\bibinfo  {journal} {Phys. Rev. D}\ }\textbf {\bibinfo {volume}
  {101}},\ \bibinfo {pages} {094012} (\bibinfo {year} {2020})},\ \Eprint
  {http://arxiv.org/abs/1911.01909} {arXiv:1911.01909 [hep-ph]} \BibitemShut
  {NoStop}%
\bibitem [{\citenamefont {Williams}\ \emph {et~al.}(2009)\citenamefont
  {Williams} \emph {et~al.}}]{Williams:2009ab}%
  \BibitemOpen
  \bibfield  {author} {\bibinfo {author} {\bibfnamefont {M.}~\bibnamefont
  {Williams}} \emph {et~al.} (\bibinfo {collaboration} {CLAS}),\ }\href
  {\doibase 10.1103/PhysRevC.80.065208} {\bibfield  {journal} {\bibinfo
  {journal} {Phys. Rev.}\ }\textbf {\bibinfo {volume} {C80}},\ \bibinfo {pages}
  {065208} (\bibinfo {year} {2009})},\ \Eprint {http://arxiv.org/abs/0908.2910}
  {arXiv:0908.2910 [nucl-ex]} \BibitemShut {NoStop}%
\bibitem [{\citenamefont {Dey}\ \emph {et~al.}(2014)\citenamefont {Dey},
  \citenamefont {Meyer}, \citenamefont {Bellis},\ and\ \citenamefont
  {Williams}}]{Dey:2014tfa}%
  \BibitemOpen
  \bibfield  {author} {\bibinfo {author} {\bibfnamefont {B.}~\bibnamefont
  {Dey}}, \bibinfo {author} {\bibfnamefont {C.~A.}\ \bibnamefont {Meyer}},
  \bibinfo {author} {\bibfnamefont {M.}~\bibnamefont {Bellis}}, \ and\ \bibinfo
  {author} {\bibfnamefont {M.}~\bibnamefont {Williams}} (\bibinfo
  {collaboration} {CLAS}),\ }\href {\doibase 10.1103/PhysRevC.90.019901,
  10.1103/PhysRevC.89.055208} {\bibfield  {journal} {\bibinfo  {journal} {Phys.
  Rev.}\ }\textbf {\bibinfo {volume} {C89}},\ \bibinfo {pages} {055208}
  (\bibinfo {year} {2014})},\ \bibinfo {note} {[Addendum: Phys.
  Rev.C90,no.1,019901(2014)]},\ \Eprint {http://arxiv.org/abs/1403.2110}
  {arXiv:1403.2110 [nucl-ex]} \BibitemShut {NoStop}%
\bibitem [{\citenamefont {Ballam}\ \emph {et~al.}(1973)\citenamefont {Ballam}
  \emph {et~al.}}]{Ballam:1972eq}%
  \BibitemOpen
  \bibfield  {author} {\bibinfo {author} {\bibfnamefont {J.}~\bibnamefont
  {Ballam}} \emph {et~al.},\ }\href {\doibase 10.1103/PhysRevD.7.3150}
  {\bibfield  {journal} {\bibinfo  {journal} {Phys. Rev.}\ }\textbf {\bibinfo
  {volume} {D7}},\ \bibinfo {pages} {3150} (\bibinfo {year}
  {1973})}\BibitemShut {NoStop}%
\bibitem [{\citenamefont {Barber}\ \emph {et~al.}(1984)\citenamefont {Barber}
  \emph {et~al.}}]{Barber:1985fr}%
  \BibitemOpen
  \bibfield  {author} {\bibinfo {author} {\bibfnamefont {D.~P.}\ \bibnamefont
  {Barber}} \emph {et~al.} (\bibinfo {collaboration} {LAMP2 Group}),\ }\href
  {\doibase 10.1007/BF01452559} {\bibfield  {journal} {\bibinfo  {journal} {Z.
  Phys.}\ }\textbf {\bibinfo {volume} {C26}},\ \bibinfo {pages} {343} (\bibinfo
  {year} {1984})}\BibitemShut {NoStop}%
\bibitem [{\citenamefont {Busenitz}\ \emph {et~al.}(1989)\citenamefont
  {Busenitz} \emph {et~al.}}]{Busenitz:1989gq}%
  \BibitemOpen
  \bibfield  {author} {\bibinfo {author} {\bibfnamefont {J.}~\bibnamefont
  {Busenitz}} \emph {et~al.},\ }\href {\doibase 10.1103/PhysRevD.40.1}
  {\bibfield  {journal} {\bibinfo  {journal} {Phys. Rev. D}\ }\textbf {\bibinfo
  {volume} {40}},\ \bibinfo {pages} {1} (\bibinfo {year} {1989})}\BibitemShut
  {NoStop}%
\bibitem [{\citenamefont {Derrick}\ \emph
  {et~al.}(1996{\natexlab{a}})\citenamefont {Derrick} \emph
  {et~al.}}]{Derrick:1996yt}%
  \BibitemOpen
  \bibfield  {author} {\bibinfo {author} {\bibfnamefont {M.}~\bibnamefont
  {Derrick}} \emph {et~al.} (\bibinfo {collaboration} {ZEUS}),\ }\href
  {\doibase 10.1007/s002880050297} {\bibfield  {journal} {\bibinfo  {journal}
  {Z. Phys.}\ }\textbf {\bibinfo {volume} {C73}},\ \bibinfo {pages} {73}
  (\bibinfo {year} {1996}{\natexlab{a}})},\ \Eprint
  {http://arxiv.org/abs/hep-ex/9608010} {arXiv:hep-ex/9608010 [hep-ex]}
  \BibitemShut {NoStop}%
\bibitem [{\citenamefont {Derrick}\ \emph
  {et~al.}(1996{\natexlab{b}})\citenamefont {Derrick} \emph
  {et~al.}}]{Derrick:1996af}%
  \BibitemOpen
  \bibfield  {author} {\bibinfo {author} {\bibfnamefont {M.}~\bibnamefont
  {Derrick}} \emph {et~al.} (\bibinfo {collaboration} {ZEUS}),\ }\href
  {\doibase 10.1016/0370-2693(96)00172-4} {\bibfield  {journal} {\bibinfo
  {journal} {Phys. Lett.}\ }\textbf {\bibinfo {volume} {B377}},\ \bibinfo
  {pages} {259} (\bibinfo {year} {1996}{\natexlab{b}})},\ \Eprint
  {http://arxiv.org/abs/hep-ex/9601009} {arXiv:hep-ex/9601009 [hep-ex]}
  \BibitemShut {NoStop}%
\bibitem [{\citenamefont {Breitweg}\ \emph {et~al.}(2000)\citenamefont
  {Breitweg} \emph {et~al.}}]{ZEUS:1999ptu}%
  \BibitemOpen
  \bibfield  {author} {\bibinfo {author} {\bibfnamefont {J.}~\bibnamefont
  {Breitweg}} \emph {et~al.} (\bibinfo {collaboration} {ZEUS}),\ }\href
  {\doibase 10.1007/s100520000374} {\bibfield  {journal} {\bibinfo  {journal}
  {Eur. Phys. J. C}\ }\textbf {\bibinfo {volume} {14}},\ \bibinfo {pages} {213}
  (\bibinfo {year} {2000})},\ \Eprint {http://arxiv.org/abs/hep-ex/9910038}
  {arXiv:hep-ex/9910038} \BibitemShut {NoStop}%
\bibitem [{\citenamefont {Aid}\ \emph {et~al.}(1995)\citenamefont {Aid} \emph
  {et~al.}}]{H1:1995hmw}%
  \BibitemOpen
  \bibfield  {author} {\bibinfo {author} {\bibfnamefont {S.}~\bibnamefont
  {Aid}} \emph {et~al.} (\bibinfo {collaboration} {H1}),\ }\href {\doibase
  10.1007/s002880050003} {\bibfield  {journal} {\bibinfo  {journal} {Z. Phys.
  C}\ }\textbf {\bibinfo {volume} {69}},\ \bibinfo {pages} {27} (\bibinfo
  {year} {1995})},\ \Eprint {http://arxiv.org/abs/hep-ex/9509001}
  {arXiv:hep-ex/9509001} \BibitemShut {NoStop}%
\bibitem [{\citenamefont {Chekanov}\ \emph {et~al.}(2002)\citenamefont
  {Chekanov} \emph {et~al.}}]{ZEUS:2001wan}%
  \BibitemOpen
  \bibfield  {author} {\bibinfo {author} {\bibfnamefont {S.}~\bibnamefont
  {Chekanov}} \emph {et~al.} (\bibinfo {collaboration} {ZEUS}),\ }\href
  {\doibase 10.1016/S0550-3213(02)00068-8} {\bibfield  {journal} {\bibinfo
  {journal} {Nucl. Phys. B}\ }\textbf {\bibinfo {volume} {627}},\ \bibinfo
  {pages} {3} (\bibinfo {year} {2002})},\ \Eprint
  {http://arxiv.org/abs/hep-ex/0202034} {arXiv:hep-ex/0202034} \BibitemShut
  {NoStop}%
\bibitem [{\citenamefont {Atkinson}\ \emph {et~al.}(1984)\citenamefont
  {Atkinson} \emph {et~al.}}]{OmegaPhoton:1983huz}%
  \BibitemOpen
  \bibfield  {author} {\bibinfo {author} {\bibfnamefont {M.}~\bibnamefont
  {Atkinson}} \emph {et~al.} (\bibinfo {collaboration} {Omega Photon,
  Bonn-CERN-Glasgow-Lancaster-Manchester-Paris-Rutherford-Sheffield}),\ }\href
  {\doibase 10.1016/0550-3213(84)90304-3} {\bibfield  {journal} {\bibinfo
  {journal} {Nucl. Phys. B}\ }\textbf {\bibinfo {volume} {231}},\ \bibinfo
  {pages} {15} (\bibinfo {year} {1984})}\BibitemShut {NoStop}%
\bibitem [{\citenamefont {Dietz}\ \emph {et~al.}(2015)\citenamefont {Dietz}
  \emph {et~al.}}]{CBELSATAPS:2015wwn}%
  \BibitemOpen
  \bibfield  {author} {\bibinfo {author} {\bibfnamefont {F.}~\bibnamefont
  {Dietz}} \emph {et~al.} (\bibinfo {collaboration} {CBELSA/TAPS}),\ }\href
  {\doibase 10.1140/epja/i2015-15006-3} {\bibfield  {journal} {\bibinfo
  {journal} {Eur. Phys. J. A}\ }\textbf {\bibinfo {volume} {51}},\ \bibinfo
  {pages} {6} (\bibinfo {year} {2015})}\BibitemShut {NoStop}%
\bibitem [{\citenamefont {Struczinski}\ \emph {et~al.}(1976)\citenamefont
  {Struczinski} \emph {et~al.}}]{Aachen-Hamburg-Heidelberg-Munich:1975jed}%
  \BibitemOpen
  \bibfield  {author} {\bibinfo {author} {\bibfnamefont {W.}~\bibnamefont
  {Struczinski}} \emph {et~al.} (\bibinfo {collaboration}
  {Aachen-Hamburg-Heidelberg-Munich}),\ }\href {\doibase
  10.1016/0550-3213(76)90123-1} {\bibfield  {journal} {\bibinfo  {journal}
  {Nucl. Phys. B}\ }\textbf {\bibinfo {volume} {108}},\ \bibinfo {pages} {45}
  (\bibinfo {year} {1976})}\BibitemShut {NoStop}%
\bibitem [{\citenamefont {Egloff}\ \emph
  {et~al.}(1979{\natexlab{a}})\citenamefont {Egloff} \emph
  {et~al.}}]{Egloff:1979xg}%
  \BibitemOpen
  \bibfield  {author} {\bibinfo {author} {\bibfnamefont {R.~M.}\ \bibnamefont
  {Egloff}} \emph {et~al.},\ }\href {\doibase 10.1103/PhysRevLett.43.1545}
  {\bibfield  {journal} {\bibinfo  {journal} {Phys. Rev. Lett.}\ }\textbf
  {\bibinfo {volume} {43}},\ \bibinfo {pages} {1545} (\bibinfo {year}
  {1979}{\natexlab{a}})},\ \bibinfo {note} {[Erratum: Phys.Rev.Lett. 44, 690
  (1980)]}\BibitemShut {NoStop}%
\bibitem [{\citenamefont {Strakovsky}\ \emph {et~al.}(2015)\citenamefont
  {Strakovsky} \emph {et~al.}}]{Strakovsky:2014wja}%
  \BibitemOpen
  \bibfield  {author} {\bibinfo {author} {\bibfnamefont {I.~I.}\ \bibnamefont
  {Strakovsky}} \emph {et~al.},\ }\href {\doibase 10.1103/PhysRevC.91.045207}
  {\bibfield  {journal} {\bibinfo  {journal} {Phys. Rev. C}\ }\textbf {\bibinfo
  {volume} {91}},\ \bibinfo {pages} {045207} (\bibinfo {year} {2015})},\
  \Eprint {http://arxiv.org/abs/1407.3465} {arXiv:1407.3465 [nucl-ex]}
  \BibitemShut {NoStop}%
\bibitem [{\citenamefont {Crouch}\ \emph {et~al.}(1967)\citenamefont {Crouch}
  \emph
  {et~al.}}]{BrownHarvardMITPadovaWeizmannInstituteBubbleChamberGroup:1967zz}%
  \BibitemOpen
  \bibfield  {author} {\bibinfo {author} {\bibfnamefont {H.~R.}\ \bibnamefont
  {Crouch}, \bibfnamefont {Jr.}} \emph {et~al.},\ }\href {\doibase
  10.1103/PhysRev.155.1468} {\bibfield  {journal} {\bibinfo  {journal} {Phys.
  Rev.}\ }\textbf {\bibinfo {volume} {155}},\ \bibinfo {pages} {1468} (\bibinfo
  {year} {1967})}\BibitemShut {NoStop}%
\bibitem [{\citenamefont {Davier}\ \emph {et~al.}(1970)\citenamefont {Davier},
  \citenamefont {Derado}, \citenamefont {Drickey}, \citenamefont {Fries},
  \citenamefont {Mozley}, \citenamefont {Odian}, \citenamefont {Villa},\ and\
  \citenamefont {Yount}}]{Davier:1969nx}%
  \BibitemOpen
  \bibfield  {author} {\bibinfo {author} {\bibfnamefont {M.}~\bibnamefont
  {Davier}}, \bibinfo {author} {\bibfnamefont {I.}~\bibnamefont {Derado}},
  \bibinfo {author} {\bibfnamefont {D.~J.}\ \bibnamefont {Drickey}}, \bibinfo
  {author} {\bibfnamefont {D.~E.~C.}\ \bibnamefont {Fries}}, \bibinfo {author}
  {\bibfnamefont {R.~F.}\ \bibnamefont {Mozley}}, \bibinfo {author}
  {\bibfnamefont {A.}~\bibnamefont {Odian}}, \bibinfo {author} {\bibfnamefont
  {F.}~\bibnamefont {Villa}}, \ and\ \bibinfo {author} {\bibfnamefont
  {D.}~\bibnamefont {Yount}},\ }\href {\doibase 10.1103/PhysRevD.1.790}
  {\bibfield  {journal} {\bibinfo  {journal} {Phys. Rev. D}\ }\textbf {\bibinfo
  {volume} {1}},\ \bibinfo {pages} {790} (\bibinfo {year} {1970})}\BibitemShut
  {NoStop}%
\bibitem [{\citenamefont {Breakstone}\ \emph {et~al.}(1981)\citenamefont
  {Breakstone}, \citenamefont {Cheng}, \citenamefont {Dorfan}, \citenamefont
  {Grillo}, \citenamefont {Heusch}, \citenamefont {Palladino}, \citenamefont
  {Schalk}, \citenamefont {Seiden},\ and\ \citenamefont
  {Smith}}]{Breakstone:1981wk}%
  \BibitemOpen
  \bibfield  {author} {\bibinfo {author} {\bibfnamefont {A.~M.}\ \bibnamefont
  {Breakstone}}, \bibinfo {author} {\bibfnamefont {D.~C.}\ \bibnamefont
  {Cheng}}, \bibinfo {author} {\bibfnamefont {D.~E.}\ \bibnamefont {Dorfan}},
  \bibinfo {author} {\bibfnamefont {A.~A.}\ \bibnamefont {Grillo}}, \bibinfo
  {author} {\bibfnamefont {C.~A.}\ \bibnamefont {Heusch}}, \bibinfo {author}
  {\bibfnamefont {V.}~\bibnamefont {Palladino}}, \bibinfo {author}
  {\bibfnamefont {T.}~\bibnamefont {Schalk}}, \bibinfo {author} {\bibfnamefont
  {A.}~\bibnamefont {Seiden}}, \ and\ \bibinfo {author} {\bibfnamefont {D.~B.}\
  \bibnamefont {Smith}},\ }\href {\doibase 10.1103/PhysRevLett.47.1782}
  {\bibfield  {journal} {\bibinfo  {journal} {Phys. Rev. Lett.}\ }\textbf
  {\bibinfo {volume} {47}},\ \bibinfo {pages} {1782} (\bibinfo {year}
  {1981})}\BibitemShut {NoStop}%
\bibitem [{\citenamefont {Barth}\ \emph {et~al.}(2003)\citenamefont {Barth}
  \emph {et~al.}}]{Barth:2003kv}%
  \BibitemOpen
  \bibfield  {author} {\bibinfo {author} {\bibfnamefont {J.}~\bibnamefont
  {Barth}} \emph {et~al.},\ }\href {\doibase 10.1140/epja/i2003-10061-y}
  {\bibfield  {journal} {\bibinfo  {journal} {Eur. Phys. J. A}\ }\textbf
  {\bibinfo {volume} {18}},\ \bibinfo {pages} {117} (\bibinfo {year}
  {2003})}\BibitemShut {NoStop}%
\bibitem [{\citenamefont {Barber}\ \emph {et~al.}(1982)\citenamefont {Barber}
  \emph {et~al.}}]{Barber:1981fj}%
  \BibitemOpen
  \bibfield  {author} {\bibinfo {author} {\bibfnamefont {D.~P.}\ \bibnamefont
  {Barber}} \emph {et~al.},\ }\href {\doibase 10.1007/BF01475724} {\bibfield
  {journal} {\bibinfo  {journal} {Z. Phys. C}\ }\textbf {\bibinfo {volume}
  {12}},\ \bibinfo {pages} {1} (\bibinfo {year} {1982})}\BibitemShut {NoStop}%
\bibitem [{\citenamefont {Atkinson}\ \emph {et~al.}(1985)\citenamefont
  {Atkinson} \emph {et~al.}}]{OmegaPhoton:1984eqn}%
  \BibitemOpen
  \bibfield  {author} {\bibinfo {author} {\bibfnamefont {M.}~\bibnamefont
  {Atkinson}} \emph {et~al.} (\bibinfo {collaboration} {Omega Photon}),\ }\href
  {\doibase 10.1007/BF01556612} {\bibfield  {journal} {\bibinfo  {journal} {Z.
  Phys. C}\ }\textbf {\bibinfo {volume} {27}},\ \bibinfo {pages} {233}
  (\bibinfo {year} {1985})}\BibitemShut {NoStop}%
\bibitem [{\citenamefont {Egloff}\ \emph
  {et~al.}(1979{\natexlab{b}})\citenamefont {Egloff} \emph
  {et~al.}}]{Egloff:1979mg}%
  \BibitemOpen
  \bibfield  {author} {\bibinfo {author} {\bibfnamefont {R.~M.}\ \bibnamefont
  {Egloff}} \emph {et~al.},\ }\href {\doibase 10.1103/PhysRevLett.43.657}
  {\bibfield  {journal} {\bibinfo  {journal} {Phys. Rev. Lett.}\ }\textbf
  {\bibinfo {volume} {43}},\ \bibinfo {pages} {657} (\bibinfo {year}
  {1979}{\natexlab{b}})}\BibitemShut {NoStop}%
\bibitem [{\citenamefont {Aston}\ \emph {et~al.}(1980)\citenamefont {Aston}
  \emph
  {et~al.}}]{Bonn-CERN-EcolePoly-Glasgow-Lancaster-Manchester-Orsay-Paris-Rutherford-Sheffield:1980dwy}%
  \BibitemOpen
  \bibfield  {author} {\bibinfo {author} {\bibfnamefont {D.}~\bibnamefont
  {Aston}} \emph {et~al.} (\bibinfo {collaboration} {Bonn-CERN-Ecole
  Poly-Glasgow-Lancaster-Manchester-Orsay-Paris-Rutherford-Sheffield}),\ }\href
  {\doibase 10.1016/0550-3213(80)90156-X} {\bibfield  {journal} {\bibinfo
  {journal} {Nucl. Phys. B}\ }\textbf {\bibinfo {volume} {172}},\ \bibinfo
  {pages} {1} (\bibinfo {year} {1980})}\BibitemShut {NoStop}%
\bibitem [{\citenamefont {Alekhin}\ \emph {et~al.}(1987)\citenamefont {Alekhin}
  \emph {et~al.}}]{HERAGroup:1987dng}%
  \BibitemOpen
  \bibfield  {author} {\bibinfo {author} {\bibfnamefont {S.~I.}\ \bibnamefont
  {Alekhin}} \emph {et~al.} (\bibinfo {collaboration} {HERA Group, COMPASS
  Group}),\ }\href@noop {} {\  (\bibinfo {year} {1987})}\BibitemShut {NoStop}%
\bibitem [{\citenamefont {Hand}(1963)}]{Hand:1963bb}%
  \BibitemOpen
  \bibfield  {author} {\bibinfo {author} {\bibfnamefont {L.~N.}\ \bibnamefont
  {Hand}},\ }\href {\doibase 10.1103/PhysRev.129.1834} {\bibfield  {journal}
  {\bibinfo  {journal} {Phys. Rev.}\ }\textbf {\bibinfo {volume} {129}},\
  \bibinfo {pages} {1834} (\bibinfo {year} {1963})}\BibitemShut {NoStop}%
\bibitem [{\citenamefont {Peskin}\ and\ \citenamefont
  {Schroeder}(1995)}]{Peskin:1995ev}%
  \BibitemOpen
  \bibfield  {author} {\bibinfo {author} {\bibfnamefont {M.~E.}\ \bibnamefont
  {Peskin}}\ and\ \bibinfo {author} {\bibfnamefont {D.~V.}\ \bibnamefont
  {Schroeder}},\ }\href@noop {} {\emph {\bibinfo {title} {{An Introduction to
  quantum field theory}}}}\ (\bibinfo  {publisher} {Addison-Wesley},\ \bibinfo
  {address} {Reading, USA},\ \bibinfo {year} {1995})\BibitemShut {NoStop}%
\bibitem [{\citenamefont {Binosi}\ and\ \citenamefont
  {Theussl}(2004)}]{Binosi:2003yf}%
  \BibitemOpen
  \bibfield  {author} {\bibinfo {author} {\bibfnamefont {D.}~\bibnamefont
  {Binosi}}\ and\ \bibinfo {author} {\bibfnamefont {L.}~\bibnamefont
  {Theussl}},\ }\href {\doibase 10.1016/j.cpc.2004.05.001} {\bibfield
  {journal} {\bibinfo  {journal} {Comput. Phys. Commun.}\ }\textbf {\bibinfo
  {volume} {161}},\ \bibinfo {pages} {76} (\bibinfo {year} {2004})},\ \Eprint
  {http://arxiv.org/abs/hep-ph/0309015} {arXiv:hep-ph/0309015} \BibitemShut
  {NoStop}%
\bibitem [{\citenamefont {Shtabovenko}\ \emph {et~al.}(2016)\citenamefont
  {Shtabovenko}, \citenamefont {Mertig},\ and\ \citenamefont
  {Orellana}}]{Shtabovenko:2016sxi}%
  \BibitemOpen
  \bibfield  {author} {\bibinfo {author} {\bibfnamefont {V.}~\bibnamefont
  {Shtabovenko}}, \bibinfo {author} {\bibfnamefont {R.}~\bibnamefont {Mertig}},
  \ and\ \bibinfo {author} {\bibfnamefont {F.}~\bibnamefont {Orellana}},\
  }\href {\doibase 10.1016/j.cpc.2016.06.008} {\bibfield  {journal} {\bibinfo
  {journal} {Comput. Phys. Commun.}\ }\textbf {\bibinfo {volume} {207}},\
  \bibinfo {pages} {432} (\bibinfo {year} {2016})},\ \Eprint
  {http://arxiv.org/abs/1601.01167} {arXiv:1601.01167 [hep-ph]} \BibitemShut
  {NoStop}%
\bibitem [{\citenamefont {Shtabovenko}\ \emph {et~al.}(2020)\citenamefont
  {Shtabovenko}, \citenamefont {Mertig},\ and\ \citenamefont
  {Orellana}}]{Shtabovenko:2020gxv}%
  \BibitemOpen
  \bibfield  {author} {\bibinfo {author} {\bibfnamefont {V.}~\bibnamefont
  {Shtabovenko}}, \bibinfo {author} {\bibfnamefont {R.}~\bibnamefont {Mertig}},
  \ and\ \bibinfo {author} {\bibfnamefont {F.}~\bibnamefont {Orellana}},\
  }\href {\doibase 10.1016/j.cpc.2020.107478} {\bibfield  {journal} {\bibinfo
  {journal} {Comput. Phys. Commun.}\ }\textbf {\bibinfo {volume} {256}},\
  \bibinfo {pages} {107478} (\bibinfo {year} {2020})},\ \Eprint
  {http://arxiv.org/abs/2001.04407} {arXiv:2001.04407 [hep-ph]} \BibitemShut
  {NoStop}%
\end{thebibliography}%

\end{document}